\documentclass[twoside,11pt]{article}
\pdfoutput=1

\usepackage{jmlr2e}
\usepackage[utf8]{inputenc} 
\usepackage{textcomp}
\usepackage{amsmath}
\usepackage{epstopdf}
\usepackage{xcolor}
\usepackage{comment}
\usepackage{caption}
\usepackage{subcaption}
\usepackage{framed}
\usepackage[noend]{algorithmic}[1]
\usepackage[english]{babel}

\newcommand{\cA}{\mathcal{A}}
\newcommand{\cC}{\mathcal{C}}

\newcommand{\cT}{\mathcal{T}}
\newcommand{\cZ}{\mathcal{Z}}

\newtheorem{notation}[theorem]{Notation}
\newtheorem{problem}[theorem]{Problem}

\makeatother
\allowdisplaybreaks

\ShortHeadings{HC selection for MCTS in Simultaneous Move Games }{ Kova\v{r}\'{i}k and Lis\'{y}}
\firstpageno{1}

\usepackage{tikz}
\usetikzlibrary{chains,scopes}

\newsavebox{\AntiL}

\savebox{\AntiL}{
\begin{tikzpicture}[node distance = 5mm and 1cm,start chain=going right, leaf/.style={on chain,join}, dot/.style={fill=black,circle,scale=0.5,on chain, join}]
{[start chain = trunk]
\node [dot] {};
{[start branch = 1 going below]
  \node [leaf]{$\frac{1}{2}$}; 
}
\node [dot]{};
{[start branch = 2 going below]
  \node [leaf]{$\frac{D-2}{2(D-1)}$}; 
}
\node [dot]{};
{[start branch = 3 going below]
  \node [leaf]{$\frac{D-3}{2(D-1)}$}; 
}
\node [dot]{};
{[start branch = 2 going below]
  \node [leaf,minimum size = 7mm]{$\dots$}; 
}
\node [dot]{};
{[start branch = 4 going below]
  \node [leaf,minimum size = 6mm]{$0$}; 
}
\node [leaf]{1};
}
\end{tikzpicture}
}

\newsavebox{\LinBound}

\savebox{\LinBound}{
\begin{tikzpicture}[node distance = 5mm and 1cm,start chain=going right, leaf/.style={on chain,join}, dot/.style={fill=black,circle,scale=0.5,on chain, join}]
{[start chain = trunk]
\node [dot] {};
{[start branch = 1 going above]
  \node [leaf]{$u_4$};
 }
{[start branch = 1 going below]
  \node [leaf]{$0$};  
}
\node [dot]{};
{[start branch = 2 going above]
  \node [leaf]{$u_3$};
 }
{[start branch = 2 going below]
  \node [leaf]{$0$}; 
}
\node [dot]{};
{[start branch = 3 going above]
  \node [leaf]{$u_2$};
 }
{[start branch = 3 going below]
  \node [leaf]{$0$}; 
}
\node [dot]{};
{[start branch = 4 going above]
  \node [leaf]{$u_1$};
 }
{[start branch = 4 going below]
  \node [leaf]{$0$}; 
}
\node [dot]{};
{[start branch = 5 going below]
  \node [leaf,minimum size = 6mm]{$0$}; 
}
\node [leaf]{1};
}
\end{tikzpicture}
}

\begin{document}

\title{Analysis of Hannan Consistent Selection for Monte\\ Carlo Tree Search in Simultaneous Move Games}

\author{\name Vojt\v{e}ch Kova\v{r}\'{i}k \email vojta.kovarik@gmail.com \\
       \name Viliam Lis\'{y} \email viliam.lisy@agents.fel.cvut.cz\\ 
       \addr Agent Technology Center, Department of Computer Science\\
       Faculty of Electrical Engineering, Czech Technical University in Prague\\
       Zikova 1903/4, Prague 6, 166 36, Czech Republic}

\editor{Editor name}

\maketitle

\begin{abstract}
Monte Carlo Tree Search (MCTS) has recently been successfully used to create strategies for playing imperfect-information games. Despite its popularity, there are no theoretic results that guarantee its convergence to a well-defined solution, such as Nash equilibrium, in these games. We partially fill this gap by analysing MCTS in the class of zero-sum extensive-form games with simultaneous moves but otherwise perfect information. The lack of information about the opponent's concurrent moves already causes that optimal strategies may require randomization.
We present theoretic as well as empirical investigation of the speed and quality of convergence of these algorithms to the Nash equilibria. Primarily, we show that after minor technical modifications, MCTS based on any (approximately) Hannan consistent selection function always converges to an (approximate) subgame perfect Nash equilibrium. Without these modifications, Hannan consistency is not sufficient to ensure such convergence and the selection function must satisfy additional properties, which empirically hold for the most common Hannan consistent algorithms.
\end{abstract}

\begin{keywords}
 Nash Equilibrium, Extensive Form Games, Simultaneous Moves, Zero Sum, Hannan Consistency 
\end{keywords}

\section{Introduction\label{sec: Intro}}

Monte Carlo tree search (MCTS) is a very popular algorithm
which recently caused a significant jump in performance of the state-of-the-art
solvers for many perfect information problems, such as the game of
Go \citep{gelly2011monte}, or domain-independent planning under uncertainty \citep{keller2012prost}. The main idea of Monte Carlo tree search is running a large number of
randomized simulations of the problem and learning the best actions
to choose based on this data. It generally uses the earlier simulations to create statistics that help guiding the latter simulations
to more important parts of the search space. After the success in
domains with perfect information, the following research applied the
principles of MCTS also to games with imperfect information, such
as an imperfect information variant of Chess \citep{Ciancarini2010}, or
imperfect information board games \citep{powley2014information,nijssen2012monte}. The same type of algorithms
can also be applied to real-world domains, such as robotics \citep{lisy2012anytime} or network security \citep{lisy2012ecai}.

While all these applications show that MCTS is a promising technique
also for playing imperfect information games, very little research
has been devoted to understanding the fundamental principles behind
the success of these methods in practice. In this paper, we aim to
partially fill this gap. We focus on the simplest class of imperfect
information games, which are games with simultaneous moves, but otherwise
perfect information. MCTS algorithms has been successfully applied
to many games form this class, including card games \citep{teytaud2011upper,lanctot2013goof},
variants of simple computer games \citep{Perick12Comparison}, or in
the most successful agents for General Game Playing \citep{Finnsson08}.
This class of games is simpler than the generic imperfect
information games, but it already includes one of the most fundamental
complication caused by the imperfect information, which is the need
for randomized (mixed) strategies. 
This can be demonstrated on the well-known game of Rock-Paper-Scissors.
Any deterministic strategy for playing the game can be easily exploited by the opponent and the optimal strategy is to randomize uniformly over all actions.

Game theory provides fundamental concepts and results that describe the optimal behaviour in games. In zero-sum simultaneous-move games, the optimal strategy is a subgame perfect Nash equilibrium. For each possible situation in the game, it prescribes a strategy, which is optimal in
several aspects. It is a strategy that gains the highest expected reward
against its worst opponent, even if the opponent knows the strategy
played by the player in advance. Moreover, in the zero-sum setting,
even if the opponent does not play rationally, the strategy still
guarantees at least the reward it would gain against a rational opponent.

While computing a Nash equilibrium in a zero-sum game is a polynomially
solvable problem \citep{koller92the}, the games where MCTS is commonly
applied are too large to allow even representing the Nash equilibrium
strategy explicitly, which is generally required by exact algorithms
for computing NE. Therefore, we cannot hope that MCTS will compute
the equilibrium strategy for the complete game in the given time and
space, but we still argue that eventual convergence to the Nash equilibrium,
or some other well understood game theoretic concept, is a desirable
property of MCTS algorithms in this class of games: first, an algorithm which converges to NE is more suitable in the anytime setting, where MCTS algorithms are most commonly used.
The more time it has available for the computation, the closer it will be to the optimal solution.
This does not always hold for MCTS algorithms in this class of games, which can stabilize in a fixed distance from an equilibrium \citep{Shafiei09, Ponsen11Computing} or even start diverging at some point \citep{lanctot2013goof}.
Second, if the game is close to its end, it may already be small
enough for an algorithm with guaranteed convergence to converge to almost exact NE and play optimally.
Non-convergent MCTS algorithms can exhibit various pathologies in these situations.
Third, understanding the fundamental game theoretic properties
of the strategies the algorithm converge to can lead to developing
better variants of MCTS for this class of games.

\subsection{Contributions}
We focus on two-player zero-sum extensive form games with simultaneous moves but otherwise perfect information. We denote the standard MCTS algorithm applied in this setting as SM-MCTS.
We present a modified SM-MCTS algorithm (SM-MCTS-A), which updates the selection functions by averages of the sampled values, rather than the current values themselves.
We show that SM-MCTS-A combined with any (approximate) Hannan consistent (HC) selection function with guaranteed exploration converges to (approximate) subgame-perfect Nash equilibrium in this class of games.
We present bounds on the relation of the convergence rate and the eventual distance from the Nash equilibrium on the main parameters of the games and the selection functions.
We then highlight the fact that without the ``-A'' modification, Hannan consistency of the selection function is not sufficient for a similar result.
We present a Hannan consistent selection function that causes the standard SM-MCTS algorithm to converge to a solution far from the equilibrium.
However, additional requirements on the selection function used in SM-MCTS can guarantee the convergence.
As an example, we define the property of having unbiased payoff observations (UPO), and show that it is a sufficient condition for convergence of SM-MCTS. We then empirically confirm that the two commonly used Hannan consistent algorithms, Exp3 and regret matching, satisfy this property, thus justifying their use in practice.
We further show that the empirical speed of convergence as well as the eventual distance from the equilibrium is typically much better than the guarantees given by the presented theory.
We empirically show that SM-MCTS generally converges to the same equilibrium as SM-MCTS-A, but does it slightly faster. 

We also give theoretical grounds for some practical improvements, which are often used with SM-MCTS, but have not been formally justified. These include removal of exploration samples from the resulting strategy and the use of average strategy instead of empirical frequencies of action choices.
All presented theoretic results trivially apply also to perfect information games with sequential moves.

\subsection{Article outline}
In Section~\ref{sec: background} we describe simultaneous-move games, the standard SM-MCTS algorithm and its modification SM-MCTS-A. We follow with the multi-armed bandit problem and show how it applies in our setting. Lastly we recall the definition of Hannan consistency and explain Exp3 and regret matching, two of the common Hannan consistent bandit algorithms. In Section \ref{sec: convergence}, we present the main theoretical results. First, we consider the modified SM-MCTS-A algorithm and present the asymptotic and finite time bounds on its convergence rate.
We follow by defining the unbiased payoff observations property and proving the convergence of SM-MCTS based on HC selection functions with this property.
In Section \ref{sec:Counterexample} we provide a counterexample showing that for general Hannan consistent algorithms, SM-MCTS does not necessarily converge and thus the result about SM-MCTS-A from Section \ref{sec: convergence} is optimal in the sense that it does not hold for SM-MCTS. We then present an example which gives a lower bound on the quality of a strategy to which SM-MCTS(-A) converges.
In Section \ref{sec: exploitability}, we discuss the notion of exploitability, which measures the quality of a strategy, and we make a few remarks about which strategy should be considered as the output of SM-MCTS(-A).
In Section \ref{sec:Experimental}, we present empirical investigation of convergence of SM-MCTS and SM-MCTS-A, as well as empirical confirmation of the fact that the the commonly used HC-algorithms guarantee the UPO property. Finally, Section~\ref{sec:Discussion} summarizes the results and highlights open questions, which might be interesting for future research.

\section{Background\label{sec: background}}
We now introduce the game theory fundamentals and notation used throughout
the paper. We define simultaneous move games, describe
the SM-MCTS algorithm and its modification SM-MCTS-A, and afterwards, we discuss existing selection functions and their properties.

\subsection{Simultaneous move games }

A finite two-player zero-sum game with perfect information and simultaneous moves
can be described by a tuple $(\mathcal{N},\mathcal{H},\mathcal{C},\mathcal{Z},\mathcal{A},\mathcal{T},\Delta_c,u_{i},h_{0})$,
where $\mathcal{N}=\{1,2\}$ contains player labels, $\mathcal{H}$
is a set of inner states, $\mathcal{C}$ is the set of chance states and $\mathcal{Z}$ denotes the terminal states.
$\mathcal{A}=\mathcal{A}_{1}\times\mathcal{A}_{2}$ is the set of
joint actions of individual players and we denote $\mathcal{A}_{1}(h)=\{1\dots m^{h}\}$
and $\mathcal{A}_{2}(h)=\{1\dots n^{h}\}$ the actions available to
individual players in state $h\in\mathcal{H}$. The game begins in an initial state $h_{0}$. The transition function
$\mathcal{T}:\mathcal{H}\times\mathcal{A}_{1}\times\mathcal{A}_{2}\mapsto\mathcal{H} \cup\mathcal{C}\cup\mathcal{Z}$
defines the successor state given a current state and actions for
both players. For brevity, we sometimes denote $\mathcal{T}(h,i,j)\equiv h_{ij}$.
The chance strategy $\Delta_c:\mathcal{C}\mapsto\mathcal{H}\times[0,1]$ determines the next states in chance nodes based on a fixed commonly known probability distribution.
The utility function $u_{1}:\mathcal{Z}\mapsto[u_{\min},u_{\max}]\subseteq\mathbb{R}$
gives the utility of player 1, with $u_{min}$ and $u_{\max}$ denoting
the minimum and maximum possible utility respectively. Without loss
of generality we assume $u_{\min}=0$ and $u_{\max}=1$. We assume
zero-sum games: $\forall z\in\mathcal{Z},u_{2}(z)=-u_{1}(z)$.

\begin{figure}
\centering \includegraphics[width=0.6\textwidth]{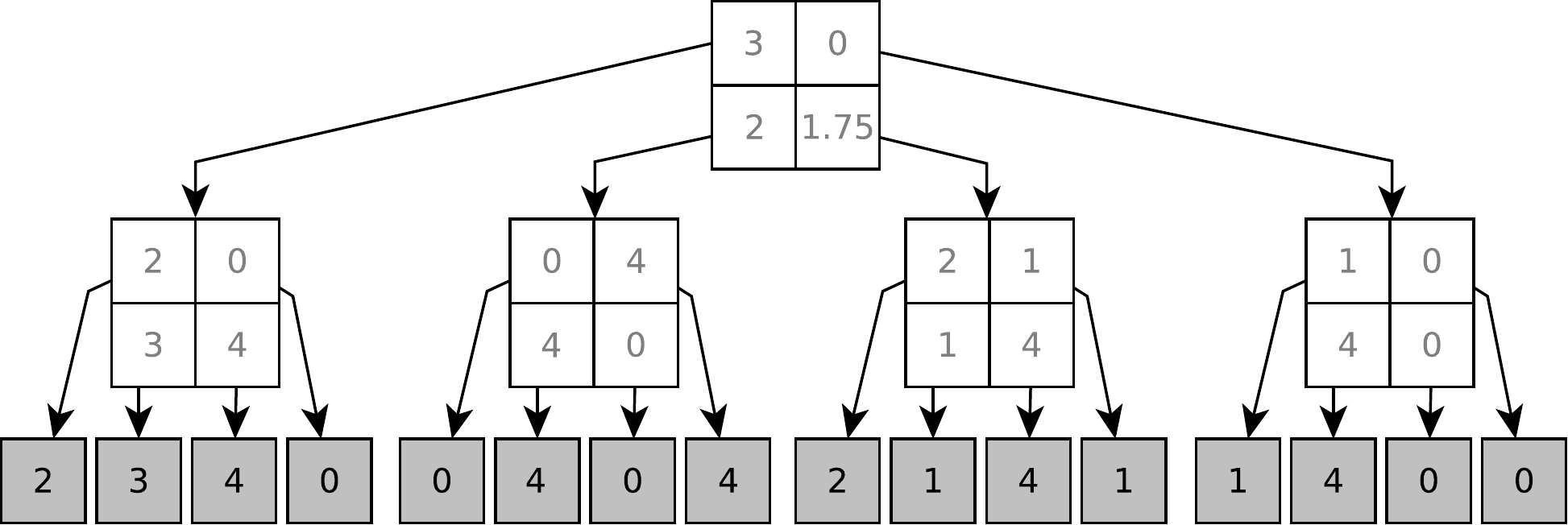}
\caption{Example game tree of a game with perfect information and simultaneous
moves. Only the leaves contain actual rewards  - the values in
the inner nodes are achieved by optimal play in the corresponding
subtree, they are not part of the definition of the game.}\label{fig:tree}
\end{figure}

A \textit{matrix game} is a single-stage simultaneous move game with
action sets $\mathcal{A}_{1}$ and $\mathcal{A}_{2}$. Each entry
in the matrix $M=(a_{ij})$ where $(i,j)\in\mathcal{A}_{1}\times\mathcal{A}_{2}$
and $a_{ij}\in[0,1]$ corresponds to a payoff (to player 1) if row
$i$ is chosen by player 1 and column $j$ by player 2. A strategy
$\sigma_{i}\in\Delta(\mathcal{A}_{i})$ is a distribution over the
actions in $\mathcal{A}_{i}$. If $\sigma_{1}$ is represented as
a row vector and $\sigma_{2}$ as a column vector, then the expected
value to player 1 when both players play with these strategies is
$u_{1}(\sigma_{1},\sigma_{2})=\sigma_{1}M\sigma_{2}$. Given a profile
$\sigma=(\sigma_{1},\sigma_{2})$, define the utilities against best
response strategies to be $u_{1}(br,\sigma_{2})=\max_{\sigma_{1}'\in\Delta(\mathcal{A}_{1})}\sigma_{1}'M\sigma_{2}$
and $u_{1}(\sigma_{1},br)=\min_{\sigma_{2}'\in\Delta(\mathcal{A}_{2})}\sigma_{1}M\sigma_{2}'$.
A strategy profile $(\sigma_{1},\sigma_{2})$ is an $\epsilon$-Nash
equilibrium of the matrix game $M$ if and only if 
\begin{equation}
u_{1}(br,\sigma_{2})-u_{1}(\sigma_{1},\sigma_{2})\leq\epsilon\hspace{1cm}\mbox{and}\hspace{1cm}u_{1}(\sigma_{1},\sigma_{2})-u_{1}(\sigma_{1},br)\leq\epsilon\label{eq:nfgNE}
\end{equation}
Two-player perfect information games with simultaneous moves are sometimes
appropriately called \textit{stacked matrix games} because at every
state $h$ there is a joint action set $\mathcal{A}_{1}(h)\times\mathcal{A}_{2}(h)$
that either leads to a terminal state or to a subgame which is itself
another stacked matrix game with a unique value, which can be determined by backward induction (see Figure~\ref{fig:tree}).

A \textit{behavioral strategy} for player $i$ is a mapping from states
$h\in\mathcal{H}$ to a probability distribution over the actions
$\mathcal{A}_{i}(h)$, denoted $\sigma_{i}(h)$. Given a profile $\sigma=(\sigma_{1},\sigma_{2})$,
define the probability of reaching a terminal state $z$ under $\sigma$
as $\pi^{\sigma}(z)=\pi^{\sigma}_{1}(z)\pi^{\sigma}_{2}(z)\pi_c(z)$, where each $\pi^{\sigma}_{i}(z)$
(resp. $\pi_c(z)$) is a product of probabilities of the actions taken by player $i$ (the chance)
along the path to $z$. Define $\Sigma_{i}$ to be the set of behavioral
strategies for player $i$. Then for any strategy profile $\sigma=(\sigma_{1},\sigma_{2})\in\Sigma_{1}\times\Sigma_{2}$
we define the expected utility of the strategy profile (for player
1) as 
\begin{equation}
u(\sigma)=u(\sigma_{1},\sigma_{2})=\sum_{z\in Z}\pi^{\sigma}(z)u_{1}(z)
\end{equation}
An $\epsilon$-Nash equilibrium profile ($\sigma_{1},\sigma_{2}$)
in this case is defined analogously to (\ref{eq:nfgNE}). In other
words, none of the players can improve their utility by more than
$\epsilon$ by deviating unilaterally. If $\sigma=(\sigma_{1},\sigma_{2})$
is an exact Nash equilibrium ($\epsilon$-NE with $\epsilon=0$),
then we denote the unique value of the game $v^{h_{0}}=u(\sigma_{1},\sigma_{2})$.
For any $h\in\mathcal{H}$, we denote $v^{h}$ the value of the subgame
rooted in state $h$.

\subsection{Simultaneous move Monte Carlo Tree Search}

Monte Carlo Tree Search (MCTS) is a simulation-based state space search
algorithm often used in game trees.
The main idea is to iteratively run simulations to a terminal state, incrementally growing a tree rooted at the current state. 
In the basic form of the algorithm, the tree is initially empty and a single leaf is added each iteration.
The nodes in the tree represent game states.
Each simulation starts by visiting nodes in the tree, selecting which actions to take based on the information maintained in the node, and then consequently transitioning to the successor node.
When a node whose immediate children are not all in the tree is visited, we expand this node by adding a new leaf to the tree.
Then we apply a rollout policy (for example, random action selection) from the new leaf to a terminal state of the game.
The outcome of the simulation is then returned as a reward to the new leaf and all its predecessors.

In Simultaneous Move MCTS (SM-MCTS), the main difference is that a joint action of both players is selected and used to transition to a following state. The algorithm has been previously
applied, for example in the game of Tron~\citep{Perick12Comparison}, Urban Rivals~\citep{teytaud2011upper}, and in general game-playing~\citep{Finnsson08}.
However, guarantees of convergence to NE remain unknown, and \citet{Shafiei09} show that the most popular selection policy (UCB) does not converge, even in a simple one-stage game.
The convergence to a NE depends critically on the selection and update policies applied, which are even more non-trivial in simultaneous-move games than in purely sequential games. We describe variants of two popular selection algorithms in Section \ref{sec: bandits}.

\begin{figure}[t]
\newcommand{\LineIf}[2]{\STATE \algorithmicif\ {#1}\ \algorithmicthen\ {#2}}
\textbf{SM-MCTS(}$h$ -- current state of the game\textbf{)}
\begin{algorithmic}[1]
\LineIf{$h \in \cZ$}{\algorithmicreturn\ $u_1(h)$}
\IF{$h \in \cC$}
  \STATE Sample $h' \sim \Delta_c(h)$
  \RETURN SM-MCTS($h'$)
\ENDIF
\IF{$h \in T$}
  \STATE $(a_1, a_2) \leftarrow$ \emph{\underline{Select}}$(h)$\label{alg:smmcts:select}
  \STATE $h' \leftarrow \cT(h,a_1,a_2)$
  \STATE $x \leftarrow $ SM-MCTS($h'$)\label{alg:smmcts:reccall}
  \STATE \emph{\underline{Update}}$(h,a_1,a_2,x)$\label{alg:smmcts:up}
  \RETURN $x$\label{alg:smmcts:return}
\ELSE
  \STATE $T \leftarrow T \cup \lbrace h \rbrace$
  \STATE $x \leftarrow$ Rollout($h$)
  \RETURN $x$
\ENDIF
\end{algorithmic}
\caption{Simultaneous Move Monte Carlo Tree Search}\label{alg:smmcts}
\end{figure}

In Figure~\ref{alg:smmcts}, we present a generic template of MCTS algorithms for simultaneous-move games (SM-MCTS). We then proceed to explain how specific algorithms are derived from this template. Figure~\ref{alg:smmcts} describes a single iteration of SM-MCTS. $T$ represents the incrementally built MCTS tree, in which each state is represented by one node. Every node $h$ maintains algorithm-specific statistics about the iterations that previously used this node. The template can be instantiated by specific implementations of the updates of the statistics on line~\ref{alg:smmcts:up} and the selection based on these statistics on line \ref{alg:smmcts:select}. In the terminal states, the algorithm returns the value of the state for the first player (line 1). In the chance nodes, the algorithm selects one of the possible next states based on the chance distribution (line 3) and recursively calls the algorithm on this state (line~4). If the current state has a node in the current MCTS tree $T$, the statistics in the node are used to select an action for each player (line~\ref{alg:smmcts:select}). These actions are executed (line 7) and the algorithm is called recursively on the resulting state (line 8). The result of this call is used to update the statistics maintained for state $h$ (line~\ref{alg:smmcts:up}). If the current state is not stored in tree $T$, it is added to the tree (line 12) and its value is estimated using the rollout policy (line 13). The rollout policy is usually uniform random action selection until the game reaches a terminal state, but it can also be based on domain-specific knowledge. Finally, the result of the Rollout is returned to higher levels of the tree.

This template can be instantiated by choosing a specific selection and update functions.
Different algorithms can be the bases for selection functions, but the most successful selection functions are based on the algorithms for multi-armed bandit problem we introduce in Section~\ref{sec: bandits}.
The action for each player in each node is selected independently, based on these algorithms and the updates update the statistics for player one by $u_1$ and for player two by $u_2=-u_1$ as if they were independent multi-armed bandit problems.

SM-MCTS algorithm does not always converge to Nash equilibrium - to guarantee convergence, additional assumptions on the selection functions are required. Therefore, we also propose a variant of the algorithm, which we denote as SM-MCTS-A. Later we show that this variant converges to NE under more reasonable assumptions on the selection function.
The difference is that for each node $h\in T$, the algorithm also stores the number $n^h$ of iterations that visit this node and the cumulative reward $X^h$ received from the recursive call in these iterations. Every time node $h$ is visited, it increases $n^h$ by one and adds $x$ to $X^h$. SM-MCTS-A then differs from SM-MCTS only on line 9, where the selection functions at $h$ are updated by $X^{h'}/n^{h'}$ instead of $x$.

We note that in our previous work \citep{lisy2013convergence} we prove a result similar to our Theorem \ref{thm: SM-MCTS-A convergence} here. However, the algorithm that we used earlier is different from SM-MCTS-A algorithm described here. In particular, SM-MCTS-A uses averaged values for decision making in each node, but propagates backwards the non-averaged values (unlike the previous version, which also updates the selection function based on the averaged values, but then it propagates backwards these averaged numbers - and on the next level, it takes averages of averages and so on). Consequently, this new version is much closer to the non-averaged SM-MCTS algorithm used in practice and it has faster empirical convergence.

\subsection{Multi-armed bandit problem} \label{sec: bandits}
Multi-armed bandit (MAB) problem is one of the most basic models in online learning. In theoretic studies, it is often the basic model for studying fundamental trade-offs between exploration and exploitation in an unknown environment \citep{auer1995gambling,auer2002finite}. In practical applications, the algorithms developed for this model has recently been used in online advertising \citep{pandey2007}, generic optimization \citep{flaxman2005online}, and most importantly for this paper in Monte Carlo tree search algorithms \citep{kocsis2006,browne2012survey,gelly2011monte,teytaud2011upper}.

The multi-armed bandit problem got its name after a simple motivating example concerning slot machines in casinos, also known as one-armed bandits. Assume you have a fixed number of coins $n$ you want to use in a casino with $K$ slot machines. Each slot machine has a hole where you can insert a coin and as a result, the machine will give you some (often zero) reward. Each of the slot machines is generally different and decides on the size of the rewards using a different mechanism. The basic task is to use the $n$ coins sequentially, one by one, to receive the largest possible cumulative reward. Intuitively, it is necessary to sufficiently explore the quality of the machines, but not to use too many coins in the machines that are not likely to be good. The following formal definitions use the notation from an extensive survey of the field by \citet{bubeck2012survey}.

\begin{definition}[Adversarial multi-armed bandit problem]\label{def: MABproblem}
Multi-armed bandit problem is a set of actions (or arms) denoted $1,\dots, K$, and a set of sequences $x_i(T)$ for each action $i$ and time step $T=1,2,\dots$. In each time step, an agent selects an action $i(T)$ and receives the payoff $x_{i(T)}(T)$. In general, the agent does not learn the values $x_i(T)$ for $i \neq i(T)$.

The \emph{adversarial} MAB problem is a MAB problem, in which in each time step an adversary selects arbitrary rewards $x_i(T) \in \left[ 0,1 \right]$ simultaneously with the agent selecting the action.
\end{definition}

The algorithms for solving the MAB problem usually optimize some notion of regret. Intuitively, the algorithms try to minimize the difference between playing the strategy given by the algorithm and playing some baseline strategy, which can possibly use information not available to the agent. For example, the most common notion of regret is the \emph{external regret}, which is the difference between playing according to the prescribed strategy and playing the fixed optimal action all the time.

\begin{definition}[External Regret]\label{def:extRegret}
The \emph{external regret} for playing a sequence of actions $i(1),\dots,i(n)$ is defined as
\[ R(t) = \max_{i=1,\dots,K}\sum_{s=1}^t x_i(s) - \sum_{s=1}^t x_{i(s)}(s). \]
By $r(t)$ we denote the average external regret $r(T):=\frac 1 t R(t)$.
\end{definition}

\subsubsection{Application to SM-MCTS(-A)}
We now explain how MAB problem applies in the setting of SM-MCTS(-A).
We focus on the situation for player 1. For a fixed node $h\in\mathcal H$, our goal is to define the MAB reward assignment $x_i(t)$ for $i\in\mathcal A_i(h)$, $t\in\mathbb N$, as they are perceived by the selection function

Firstly we introduce two auxiliary symbols $u^h(T)$ and $T^h(t)$: Let $T$ be an iteration during which the node $h$ got visited. By $u^h(T)$ we denote the value (from line 9 of the algorithm on Figure \ref{alg:smmcts}) by which the selection function was updated during iteration $T$ (this value is either $x$ for SM-MCTS, or $X^{h'}/n^{h'}$ for SM-MCTS-A). We also set $T^h(t)$ to be the iteration during which node $h$ was visited for the $t$-th time.

By $i(t)\in\mathcal A_1(h)$ and $j(t)\in\mathcal A_2(h)$ we denote the actions, which were selected in $h$ during iteration $T^h(t)$. We can now define the desired MAB reward assignment. By the definition of SM-MCTS(-A) algorithm (line 9), the reward $x_{i(t)}(t)$ has to be equal to the $t$-th observed value $u^h\left(T^h(t)\right)$, thus it remains to define the rewards $x_i(t)$ for $i\neq i(t)$. Intuitively, the rewards for these actions should be ``the values we would have seen if we chose differently''. Formally we set
\begin{eqnarray*}
x_i(t) & := & u^h\left(T^h(\tilde t)\right) \textrm{, where }T^h(\tilde t)\textrm{ is the earliest iteration during} \\
 & & \textrm{which $h$ got visited, such that } \tilde t \geq t,\ i(\tilde t)=i \textrm{ and } \ j(\tilde t)=j(t).\\
\end{eqnarray*}
We can see that for $i=i(t)$, we have $\tilde t = t$ and therefore the definition coincides with the one we promised earlier.

\emph{Technical remark}: Strictly speaking, it is not immediately obvious that $\left(x_i(t)\right)$, as defined above,  is a MAB reward assignment - in MAB problem, the rewards $x_i(t)$, $i\in\mathcal A_i(h)$ have to be defined before the $t$-th action is chosen. Luckily, this is not a problem in our case - in theory we could compute SM-MCTS($h'$) for all possible child nodes $h'$ in advance (before line 6), and keep each of them until they are selected. The overall behavior of SM-MCTS(-A) would remain the same (except that it would run much slower) and the rewards $\left(x_i(t)\right)$ would correspond to a MAB problem.

In the remainder of this section, we introduce the technical notation used throughout the paper. First, we define the notions of \emph{cumulative payoff} $G$ and \emph{maximum cumulative payoff}  $G_{\max}$ and relate these quantities to the external regret:
\begin{eqnarray*}
G(t) & := & \sum_{s=1}^{t}x_{i(s)}(s)\\
G_{\max}(t) & := & \max_{i\in\mathcal{A}_{1}}\sum_{s=1}^{t}x_{i(s)}(s),\\
R(t) & = & G_{\max}(t)-G(t).
\end{eqnarray*}
We also define the corresponding average notions and relate them to the average regret: $g(t) := G(t)/t$, $g_{\max}(t) := G_{\max}(t)/t$, $r(t) = g_{\max} (t)-g(t)$. If there is a risk of confusion as to in which node we are interested, we will add a superscript $h$ and denote these variables as $g^h(t),\ g_{\max}^h(t)$ and so on.

Focusing now on the given node $h$, let $i$  be an action of player $1$ and $j$ an action of player $2$. We denote by $t_{i}$, $t_{j}$ the number of times these actions were chosen up to the $t$-th visit of $h$ and $t_{ij}$ the number of times both of these actions has been chosen at once. By \emph{empirical frequencies} we mean the strategy profile $\hat\sigma^h(t)=\left(\hat{\sigma}^h_{1}(t),\hat{\sigma}^h_{2}(t)\right)$ given by the formulas
\[ \hat{\sigma}^h_{1}(t)(i)=t_{i}/t, \ \ \ \ \hat{\sigma}^h_{2}(t)(j)=t_{j}/t \]
By \emph{average strategies}, we mean the strategy profile $\left(\bar{\sigma}^h_{1}(t),\bar{\sigma}^h_{2}(t)\right)$ given by the formulas
\[ \bar{\sigma}^h_{1}(t)(i)=\sum_{s=1}^{t}\sigma^h_{1}(s)(i)/t, \ \ \ \  \bar{\sigma}^h_{2}(t)(j)=\sum_{s=1}^{t}\sigma^h_{2}(s)(j)/t,\]
where $\sigma^h_{1}(s)$, $\sigma^h_{2}(s)$ are the strategies used at $h$ at time $s$.

Lastly, by $\hat\sigma(T)$ we denote the collection $\left(\hat\sigma^h(t^h(T))\right)_{h\in\mathcal{H}}$ of empirical frequencies at all nodes $h\in\mathcal H$, where $t^h(T)$ denotes, for the use of this definition, the number of visits of $h$ up to the $T$-th iteration of SM-MCTS(-A). Similarly we define the average strategy $\bar\sigma(T)$. The following lemma says there eventually is no difference between these two strategies.

\begin{lemma} \label{emp a avg}
As $t$ approaches infinity, the empirical frequencies and average strategies will almost surely be equal. That is, 
$\limsup_{t \rightarrow \infty} \max_{i \in \cA_1}\,|\hat{\sigma}_1(t, i)-\bar{\sigma}_1(t, i)| = 0$
 holds with probability $1$.
\end{lemma}

The proof is a consequence of the Strong Law of Large Numbers (and it can be found in the appendix).

\subsection{Hannan consistent algorithms}
A desirable goal for an algorithm in MAB setting is the classical notion of Hannan consistency (HC). Having this property means that for high enough $t$, the algorithm performs nearly as well as it would if it played the optimal constant action since the beginning.
\begin{definition}[Hannan consistency]\label{def:HC}
An algorithm is $\epsilon$-Hannan consistent for some $\epsilon\geq0$ if $\limsup_{t\rightarrow\infty} r(t)\leq \epsilon$ holds with probability 1, where the ``probability'' is understood with respect to the randomization of the algorithm. Algorithm is Hannan consistent if it is 0-Hannan consistent.
\end{definition}

We now present regret matching and Exp3, two of the $\epsilon$-Hannan consistent algorithms previously used in MCTS context. The proofs of Hannan consistency of variants of these two algorithms, as well as more related results, can be found in a survey by \citet[Section 6]{cesa2006prediction}. The fact that the variants presented here are $\epsilon$-HC is not explicitly stated there, but it immediately follows from the last inequality in the proof of Theorem 6.6 in the survey.

\subsubsection{Exponential-weight algorithm for Exploration and Exploitation \label{sec:exp3}}

\begin{figure}
\begin{algorithmic}[1]
\REQUIRE{$K$ - number of actions; $\gamma$ - exploration parameter}
\STATE $\forall_i G_i \leftarrow 0$
\FOR {$t \leftarrow 1,2,\dots$}
  \STATE $\forall_i p_{i} \leftarrow \frac{exp(\tfrac{\gamma}{K} G_i)}{\sum_{j=1}^K exp(\tfrac{\gamma}{K} G_j)}$
  \STATE $p'_i \leftarrow (1-\gamma)p_i+\tfrac{\gamma}{K}$
  \STATE Use action $I_t$ from distribution $p'$ and receive reward $r$
  \STATE $G_{I_t} \leftarrow G_{I_t} + \frac{r}{p'_{I_t}}$
\ENDFOR
\end{algorithmic}
\caption{Exponential-weight algorithm for Exploration and Exploitation (Exp3) algorithm for regret minimization in adversarial bandit setting}
\label{alg:exp3}
\end{figure}

The most popular algorithm for minimizing regret in adversarial bandit setting is the Exponential-weight algorithm for Exploration and Exploitation (Exp3) proposed by \cite{Auer2003Exp3} and further improved by \cite{stoltz2005incomplete}. The algorithm has many different variants for various modifications of the setting and desired properties. We present a formulation of the algorithm based on the original version in Figure~\ref{alg:exp3}.

Exp3 stores the estimates of cumulative reward of each action over all iterations, even those in which the action was not selected. In the pseudo-code in Figure~\ref{alg:exp3}, we denote this value for action $i$ by $G_i$. It is initially set to $0$ on line~1. In each iteration, a probability distribution $p$ is created proportionally to the exponential of these estimates. The distribution is combined with a uniform distribution with probability $\gamma$ to ensure sufficient exploration of all actions (line~4). After an action is selected and the reward is received, the estimate for the performed action is updated using \emph{importance sampling} (line~6): the reward is weighted by one over the probability of using the action.
As a result, the expected value of the cumulative reward estimated only from the time steps where the agent selected the action is the same as the actual cumulative reward over all the time steps.

\subsubsection{Regret matching \label{sec:rm}}
An alternative learning algorithm that allows minimizing regret in stochastic bandit setting is regret matching \citep{hart2001reinforcement}, later generalized as \emph{polynomially weighted average forecaster} \citep{cesa2006prediction}. Regret matching (RM) corresponds to selection of the parameter $p=2$ in the more general formulation. It is a general procedure originally developed for playing known general-sum matrix games in \citep{hart2000simple}. The algorithm computes, for each action in each step, the regret for not playing another fixed action every time the action has been played in the past. The action to be played in the next round is selected randomly with probability proportional to the positive portion of the regret for not playing the action. This procedure has been shown to converge arbitrarily close to the set of correlated equilibria in general-sum games. As a result, it converges to a Nash equilibrium in a zero-sum game. The regret matching procedure in \cite{hart2000simple} requires the exact information about all utility values in the game, as well as the action selected by the opponent in each step. In \cite{hart2001reinforcement}, the authors modify the regret matching procedure and relax these requirements. Instead of computing the exact values for the regrets, the regrets are estimated in a similar way as the cumulative rewards in Exp3. As a result, the modified regret matching procedure is applicable in MAB.

\begin{figure}
\begin{algorithmic}[1]
\REQUIRE{$K$ - number of actions; $\gamma$ - the amount of exploration}
\STATE $\forall i \; R_i \leftarrow 0$
\FOR {$t \leftarrow 1,2,\dots$}
  \STATE $\forall i \; R_i^+ \leftarrow \max\{0,R_i\}$ 
  \IF {$\sum_{j=1}^K R_j^+ = 0$}
    \STATE $\forall i \; p_i \leftarrow 1/K$
  \ELSE
    \STATE $\forall i \; p_i \leftarrow (1-\gamma_t)\frac{R_i^+}{\sum_{j=1}^K R_j^+} + \frac{\gamma}{K}$
  \ENDIF
  \STATE Use action $I_t$ from distribution $p$ and receive reward $r$
  \STATE $\forall i\; R_i \leftarrow R_i - r$
  \STATE $R_{I_t} \leftarrow R_{I_t} + \frac{r}{p_{I_t}}$
\ENDFOR
\end{algorithmic}
\caption{regret matching variant for regret minimization in adversarial bandit setting.}
\label{alg:banditRM}
\end{figure}

We present the algorithm in Figure~\ref{alg:banditRM}.
The algorithm stores the estimates of the regrets for not playing action $i$ in all time steps in the past in variables $R_i$.
On lines 3-7, it computes the strategy for the current time step.
If there is no positive regret for any action, a uniform strategy is used (line~5).
Otherwise, the strategy is chosen proportionally to the positive part of the regrets (line~7).
The uniform exploration with probability $\gamma$ is added to the strategy as in the case of Exp3. 
It also ensures that the addition on line 10 is bounded.

\cite{cesa2006prediction} prove that regret matching eventually achieves zero regret in the adversarial MAB problem, but they provide the exact finite time bound only for the perfect-information case, where the agent learns rewards of all arms.

\section{Convergence of SM-MCTS and SM-MCTS-A}\label{sec: convergence}

\begin{table}[t] 
\begin{framed}
\small
\begin{eqnarray*}
h\in \mathcal H, \mathcal A, D & \text{game nodes, action space, depth of the game tree}\\
u, v, v^h, d_h & \text{utility, game value, subgame value, node depth }\\
\sigma, \hat \sigma, \bar \sigma, br & \text{strategy, empirical st., average st., best response }\\
\text{NE, HC} & \text{Nash equilibrium, Hannan consistent }\\
\text{UPO} & \text{Unbiased payoff observations }\\
\text{SM-MCTS(-A)} & \text{(averaged) simultaneous-move Monte Carlo tree search }\\
\text{MAB} & \text{multi-armed bandit }\\
i(t) \text{ (or also }a(t)\text{)} & \text{action chosen at time t }\\
t_i, t_{ij} & \text{uses of action i (joint action (i,j)) up to time t }\\
x_i(t) & \text{reward assigned to an action i at time t }\\
r(t), R(t) & \text{(average) external regret at time t}\\
G,g,G_{\max},g_{\max} & \text{ cumulative payoff (average, maximum, maximum average)}\\
\text{Exp3} & \text{Exponential-weight algorithm for Exploration and Exploitation }\\
\text{RM} & \text{regret matching algorithm }\\
\text{CFR} & \text{an algorithm for counterfactual regret minimization} \\
\gamma & \text{exploration rate }\\
C, c & \text{positive constants }\\
\eta & \text{arbitrarily small positive number }\\
expl & \text{exploitability of a strategy }\\
\hat p & \text{empirical strategy with removed exploration }\\
\mathbb I & \text{indicator function}
\end{eqnarray*}
\end{framed}
\caption{The most common notation for quick reference}\label{tab:notation}
\end{table}

In this section, we present the main theoretic results. Apart from a few cases, we will only present the key ideas of the proofs here, while the full proofs can be found in the appendix. We will assume without loss of generality that the game does not contain chance nodes (that is, $\mathcal{C}=\emptyset$); all of our results (apart from those in Section \ref{sub: SM-MCTS-A bound}) are of an asymptotic nature, and so they hold for general nonempty $\mathcal{C}$, since we can always use the law of large numbers to make the impact of chance nodes negligible after sufficiently high number of iterations. We choose to omit the chance nodes in our analysis, since their introduction would only require additional, purely technical, steps in the proofs, without shedding any new light on the subject. For an overview of the notation we use, see Table \ref{tab:notation}.

In order to ensure that the SM-MCTS(-A) algorithm will eventually visit each node we need the selection function to satisfy the following property. 
\begin{definition}
We say that $A$ is an \emph{algorithm with guaranteed exploration}
if, for players $1$ and $2$ both using $A$ for action selection,
$\lim_{t\rightarrow\infty}t_{ij}=\infty$ holds almost surely for
each $(i,j)\in\mathcal{A}_{1}\times\mathcal{A}_{2}.$ 
\end{definition}
It is an immediate consequence of this definition that when an algorithm with guaranteed exploration is used in SM-MCTS(-A), every node of the game tree will be visited indefinitely. From now on, we will therefore assume that, at the start of our analysis, the full game tree is already built - we do this, because it will always happen after a finite number of iterations and, in most cases, we are only interested in the limit behavior of SM-MCTS(-A) (which is not affected by the events in the first finitely many steps). 

Note that most of the HC algorithms, namely RM and Exp3, guarantee exploration without the need for any modifications. There exist some (mostly artificial) HC algorithms, which do not have this property. However, they can always be adjusted in the following way. 
\begin{definition}
Let $A$ be an algorithm used for choosing action in a matrix game
$M$. For fixed exploration parameter $\gamma\in\left(0,1\right)$
we define modified algorithm $A^{*}$ as follows: For time $s=1,2,...$:
either explore with probability $\gamma$ or run one iteration of
$A$ with probability $1-\gamma$, where ``explore'' means we choose
the action randomly uniformly over available actions, without updating
any of the variables belonging to $A$.
\end{definition}
Fortunately, $\epsilon$-Hannan consistency is not substantially influenced
by the additional exploration: 
\begin{lemma}
\label{lemma: A* je HC}Let $A$ be an $\epsilon$-Hannan consistent
algorithm. Then $A^{*}$ is an $(\epsilon+\gamma)$-Hannan consistent
algorithm with guaranteed exploration.
\end{lemma}

\subsection{Asymptotic convergence of SM-MCTS-A}\label{sub: SM-MCTS-A convergence}

The goal of this section is to prove the following Theorem \ref{thm: SM-MCTS-A convergence}. We will do so by backward induction, stating firstly the required lemmas
and definitions. The Theorem \ref{thm: SM-MCTS-A convergence} itself
will then follow from the Corollary \ref{cor: ind.step}.
\begin{theorem}
\label{thm: SM-MCTS-A convergence}Let $G$
be a zero-sum game with perfect information and simultaneous moves with maximal
depth $D$ and let $A$ be an $\epsilon$-Hannan consistent algorithm
with guaranteed exploration, which we use as a selection policy for SM-MCTS-A.

Then for arbitrarily small $\eta>0$, there
almost surely exists $t_{0}$, so that the empirical frequencies $(\hat{\sigma}_{1}(t),\hat{\sigma}_{2}(t))$
form a subgame-perfect
\[ 
\left(2D\left(D+1\right)\epsilon+\eta\right)\mbox{-equilibrium for all }t\geq t_{0}.
\]

\end{theorem}
In other words, the average strategy will eventually get arbitrarily close to $C\epsilon$-equilibrium. In particular a Hannan-consistent algorithm ($\epsilon=0$) will eventually get arbitrarily close to Nash equilibrium. This also illustrates why we cannot remove the number $\eta$, as even a HC algorithm might not reach NE in finite time. In the
following $\eta>0$ will denote an arbitrarily small number. As $\eta$ can be chosen independently of everything else, we will not focus on the constants in front of it, writing simply $\eta$ instead of $2\eta$ etc.

It is well-known that two Hannan consistent players will eventually converge
to NE in a matrix game - see \cite{waugh09d} and \cite{Blum07}. We prove a similar result for the approximate versions of the notions.
\begin{lemma}
\label{L: HC-and-NE}Let $\epsilon\geq0$ be a real number. If both
players in a matrix game $M$ are $\epsilon$-Hannan consistent, then
the following inequalities hold for the empirical frequencies almost
surely: 
\begin{equation}
\mbox{\hspace{0.3cm}}v-\epsilon\leq\underset{t\rightarrow\infty}{\liminf}\, g(t)\leq\underset{t\rightarrow\infty}{\limsup}\, g(t)\leq v+\epsilon,\label{eq: HC a NE2}
\end{equation}
\begin{equation}
v-2\epsilon\leq\underset{t\rightarrow\infty}{\liminf}\, u\left(\hat{\sigma}_{1}(t),br\right)\mbox{\hspace{0.3cm} \ensuremath{\&}\hspace{0.3cm} }\underset{t\rightarrow\infty}{\limsup}\, u\left(br,\hat{\sigma}_{2}(t)\right)\leq v+2\epsilon.\label{eq: HC a NE 1}
\end{equation}
\end{lemma}

The inequalities (\ref{eq: HC a NE2}) are a consequence of the definition
of $\epsilon$-HC and the game value $v$. The proof of inequality (\ref{eq: HC a NE 1})
then shows that if the value caused by the empirical frequencies was
outside of the interval infinitely many times with positive probability,
it would be in contradiction with definition of $\epsilon$-HC. Next, we present the induction hypothesis around which the proof of Theorem~\ref{thm: SM-MCTS-A convergence} revolves.

\paragraph{Induction hypothesis $\left(IH_{d}\right):$\label{par: (IH)}} For a node $h$ in the game tree, we denote by $d_{h}$ the depth
of the tree rooted at $h$ (not including the terminal states - therefore
when $d_{h}=1$, the node is a matrix game). Let $d\in\left\{ 1,...,d_{\mbox{root}}\right\} $.
Induction hypothesis $\left(IH_{d}\right)$ is then the claim
{} that for each node $h$ with $d_{h}=d$, there almost surely exists
$t_{0}$ such that for each $t\geq t_{0}$ 
\begin{enumerate}
\item the payoff $g^{h}(t)$ will fall into the interval $\left(v^{h}-C_{d}\epsilon,v^{h}+C_{d}\epsilon\right)$;
\item the utilities $u\left(\hat{\sigma}_{1}(t),br\right)\leq u\left(br,\hat{\sigma}_{2}(t)\right)$ with respect to the matrix game $\left(v_{ij}^{h}\right)$, will fall into the interval $\left(v^{h}-2C_{d}\epsilon,v^{h}+2C_{d}\epsilon\right)$;
\end{enumerate}
where $C_{d}=d+\eta$ and $v_{ij}^{h}$ is the value of subgame rooted
at the child node of $h$ indexed by $ij$.

Note that Lemma \ref{L: HC-and-NE} ensures that $\left(IH_{1}\right)$
holds. Our goal is to prove 2. for every $h\in\mathcal{H}$, which
then implies the main result. However, for the induction
itself to work, the condition 1. is required. We now introduce the
necessary technical tools. 
\begin{definition}
Let $M=\left(a_{ij}\right)$ be a matrix game. For $t\in\mathbb{N}$
we define $M(t)=\left(a_{ij}(t)\right)$ to be a game, in which if
players chose actions $i$ and $j$, they observe (randomized) payoffs
$a_{ij}\left(t,(i(1),...i(t-1)),(j(1),...j(t-1))\right)$. We
will denote these simply as $a_{ij}(t)$, but in fact they are random
variables with values in $[0,1]$ and their distribution in time $t$
depends on the previous choices of actions.

We say that $M(t)=\left(a_{ij}(t)\right)$ is a \emph{repeated game
with error} $e$, if there almost surely exists $t_{0}\in\mathbb{N}$,
such that $\left|a_{ij}(t)-a_{ij}\right|<e$ holds for some matrix $(a_{ij})$ and all $t\geq t_{0}$. By symbols $G(t),\ R(T),\ r(t)$ (and so on) we will denote the payoffs, regrets and other variables related to the distorted payoffs $a_{ij}(t)$. On the other hand, by symbol $u(\sigma)$ we will refer to the utility of strategy $\sigma$ with respect to the matrix game $(a_{ij})$.
\end{definition}

The intuition behind this definition is that the players are repeatedly playing the original matrix game $M$ - but for some reason, they receive imprecise information about their payoffs. The application we are interested in is the following: we take a node $h$ inside the game tree. The matrix game without error is the matrix game $M=\left(v_{ij}\right)$, where $v_{ij}$ are the values of subgames nested at $h$. By $\left(IH_{d_{h}-1}\right)$, the payoffs received in $h$ during SM-MCTS-A can be described as a repeated game with error, where the observed payoffs are $g^{h_{ij}}$.

The following proposition is an analogy of Lemma \ref{L: HC-and-NE}
for repeated games with error. It shows that an $\epsilon$-HC algorithms
will still perform well even if they observe slightly perturbed rewards. 
\begin{proposition}
\label{Prop: hry s chybou}Let $M=\left(v_{ij}\right)$
be a matrix game with value $v$ and $\epsilon,c\geq0$. If $M(t)$
is corresponding repeated game with error $c\epsilon$ and both players
are $\epsilon$-Hannan consistent, then the following inequalities
hold almost surely: 
\begin{equation}
\mbox{\hspace{0.3cm}}v-(c+1)\epsilon\leq\underset{t\rightarrow\infty}{\liminf}\, g(t)\leq\underset{t\rightarrow\infty}{\limsup}\, g(t)\leq v+(c+1)\epsilon,\label{eq: hra s chybou2}
\end{equation}
\begin{equation}
v-2(c+1)\epsilon\leq\underset{t\rightarrow\infty}{\liminf}\, u\left(\hat{\sigma}_{1},br\right)\leq\underset{t\rightarrow\infty}{\limsup}\, u\left(br,\hat{\sigma}_{2}\right)\leq v+2(c+1)\epsilon.\label{eq: hra s chybou1}
\end{equation}
\end{proposition}
The proof is similar to the proof of Lemma~\ref{L: HC-and-NE}. It
needs an additional claim that if the algorithm is $\epsilon$-HC
with respect to the observed values with errors, it still has a bounded
regret with respect to the exact values.
\begin{corollary}
\label{cor: ind.step}$\left(IH_{d}\right)\implies\left(IH_{d+1}\right)$.\end{corollary}
\begin{proof}
Property 1. of $\left(IH_{d}\right)$ implies that every node $h$ with $d_{h}\leq d$
is a repeated game with error $d\epsilon+\eta$. Proposition \ref{Prop: hry s chybou}
then implies that any $h$ with $d_{h}=d+1$ is again a repeated
game with error, and by inequality \eqref{eq: hra s chybou2} the value
of error increases to $\left(d+1\right)\epsilon+\eta$, which gives
$\left(IH_{d+1}\right)$. 
\end{proof}

Recall here the following well-known fact:
\begin{remark}\label{Rem: u(br) and NE}
In a zero-sum game with value $v$ the following implication holds: 
\[
\left(u_{1}(br,\hat{\sigma}_{2})<v+\frac{\epsilon}{2}\,\textrm{\,\ and\,\,}\, u_{1}(\hat{\sigma}_{1},br)>v-\frac{\epsilon}{2}\right)\Longrightarrow
\]
\[
\left(u_{1}(br,\hat{\sigma}_{2})-u_{1}(\hat{\sigma}_{1},\hat{\sigma}_{2})<\epsilon\,\textrm{\,\ and\,\,}\, u_{2}(\hat{\sigma}_{1},br)-u_{2}(\hat{\sigma}_{1},\hat{\sigma}_{2})<\epsilon\right)\overset{\textrm{def}}{\iff}
\]
\[
(\hat{\sigma}_{1},\hat{\sigma}_{2})\textrm{ is an }\mbox{\ensuremath{\epsilon}}\textrm{-equilibrium.}
\]

The following example demonstrates that the above implication would not hold if we replaced $\epsilon/2$ by $\epsilon$. Consider the following game
\begin{center} \begin{tabular}{|c|c|}
\hline 
0.4  & 0.5\tabularnewline
\hline 
0.6  & 0.5\tabularnewline
\hline 
\end{tabular}\end{center}
with a strategy profile (1,0), (1,0). The value of the
game is $v=0.5$, $u(br,(1,0))=0.6$ and $u((1,0),br)=0.4$. The best
responses to the strategies of both players are $0.1$ from the game
value, but $(1,0),(1,0)$ is a $0.2$-NE, since player 1 can improve
by $0.2$. 
\end{remark}

\begin{proof}[Proof of Theorem \ref{thm: SM-MCTS-A convergence}]
First, we observe that by Lemma \ref{L: HC-and-NE}, $\left(IH_{1}\right)$ holds, and consequently by Corollary \ref{cor: ind.step}, $\left(IH_{d}\right)$ holds for every $d=1,...,D$. Denote by $u^{h}(\sigma)$ (resp. $u_{ij}^{h}(\sigma)$) the expected payoff corresponding to the strategy $\sigma$ used in the subgame rooted at node $h\in\mathcal{H}$ (resp. its child). Remark \ref{Rem: u(br) and NE} then states that, in order to prove Theorem \ref{thm: SM-MCTS-A convergence}, it is enough to show that for every $h\in\mathcal{H}$, the strategy $\hat{\sigma}\left(t\right)$ will eventually satisfy
\begin{equation}
u^{h}\left(br,\hat{\sigma}_{2}\left(t\right)\right)\leq v^{h}+\left(d_h+1\right)d_h\epsilon+\eta. \label{eq: T8}
\end{equation}
We will do this by backward induction.
The property 1. from $\left(IH_{1}\right)$ implies that the inequality \eqref{eq: T8} holds for nodes $h$ with $d_{h}=1$. Let $1<d\leq D$, $h\in\mathcal{H}$ be such that $d_h=d$ and assume, as a hypothesis for backward induction, that the inequality \eqref{eq: T8} holds for each $h'$ with $d_{h'}<d$. We observe that 
\begin{eqnarray*}
u^{h}\left(br,\hat{\sigma}_{2}\left(t\right)\right) & = & \max_{i}\sum_{j}\hat{\sigma}_{2}\left(t\right)\left(j\right)u^{h_{ij}}\left(br,\hat{\sigma}_{2}\left(t\right)\right)\\
 & \leq & v^{h}+\left(\max_{i}\sum_{j}\hat{\sigma}_{2}\left(t\right)\left(j\right)v_{ij}^{h}-v^{h}\right)+\\
 &  & +\max_{i}\sum_{j}\hat{\sigma}_{2}\left(t\right)\left(j\right)\left(u_{ij}^{h}\left(br,\hat{\sigma}_{2}\left(t\right)\right)-v_{ij}^{h}\right).
\end{eqnarray*}
 By property 2. in $\left(IH_{d}\right)$ the first term in the brackets is at most $2d\epsilon+\eta$. By the backward induction hypothesis we have
\[ u_{ij}^{h}\left(br,\hat{\sigma}_{2}\left(t\right)\right)-v_{ij}^{h}\leq d\left(d-1\right)\epsilon+\eta \]
Therefore we have
\[ u^{h}\left(br,\hat{\sigma}_{2}\left(t\right)\right)\leq v^{h}+2d\epsilon+d\left(d-1\right)\epsilon+\eta=v^{h}+\left(d+1\right)d\epsilon+\eta. \]
For $d=D$ and $h=\textrm{root}$, Remark~\ref{Rem: u(br) and NE} implies that $\left(\hat{\sigma}_{1}\left(t\right),\hat{\sigma}_{2}\left(t\right)\right)$ will form $\left(2D\left(D+1\right)\epsilon+\eta\right)$-equilibrium of the whole game.
\end{proof}

\subsection{SM-MCTS-A finite time bound\label{sub: SM-MCTS-A bound}}

In this section, we find a probabilistic finite time bound on the performance of HC algorithms in SM-MCTS-A. We do this by taking the propositions from Section \ref{sub: SM-MCTS-A convergence} and working with their quantified versions.
\begin{theorem}[Finite time bound for SM-MCTS-A]
\label{thm:Bound}Consider the following setting: A game with at most $b$ actions at each node $h\in\mathcal{H}$ and depth $D$, played by SM-MCTS-A
using an $\epsilon$-Hannan consistent algorithm $A$ with exploration
$\gamma$. Fix $\delta>0$. Then with probability at least $1-\left(2\left|\mathcal{H}\right|+D\right)\delta$,
the empirical frequencies will form an $4D\left(D+1 \right)\epsilon$-equilibrium
for every $t\geq T_{0}$, where 
\[
T_{0}=16^{D-1}\epsilon^{-\left(D-1\right)}\left(\frac{b}{\gamma}\right)^{\frac{D}{2}\left(D-1\right)}\log\left(2\left|\mathcal{H}\right|-2\right)T_{A}\left(\epsilon,\delta\right)
\]
and $T_{A}\left(\epsilon,\delta\right)$ is the time needed for $A$
to have with probability at least $1-\delta$ regret below $\epsilon$
for all $t\geq T_{A}(\epsilon,\delta)$.
\end{theorem}
We obtain this bound by going through the proof of Theorem \ref{thm: SM-MCTS-A convergence} in more detail, replacing statements of the type ``inequality of limits holds'' by ``for all $t\geq t_0$ a slightly worse inequality holds with probability at least $1-\delta$''. We also note that the actual convergence will be faster than the one stated above, because the theorem relies on quantification of the guaranteed exploration property (necessary for our proof), rather than the fact that MCTS attempts to solve the exploration-exploitation problem (the major reason for its popularity in practice).

\subsection{Asymptotic convergence of SM-MCTS}

We would like to prove an analogy of  Theorem \ref{thm: SM-MCTS-A convergence} for SM-MCTS. Unfortunately, such a goal is unattainable in general
- in Section \ref{sec:Counterexample} we present a counterexample, showing that a such a theorem with no additional assumptions does not hold. Instead we define, for an algorithm A, the property of having $\epsilon$-unbiased payoff observations ($\epsilon$-UPO, see Definition \ref{def: UPO}) and prove the following Theorem \ref{thm: SM-MCTS convergence} for $\epsilon$-HC algorithms with this property. We were unable to prove that specific $\epsilon$-HC algorithms have this $\epsilon$-UPO property, but instead, later in Section \ref{sec:Experimental}, we provide empirical evidence supporting our hypothesis that the ``typical'' algorithms, such as regret matching or Exp3, indeed do have $\epsilon$-unbiased payoff observations.
	
\begin{theorem}\label{thm: SM-MCTS convergence}
Let $A$ be an $\epsilon$-HC algorithm with guaranteed exploration that has $\epsilon$-UPO. If $A$ is used as selection policy for SM-MCTS, then the average strategy of $A$ will eventually get arbitrarily close to $C\epsilon$-NE of the whole game, where $C=12\left(2^{D}-1\right)-8D$.\end{theorem}

We now present the notation required for the definition of the $\epsilon$-UPO property, and then proceed to the proof of Theorem \ref{thm: SM-MCTS convergence}. As we will see, the structure of this proof is similar to the structure of Section \ref{sub: SM-MCTS-A convergence}, but some of the propositions have slightly different form.

\subsubsection{Definition of the UPO property}
\begin{notation}
\label{not: property}Let $h\in\mathcal{H}$ be a node. We will take a closer
look at what is happening at $h$. Let $h_{ij}$ be the children of
$h$. Since the events in $h$ and above do not affect what happens
in $h_{ij}$ (only the time when does it happen), we denote by $s_{ij}\left(1\right),\, s_{ij}\left(2\right),...$
the sequence of payoffs we get for sampling $h_{ij}$ for the first
time, the second time and so on. The correspondence between these
numbers $s_{ij}$ and the payoffs $x_{ij}$ observed in $h$ is $x_{ij}\left(t\right)=s_{ij}\left(\left(t-1\right)_{ij}+1\right)$, where $(t-1)_{ij}$ is the number of uses of joint action $(i,j)$ up to time $t-1$.

Note that all of these objects are, in fact, random variables and their distribution depends on the used selection policy.
By $\bar{s}_{ij}\left(n\right)=\frac{1}{n}\sum_{m=1}^{n}s_{ij}\left(m\right)$
we denote the standard arithmetical average of $s_{ij}$. Finally, setting $t^*_{ij}\left(k\right)=\min\left\{ t\in\mathbb{N}|\, t_{ij}=k\right\} $,
we define the weights $w_{ij}\left(k\right)$ and the weighted
average $\tilde{s}_{ij}\left(k\right)$: 
\[
w_{ij}\left(n\right)=1+\left|\left\{ t\in\mathbb{N}|\,t^*_{ij}\left(n-1\right)\leq t\leq t^*_{ij}\left(n\right)\ \& \ t\mbox{ satisfies }j(t)=j\mbox{ but }i(t)\neq i\right\}\right|,
\]
\[
\tilde{s}_{ij}\left(n\right)=\frac{1}{\sum_{m=1}^{n}w_{ij}\left(m\right)}\sum_{m=1}^{n}w_{ij}\left(m\right)s_{ij}\left(m\right).
\]
\end{notation}
\begin{remark}[Motivation for the definition of UPO property]
If our algorithm $A$ is $\epsilon$-HC, we know that if $h_{ij}$ is a node with $d_{h_{ij}}=1$ and value $v_{ij}$, then $\limsup_{n}\left|\bar{s}_{ij}\left(n\right)-v_{ij}\right|\leq\epsilon$
(Lemma \ref{L: HC-and-NE} \eqref{eq: HC a NE2}, where $g\left(n\right)=\bar{s}_{ij}\left(n\right)$). In more vague words, ``we have some information about $\bar{s}_{ij}$'', therefore, we would prefer to work with these ``simple'' averages. Unfortunately, the variables, which naturally appear in the context of SM-MCTS are the ``complicated'' averages $\tilde{s}_{ij}$ - we will see this in the proof of Theorem \ref{thm: SM-MCTS convergence} and it also follows from the fact that, in general, there is no relation between quality the performance of SM-MCTS and the value of differences $\bar{s}_{ij}\left(n\right)-v_{ij}$ (see Section \ref{sub: Counterexample} for a counterexample). This leads to the following definition:
\end{remark}

\begin{definition}[UPO] \label{def: UPO}
We say that an algorithm $A$ guarantees $\epsilon$-unbiased payoff observations, if for every (simultaneous-move zero-sum perfect information) game $G$, every node $h$ and actions $i$, $j$, the arithmetic averages $\bar{s}_{ij}$ and weighted averages $\tilde{s}_{ij}$ almost surely satisfy
\[ \underset{t\rightarrow\infty}{\limsup}\left|\tilde{s}_{ij}\left(n\right)-\bar{s}_{ij}\left(n\right)\right|\leq\epsilon.\]
We will sometimes abbreviate this by saying that ``$A$ is $\epsilon$-UPO algorithm''.
\end{definition}
Observe that this in particular implies that if, for some $c>0$,
\[
\limsup_{n\rightarrow\infty}\,\left|\bar{s}_{ij}\left(n\right)-v_{ij}\right|\leq c\epsilon
\]
holds almost surely, then we also have 
\[
\limsup_{n\rightarrow\infty}\,\left|\tilde{s}_{ij}\left(n\right)-v_{ij}\right|\leq\left(c+1\right)\epsilon\mbox{ a.s..}
\]

Next, we present a few examples which motivate the above definition and support the discussion that follows.
\begin{example}[Examples related to the UPO property]\label{ex: UPO}$\ $
\begin{enumerate} \item Suppose that $w_{ij}(n), s_{ij}(n), n\in\mathbb{N}$ do not necessarily originate from SM-MCTS algorithm, but assume they satisfy:
\begin{enumerate}
\item $w_{ij}(n),\ s_{ij}(n),\ n\in\mathbf{N}$ are independent
\item $\exists C>0\ \forall n\in\mathbb{N}\ :\  w_{ij}(n)\in[0,C]$
\item $\forall n\in\mathbb{N}:\  s_{ij}(n)\in[0,1]\ \&\ \mathbf{E}[|s_{ij}(n)-v_{ij}|]\leq \frac \epsilon 2$ for some $v_{ij}\in[0,1]$.
\end{enumerate}
Then, by strong law of large numbers, we almost surely have
\begin{equation}
\underset {n\rightarrow\infty} \limsup |\bar s_{ij}(n)-\tilde s_{ij}(n)|\leq\epsilon. \label{eq: X}
\end{equation}
\item The previous case can be generalized in many ways - for example it is sufficient to replace bounded $w_{ij}(n)$ by ones satisfying
\[ \exists q\in(0,1)\ \forall n\ \forall i,j:\ \mathbf{Pr}[w_{ij}(n)\geq k]\leq q^k \]
(an assumption which holds with $q=\gamma/\left|\mathcal{A}_1(h)\right|$ when $w_{ij}(n), s_{ij}(n)$  originate from SM-MCTS with fixed exploration). Also, the variables $w_{ij}(n)$, $s_{ij}(n)$ do not have to be fully independent - it might be enough if the correlation between each $s_{ij}(n)$ and $w_{ij}(n)$ was ``low enough for most $n\in\mathbb{N}$''.
\item In Section \ref{sec:Experimental} we provide empirical evidence, which suggests that when the variables $s_{ij}(n),\ w_{ij}(n)$ originate from SM-MCTS with Exp3 or RM selection policy, then the assertion \eqref{eq: X} of 1. holds as well (and thus these two $\epsilon$-HC algorithms are $\epsilon$-UPO).
\item Assume that $(s_{ij}(n))_{n=1}^\infty=(1,0,1,0,1,...)$ and $(w_{ij}(n))_{n=1}^\infty=(1,3,1,3,1,...)$. Then we have $\bar s_{ij}(n)\rightarrow\frac 1 2$, but $\tilde s_{ij}(n)\rightarrow\frac 1 4$.
\item In Section \ref{sub: Counterexample} we construct an example of $\epsilon$-HC algorithm, based on 4., such that when it is used as a selection policy in a certain game, we have $\underset {n\rightarrow\infty} \limsup |\bar s_{ij}(n)-\tilde s_{ij}(n)|\geq \frac 1 4$.
\end{enumerate}
\end{example}
The cases 2. and 3. from Example \ref{ex: UPO} suggest that it is possible to prove that specific $\epsilon$-HC algorithms are $\epsilon$-UPO. On the other hand, 5. shows that the implication $(\textrm{A is }\epsilon\textrm{-HC}\implies \textrm{A is } C\epsilon\textrm{-UPO})$ does not hold, no matter how high $C>0$ we choose. Also, the guarantees we have about the behavior of, for example, Exp3 are much weaker than the assumptions made in 1 - there is no independence between $w_{ij}(n), s_{ij}(m), \ m,n\in\mathbb{N}$, at best we can use some martingale theory. Moreover, even in nodes $h\in\mathcal{H}$ with $d_h=1$, we have $\underset {n\rightarrow\infty} \limsup |\bar s_{ij}(n)-v_{ij}|\leq \epsilon$, instead of assumption  $(c)$ from 1.. This implies that the proof that specific $\epsilon$-HC algorithms are $\epsilon$-UPO will not be trivial.

\subsubsection{The proof of Theorem \ref{thm: SM-MCTS convergence}}
The following proposition shows that if the assumption holds, then having low regret in some
$h\in\mathcal{H}$ with respect to observed rewards is sufficient to bound the regret with respect to the rewards originating from
the matrix game $\left(v_{ij}\right)$.
\begin{proposition}
\label{prop: when tilde s = bar s}Let $h\in\mathcal{H}$ and
$\epsilon,c\geq0$. Let $A$ be an $\epsilon$-HC algorithm which generates the sequence of actions $\left(i(t)\right)$ at $h$ and suppose that the adversary chooses actions $\left(j(t)\right)$. If $\underset{n\rightarrow\infty}{\limsup}\,\left|\bar{s}_{ij}\left(n\right)-v_{ij}\right|\leq c\epsilon$
holds a.s. for each $i,j$ and $A$ is $\epsilon$-UPO, then we almost
surely have 
\begin{equation}
\limsup_{t\rightarrow\infty}\frac{1}{t}\left(\max_{i(0)}\sum_{s=1}^{t}v_{i(0)j(s)}-\sum_{s=1}^{t}v_{i(s)j(s)}\right)\leq2\left(c+1\right)\epsilon.\label{eq: regret}
\end{equation}
Consequently the choice of actions $\left(i(t)\right)$ made by the
algorithm $A$ is $2\left(c+1\right)\epsilon$-HC with respect to
the matrix game $\left(v_{ij}\right)$.
\end{proposition}
The proof of this proposition consists of rewriting the sums in inequality \eqref{eq: regret}
and using the fact that the weighted averages $\tilde{s}_{ij}$ are
close to the standard averages $\bar{s}_{ij}$. Denote by $\left(IH_{d}^{'}\right)$
the claim, which is the same as $\left(IH_{d}\right)$ from paragraph
\ref{par: (IH)} except that it concerns SM-MCTS algorithm rather than
SM-MCTS-A and $C_{d}=3\cdot2^{d-1}-2$. Lemma \ref{L: HC-and-NE} then
immediately gives the following corollary. Analogously to the Section
\ref{sub: SM-MCTS-A convergence}, this in turn implies the main theorem
of this section, the proof of which is similar to the proof of Theorem \ref{thm: SM-MCTS-A convergence}.
\begin{corollary}
\label{cor: ind.step SM-MCTS}$\left(IH_{d}^{'}\right)\implies\left(IH_{d+1}^{'}\right)$.\end{corollary}
\begin{proof}
By Lemma \ref{L: HC-and-NE}, the implication holds for \emph{some} constants  $C_{d}$. It remains to show that $C_{d}=3\cdot2^{d-1}-2$. We proceed by backward induction - since the algorithm $A$ is $\epsilon$-HC, we know that, by Lemma \ref{L: HC-and-NE}, $\left(IH_{1}^{'}\right)$ holds with $C_{1}=1$.
For $d\geq2$, Proposition \ref{prop: when tilde s = bar s} implies $C_{d+1}=2\left(C_{d}+1\right)$. A classical induction then gives the result.
\end{proof}

\begin{proof}[Proof of theorem {\ref{thm: SM-MCTS convergence}}]
Using Corollary \ref{cor: ind.step SM-MCTS}, the proof is identical to the proof of Theorem \ref{thm: SM-MCTS-A convergence} - it remains to determine the new value of $C$. As in the proof of Theorem \ref{thm: SM-MCTS-A convergence} we have $C=2\cdot2\sum_{d=1}^{D}C_d$, and we need to calculate this sum: 
\[
\sum_{d=1}^{D}C_{d}=\sum_{d=1}^{D}\left(3\cdot2^{d-1}-2\right)=3\left(1+...+2^{D-1}\right)-2D=3\left(2^{D}-1\right)-2D.
\]
\end{proof}

\section{Exploitability and exploration removal\label{sec: exploitability}}

One of the most common measure of the quality of a strategy in imperfect information games is the notion of \emph{exploitability} \citep[for example,][]{Johanson2011}. It will be useful for the empirical evaluation of our main result  in Section \ref{sec:Experimental}, as well as for the discussion of lower bounds in Section \ref{sec:Counterexample}. In this section, we first recall the definition of this notion and we follow with few observations concerning which strategy should be considered the output of SM-MCTS(-A) algorithms.
\begin{definition}
\emph{Exploitability} of strategy $\sigma_{1}$ of player 1 is the quantity
\[ \textrm{expl}_{1}\left(\sigma_{1}\right):=v-u\left(\sigma_{1},\textrm{br}\right),
\]
where $v$ is the value of the game and $\textrm{br}$ is a second player's
best response strategy to $\sigma_{1}$. Analogously we define
$\textrm{expl}_{2}$ for the second player's strategies.
\end{definition}
Clearly we always have $\textrm{expl}_{i}\left(\sigma_{i}\right)\ge0$, $i=1,2$ and a strategy
profile $ $$\sigma=\left(\sigma_{1},\sigma_{2}\right)$ is a Nash
equilibrium iff $\textrm{expl}_{1}\left(\sigma_{1}\right)=\textrm{expl}_{2}\left(\sigma_{2}\right)=0$.

\begin{remark}[Removing the exploration]
In SM-MCTS(-A) we often use a selection function with fixed exploration parameter $\gamma>0$, such that the algorithm is guaranteed to converge to $C\gamma$-equilibrium for some constant $C>0$ (for example Exp3 or regret matching). \citet{teytaud2011upper} suggest removing the random noise caused by this exploration from the resulting strategies, but they do it heuristically and do not formally analyze this procedure.
By definition of exploration, the average strategy $(\bar{\sigma}_{1}(t),\bar{\sigma}_{2}(t))$ produced by SM-MCTS(-A) algorithms is of the form 
\[ \bar{\sigma}_{i}\left(t\right)=\left(1-\gamma\right)\bar{p}_{i}\left(t\right)+\gamma\cdot \textrm{rnd} \]
for some strategy $\bar{p}_{i}\left(t\right)$, where $\textrm{rnd}$ is the strategy used when exploring, assigning to each action the same probability.

In general, $\textrm{rnd}$ will not be an equilibrium strategy of our game. This means that for small values of $\gamma$ and high enough $t$,
so that the algorithms have time to converge (that is when $\bar{\sigma}_{i}\left(t\right)$ is reasonably good), we have 
\[ \textrm{expl}_{i}\left(\textrm{rnd}\right)> C\gamma\geq \textrm{expl}_{i}\left(\bar{\sigma}_{i}\left(t\right)\right). \]
And finally since the function $\textrm{expl}_{i}$ is linear, we have
\begin{eqnarray*}
C\gamma & \geq &  \textrm{expl}_{i}\left(\bar{\sigma}_{i}\left(t\right)\right)\\
 & = & \left(1-\gamma\right)\textrm{expl}_{i}\left(\bar{p}_{i}\left(t\right)\right)+\gamma\cdot\textrm{expl}_{i}\left(rnd\right)\\
 & \geq & \left(1-\gamma\right)\textrm{expl}_{i}\left(\bar{p}_{i}\left(t\right)\right)+\gamma\cdot C\gamma.
\end{eqnarray*}
 This necessarily implies that $\textrm{expl}_{i}\left(\bar{p}_{i}\left(t\right)\right)\leq \textrm{expl}_{i}\left(\bar{\sigma}_{i}\left(t\right)\right)$.
\end{remark}
We can summarize this remark by the following proposition (the proof of which
consists of using the fact that utility is a linear function):
\begin{proposition}
Let $\bar{\sigma}\left(t\right)=\left(\bar{\sigma}_{1}\left(t\right),\bar{\sigma}_{2}\left(t\right)\right)$
be the average strategy. Let $\gamma>0$ and set 
\[
\bar{p}_{i}\left(t\right):=\frac{1}{\left(1-\gamma\right)}\bar{\sigma}_{i}\left(t\right)-\frac{\gamma}{\left(1-\gamma\right)}\textup{\textrm{rnd}}.
\]
 Then the following holds: \\
 {$\left(1\right)$} \textup{$\textrm{expl}_{i}\left(\bar{p}_{i}\left(t\right)\right)\leq \textrm{expl}_{i}\left(\bar{\sigma}_{i}\left(t\right)\right)+\gamma/\left(1-\gamma\right)$}.\\
{$\left(2\right)$} If \textup{$\textrm{expl}_{i}\left(\textrm{rnd}\right)>C\gamma\geq \textrm{expl}_{i}\left(\bar{\sigma}_{i}\left(t\right)\right)$} \textup{\emph{holds for some $C>0$, then the strategy $\bar{p}_{i}\left(t\right)$ satisfies }$\textrm{expl}_{i}\left(\bar{p}_{i}\left(t\right)\right)<\textrm{expl}_{i}\left(\bar{\sigma}_{i}\left(t\right)\right)$.}
\end{proposition}

Less formally speaking, there are two possibilities. First is that our algorithm had so little time to converge that it is better to disregard its output $\bar{\sigma}_{i}\left(t\right)$ and play randomly instead. If this is not the case, then by (2) it is always better to remove the exploration and use the strategy $\bar{p}_i(t)$ instead of $\bar{\sigma}_{i}\left(t\right)$. And by (1), even if we remove the exploration, we cannot increase the exploitability of $\bar{p}_i(t)$ by more than $\gamma/(1-\gamma)$. We illustrate this by experiments presented in Section \ref{sec:Experimental},  where we compare the quality of strategies $\bar{p}_{1}\left(t\right)$ and $\bar{\sigma}_{1}\left(t\right)$.

\section{Counterexample and lower bounds\label{sec:Counterexample}}
In this section we first show that the dependence of constant $C$ from Theorems \ref{thm: SM-MCTS-A convergence} and \ref{thm: SM-MCTS convergence} on the depth $D$ of the game tree cannot be improved below linear dependence. Main result of this section is then an example showing that, without the $\epsilon$-UPO property, Theorem \ref{thm: SM-MCTS convergence} does not hold.

\subsection{Dependence of the eventual NE distance on the game depth}
\begin{proposition}
There exists $k>0$, such that none of the Theorems \ref{thm: SM-MCTS-A convergence} and \ref{thm: SM-MCTS convergence} hold if the constant $C$ is replaced by $\tilde C=kD\epsilon$. This remains true even when the exploration is removed from the strategy $\hat \sigma$.
\end{proposition}
The proposition above follows from Example \ref{example: upper bound}.

\begin{figure}[t]
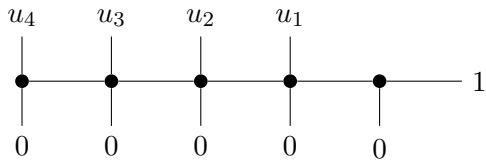

\centering
\usebox{\LinBound}
\caption{A single-player game where the quality of a strategy has linear dependence on the exploration parameter and on the game depth $D$. The numbers $u_d$ satisfy $0<u_1<u_2<\dots<u_D<1$.}\label{fig: C is linear in D}
\end{figure}

\begin{example}\label{example: upper bound}
Let $G$ be the single player game\footnote{The other player always has only a single \emph{no-op} action.} from Figure \ref{fig: C is linear in D}, $\eta>0$ some small number, and $D$ the depth of the game tree. Let Exp3 with exploration parameter $\gamma=k\epsilon$ be our $\epsilon$-HC algorithm (for a suitable choice of $k$). We recall that this algorithm will eventually identify the optimal action and play it with frequency $1-\gamma$, and it will choose randomly otherwise. Denote the available actions at each node as (up, right, down), resp. (right, down) at the rightmost inner node. We define each of the rewards $u_d$, $d=1,...,D-1$ in such a way that Exp3 will always prefer to go up, rather than right. By induction over $d$, we can see that the choice $u_1=1-\gamma/2+\eta$, $u_{d+1}=(1-\gamma/3)u_d$ is sufficient and for $\eta$ small enough, we have
\[ u_{D-1}=(1-\frac \gamma 2+\eta)(1-\frac \gamma 3)\dots (1-\frac \gamma 3)\leq \left(1-\frac \gamma 3\right)^{D-1} \doteq 1-\frac {D-1} 3 \gamma \]
(where by $\doteq$ we mean that for small $\gamma$, in which we are interested, the difference between the two terms is negligible).
Consequently in each of the nodes, Exp3 will converge to the strategy $\left(1-\frac 2 3 \gamma, \frac 1 3 \gamma, \frac 1 3 \gamma\right)$  (resp.  $\left(1-\frac \gamma 2,\frac \gamma 2\right)$), which yields the payoff of approximately $(1-\gamma/3)u_{D-1}$. Clearly, the expected utility of such a strategy is approximately
\[ u={\left(1-\gamma/3\right)}^D \doteq 1-\frac D 3\gamma. \]
On the other hand, the optimal strategy of always going right leads to utility 1, and thus our strategy $\hat \sigma$ is $\frac{D}{3}\gamma$-equilibrium.

Note that in this particular example, it makes no difference whether SM-MCTS or SM-MCTS-A is used. We also observe that when the exploration is removed, the new strategy is to go up at the first node with probability 1, which again leads to regret of approximately $\frac D 3 \gamma$.
\end{example}
 By increasing the branching factor of the game in the previous example from 3 to $b$ (adding more copies of the ``$0$'' nodes) and modifying the values of $u_d$ accordingly, we could make the above example converge to $2 \frac {b-2} b D\gamma$-equilibrium (resp. $\frac {b-2} b D\gamma$ once the exploration is removed).
 
 In fact, we were able to construct a game of depth $D$ and $\epsilon$-HC algorithms, such that the resulting strategy $\hat \sigma$ converged to $3D\epsilon$-equilibrium ($2D\epsilon$ after removing the exploration). However, the $\epsilon$-HC algorithms used in this example are non-standard and would require the introduction of more technical notation. Therefore, since in our main theorem we use quadratic dependence $C=kD^2$, we instead choose to highlight the following open question:
\begin{problem}\label{prob: linearity in D}
Does Theorem \ref{thm: SM-MCTS-A convergence} (and Theorem \ref{thm: SM-MCTS convergence}) hold with $C=kD$ for some $k>0$ (or is the presented bound tight)?
\end{problem}
It is our hypothesis that the answer is affirmative (and possibly the values $k=3$, resp. $k=2$ after exploration removal, are optimal), but the proof of such proposition would require techniques different from the one used in the proof of Theorem \ref{thm: SM-MCTS-A convergence}.

\subsection{\label{sub: Counterexample}Counterexample for Theorem \ref{thm: SM-MCTS convergence}}

Recall that in Section \ref{sec: convergence} we proved two theorems of the following form:
\begin{proposition} \label{prop: dummy}
Let $A$ be an $\epsilon$-HC algorithm with guaranteed exploration and let $G$ be (zero-sum simultaneous moves perfect information) game. If $A$ is used as selection policy for SM-MCTS(-A), then the empirical frequencies will eventually get arbitrarily close to $C\epsilon$-NE of the whole game, for some $C>0$.
\end{proposition}
The goal of this subsection is to prove the following theorem:
\begin{theorem} \label{thm: counterexample}
There exists a simultaneous move zero-sum game $G$ with perfect information and a 0-HC algorithm $A$ with guaranteed exploration, such that when $A$ is used as a strategy for SM-MCTS algorithm (rather than SM-MCTS-A), then the average strategy $\hat{\sigma}\left(t\right)$ almost surely does not converge to the set of $\frac 1 4$-Nash equilibria of $G$.
\end{theorem}
This in particular implies that no theorem similar to Proposition \ref{prop: dummy} holds for SM-MCTS, unless $A$ satisfies some additional assumptions, such as being an $\epsilon$-UPO algorithm.

We now present some observations regarding the proof of Theorem \ref{thm: counterexample}.
\begin{remark}\label{rem: WLOG}
Firstly, it is enough to find the game $G$ and construct for each $\epsilon>0$
an algorithm $A_{\epsilon}$ which is $\epsilon$-HC, but $\hat{\sigma}\left(t\right)$
does not converge to the set of $\frac{1}{4}$-NE. From these $\epsilon$-HC
algorithms $A_{\epsilon}$, the desired $0$-HC algorithm can be constructed
in a standard way - that is using 1-Hannan consistent algorithm $A_{1}$ for some period $t_{1}$, then $\frac 1 2$-HC algorithm $A_{1/2}$ for a longer period $t_{2}$ and so on. By choosing a sequence $\left(t_{n}\right)_n$, which increases quickly enough, we can guarantee that the resulting combination of algorithms $\left(A_{1/n}\right)$ is 0-Hannan consistent.

Furthermore, we can assume without loss of generality that the algorithm $A$ knows if it is
playing as the first or the second player and that in each node of
the game, we can actually use a different algorithm $A$. This is true, because the algorithm always accepts a number of available actions as input. Therefore we could define the algorithm differently based on this number, and modify our game $G$ in some trivial way (such as duplicating rows or columns) which would not affect our example.
\end{remark}

The structure of the proof of Theorem \ref{thm: counterexample} is now as follows. First, in Example \ref{ex: ideal case}, we introduce game $G$ and a sequence of joint actions leading to
\[
\sum_{h\in\mathcal{H}}r^h(t^h(T))=0 \textrm{ }\&\textrm{ } r^G(T)=\frac 1 4.
\]
This behavior will serve as a basis for our counterexample. However, the ``algorithms'' generating this sequence of actions will be oblivious to the actions of opponent, which means that they will not be $\epsilon$-HC. In the second step of our proof, we modify these algorithms in such a way that the resulting sequence of joint actions stays similar to the original sequence, but the new algorithms are $\epsilon$-HC. Theorem \ref{thm: counterexample} then follows from Lemma \ref{lemma: modification of A} and Remark \ref{rem: WLOG}.

\begin{figure}[th]
\centering{}
\includegraphics[width=0.3\textwidth]{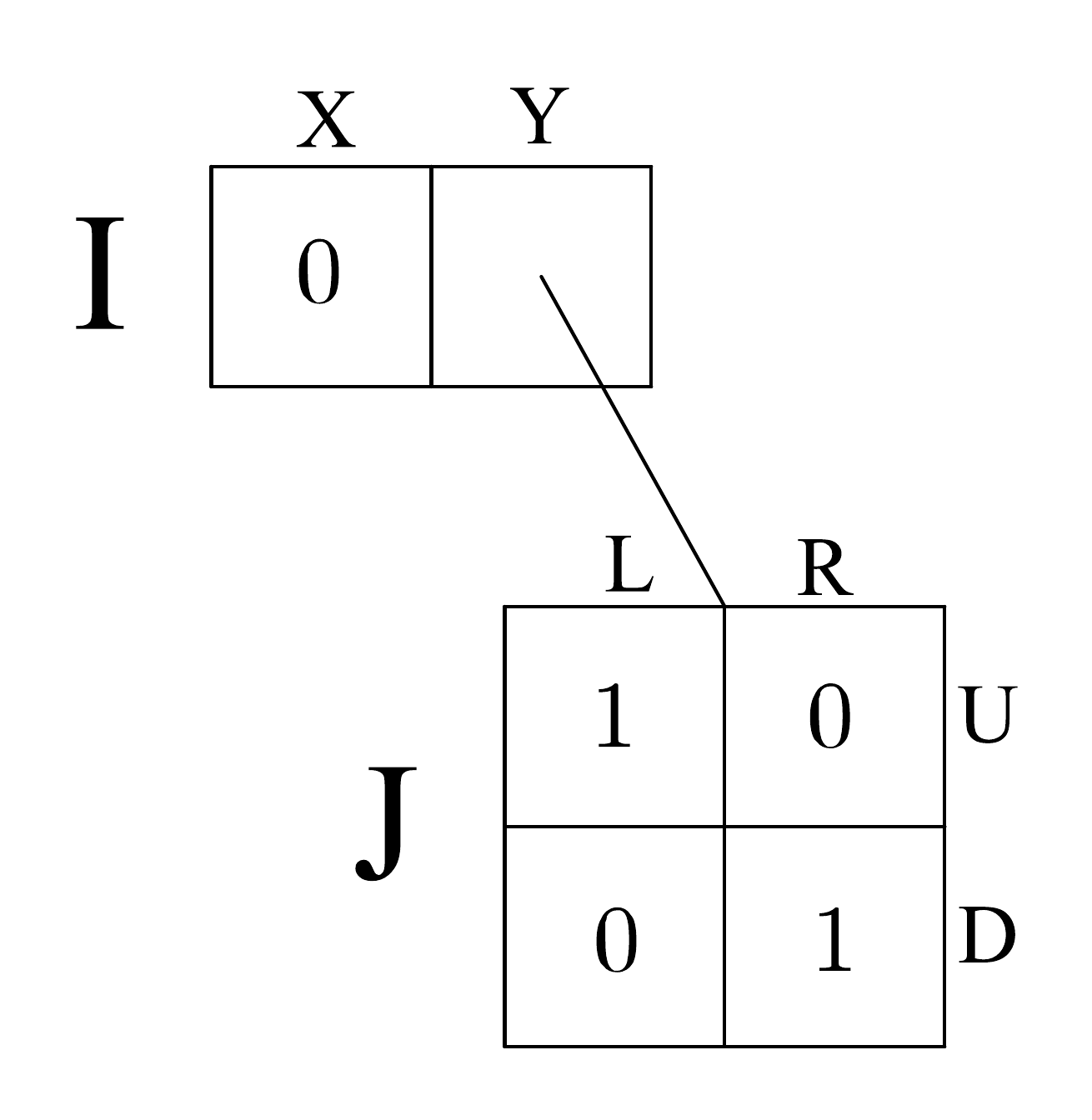}
\caption[Game $G$]{Example of a game in which it is possible to minimize regret at each of the nodes while having high overall regret.}\label{fig: hra}
\end{figure}

\begin{example}\label{ex: ideal case}
The game: Let $G$ be the game from Figure \ref{fig: hra}.

Behavior at $J$: At the node $J$, the players repeat (not counting the iterations
when the play does not reach $J$) the pattern (U,L), (U,R), (D,R), (D,L), generating payoff sequence
\[ s_{Y}\left(1\right),s_{Y}\left(2\right),...=1,0,1,0,... .\]
Looking at time steps of the form $t=4k$, $k\in\mathbb N$, the average strategy $J$ will then be $\hat{\sigma}_{1}^{J}=\hat{\sigma}_{2}^{J}=\left(\frac{1}{2},\frac{1}{2}\right)$ and the corresponding payoff of the maximizing player 1 is $\frac{1}{2}$. Note that neither of the players could improve his utility at $J$ by changing all his actions to any single action, therefore for both players, we have $r^J(t)=0$.

Behavior at $I$: Let $T=4k$. At the node $I$, player 1 repeatedly plays $Y,X,X,Y,...$. For iteration $t$ and action $a$, we denote by $x_a(t)$ the reward we would receive if we played $a$ at node $I$ at time $t$, provided we repeated the $Y,X,X,Y$ pattern up to iteration $t-1$ and used the above defined behavior at $J$ (formally we have $x_{X}(t)=0, x_{Y}(t)=s_Y\left((t-1)_Y+1\right)$, where, as always, $(t-1)_Y$ denotes the number of uses of action $Y$ up to time $t-1$). Denote by $a(t)$ the action played at time $t$. The payoffs $x_{a(t)}$ we actually do receive will then form a 4-periodic sequence
\[ x_{Y}\left(1\right)=1, x_{X}\left(2\right)=0, x_{X}\left(3\right)=0, x_{Y}\left(4\right)=0. \]
Clearly if we change the strategy from the current $\sigma^{I}=\left(\frac{1}{2},\frac{1}{2}\right)$ to $\left(0,1\right)$, we would receive an average payoff $\frac{1}{2}$. This means that the average overall regret of the whole game $G$ for player 1 is equal to $r^G(T)=\frac 1 4$. However, if we look only at the situation at node $I$ and represent it as a bandit problem, we see that the payoff sequence $x_{Y}\left(1\right),x_{Y}\left(2\right),...$ for
action $Y$ will be $1,0,0,0,1,0,0,0,...$ (while $x_{X}\left(t\right)=0$ for each $t$). At first, this might seem strange, but note that the reward for action $Y$ does not change when $X$ is chosen. This implies that, from the MAB point of view, the player believes he cannot receive an average payoff higher than $\frac{1}{4}$ and thus he observes no regret and $r^I(T)=0$.
\end{example}
\begin{remark}
Recall here the definition of $\epsilon$-UPO property of an algorithm, which requires the ``observed average payoffs'' $\tilde s_a(t)$ for all actions $a$ to be close to the real average payoffs $\bar s_a(t)$. In this case, we have $\bar s_B(t)=\frac 1 2$ and $\tilde s_B(t)=\frac 1 4$, which means that the above algorithm is far from being $\epsilon$-UPO.
\end{remark}

\begin{lemma}\label{lemma: modification of A}
Let $G$ be the game from Figure~\ref{fig: hra}. Then for each $\epsilon>0$
there exist $\epsilon$-HC algorithms $A^{I},\, A_{1}^{J},\, A_{2}^{J}$,
such that when these algorithms are used for SM-MCTS in $G$, the resulting average strategy $\hat{\sigma}\left(t\right)$
converges to $\hat{\sigma}^{I}=\hat{\sigma}_{1}^{J}=\hat{\sigma}_{2}^{I}=\left(\frac{1}{2},\frac{1}{2}\right)$.
\end{lemma}

As noted in Example \ref{ex: ideal case}, the strategy $\hat{\sigma}$ satisfies $u_{1}\left(\hat{\sigma}\right)=\frac{1}{4}$, while the equilibrium strategy $\pi$, where $\pi^{I}=\left(0,1\right)$,~$\pi_{1}^{J}=\pi_{2}^{J}=\left(\frac{1}{2},\frac{1}{2}\right)$, gives utility $u_{1}\left(\pi\right)=\frac{1}{2}$. Therefore the existence of algorithms from Lemma \ref{lemma: modification of A} proves Theorem \ref{thm: counterexample}.

The key idea behind Lemma \ref{lemma: modification of A} is the following: both players repeat the pattern from Example \ref{ex: ideal case}, but we let them perform random checks which detect any adversary who deviates enough to change the average payoff. If the players repeat the pattern, by the previous example they observe no regret at any of the nodes. On the other hand, if one of them deviates significantly, he will be detected by the other player, who then switches to a ``safe'' $\epsilon$-HC algorithm, leading again to a low regret. The definition of the modified algorithms used in Lemma \ref{lemma: modification of A}, along with the proof of their properties, can be found in the appendix.

We recall that there exists an algorithm, called CFR \citep{zinkevich2007regret}, which provably converges in our setting. The following remark explains why the proof of its convergence cannot be simply modified to work for SM-MCTS(-A), but a new proof had to be found instead.
\begin{remark}[CFR and bounding game regret by sum of node regrets]\label{rem: CFR}
The convergence of CFR algorithm relies on two facts: firstly, in each node $h\in\mathcal H$, the algorithm minimizes so called average immediate counterfactual regret, which we denote here by $R^{h,+}_\textrm{imm}(T)/T$. Secondly, the overall average regret in the whole game, which we denote by $r^G(T)$, can be bounded by the sum of ``local'' regrets in the game nodes:
\begin{equation}
r^G(T)\leq\sum_{h\in\mathcal{H}}R^{h,+}_\textrm{imm}(T)/T \label{eq: cfr}
\end{equation}
 \citep[Theorem 3 by][]{zinkevich2007regret}. It is then well known that when both players have low overall regret $r^G(T)$, the average strategy is close to an equilibrium.
 
We now look at the similarities between this situation for CFR and for SM-MCTS. $\epsilon$-HC algorithms, used by SM-MCTS, guarantee that the average  regret $r^h(t)$ is, in the limit, at most $\epsilon$ at every $h\in\mathcal H$. In other words, SM-MCTS also minimizes some kind of regret in each of the nodes $h\in\mathcal{I}$, like CFR does. It is then logical to ask whether it is also possible to bound $r^G(T)$ by the sum of ``local'' regrets $\sum_{h\in\mathcal{H}}r^h(t^h(T))$, like for counterfactual regret in CFR (where by $t^h(T)$ we denote the ``local time'' at node $h$, or more precisely the number of visits of node $h$ during SM-MCTS iterations 1,...,T). The following proposition, which is an immediate consequence of Theorem \ref{thm: counterexample}, gives a negative answer to this question.
\begin{corollary}
There exists a game $G$ and $\alpha>0$, such that for every $\beta>0$, there exists a sequence of joint actions resulting in
\[
\sum_{h\in\mathcal{H}}r^h(t^h(T))<\beta \textrm{ }\&\textrm{ } \alpha\leq r^G(T).
\]
In particular, the inequality $r^G(T)\leq\sum_{h\in\mathcal{H}}r^h(t^h(T))$ does not hold and this approach which worked for CFR cannot be applied to SM-MCTS. Intuitively, this is caused by the differences between the two distinct notions of regret used by SM-MCTS and CFR.
\end{corollary}
\end{remark}

\section{Experimental evaluation}\label{sec:Experimental}
In this section, we present the experimental data related to our theoretical results. First, we empirically evaluate our hypothesis that Exp3 and regret matching algorithms ensure the $\epsilon$-UPO property. Second, we test the empirical convergence rates of SM-MCTS and SM-MCTS-A on synthetic games as well as smaller variants of games played by people. We investigate the practical dependence of the convergence error based on the important parameters of the games and evaluate the effect of removing the samples due to exploration from the computed strategies.

\subsection{Experimental Domains}

\paragraph{\textbf{Goofspiel}}
Goofspiel is a card game that appears in many works dedicated to simultaneous-move games (for example \cite{Ross71Goofspiel,Rhoads12Computer,saffidine2012,lanctot2013goof,bosansky2013-ijcai}).
There are $3$ identical decks of cards with values $\{0,\dots, (d-1)\}$ (one for nature and one for each player).
Value of $d$ is a parameter of the game.
The deck for the nature is shuffled at the beginning of the game. In each round, nature reveals the top card from its deck. Each player selects any of their remaining cards and places it face down on the table so that the opponent does not see the card. Afterwards, the cards are turned face up and the player with the higher card wins the card revealed by nature. The card is discarded in case of a draw. At the end, the player with the higher sum of the nature cards wins the game or the game is a draw.
People play the game with $13$ cards, but we use smaller numbers in order to be able to compute the distance from the equilibrium (that is, exploitability) in a reasonable time.
We further simplify the game by a common assumption that both players know the sequence of the nature's cards in advance.

\paragraph{\textbf{Oshi-Zumo}}

Each player in Oshi-Zumo (for example, \cite{buro2003}) starts with $N$ coins, and a one-dimensional playing board with $2K+1$ locations (indexed $0, \ldots, 2K$) stretches between the players.
At the beginning, there is a stone (or a wrestler) located in the center of the board (that is, at position $K$).
During each move, both players simultaneously place their bid from the amount of coins they have (but at least one if they still have some coins).
Afterwards, the bids are revealed, the coins used for bids are removed from the game, and the highest bidder pushes the wrestler one location towards the opponent's side.
If the bids are the same, the wrestler does not move.
The game proceeds until the money runs out for both players, or the wrestler is pushed out of the board.
The player closer to the wrestler's final position loses the game.
If the final position of the wrestler is the center, the game is a draw.
In our experiments, we use a version with $K=2$ and $N=5$.

\paragraph{\textbf{Random Game}}

In order to achieve more general results, we also use randomly generated games. The games are defined by the number of actions $B$ available to each player in each decision point and a depth $D$ ($D=0$ for leaves), which is the same for all branches. The utility values in the leafs are selected randomly form a uniform distribution over $\langle 0,1\rangle$.

\begin{figure}[t]
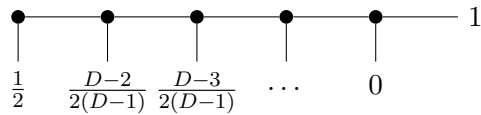

\centering
\usebox{\AntiL}
\caption{The Anti game used for evaluation of the algorithms.}\label{fig:anti}
\end{figure}

\paragraph{\textbf{Anti}}

The last game we use in our evaluation is based on the well-known single player game, which demonstrates the super-exponential convergence time of the UCT algorithm \citep{coquelin2007bandit}. The game is depicted in Figure~\ref{fig:anti}. In each stage, it deceives the MCTS algorithm to end the game while it is optimal to continue until the end.

\subsection{$\epsilon$-UPO property}
In order to be able to apply Theorem \ref{thm: SM-MCTS convergence} (that is, convergence of SM-MCTS without averaging) to Exp3 and regret matching, the selection algorithms have to assure the $\epsilon$-UPO property for some $\epsilon$.
So far, we were unable to prove this hypothesis.
Instead, we support this claim by the following numerical experiments.
Recall that having $\epsilon$-UPO property is defined as the claim that for every game node $h\in\mathcal H$ and every joint action $\left(i,j\right)$  available at $h$, the difference
$\left|\bar{s}_{ij}\left(n\right)-\tilde{s}_{ij}\left(n\right)\right|$
between the weighted and arithmetical averages decreases below $\epsilon$, as the number $n$ of uses of $(i,j)$ at $h$ increases to infinity.

\begin{figure}[t]
\includegraphics[width=\textwidth]{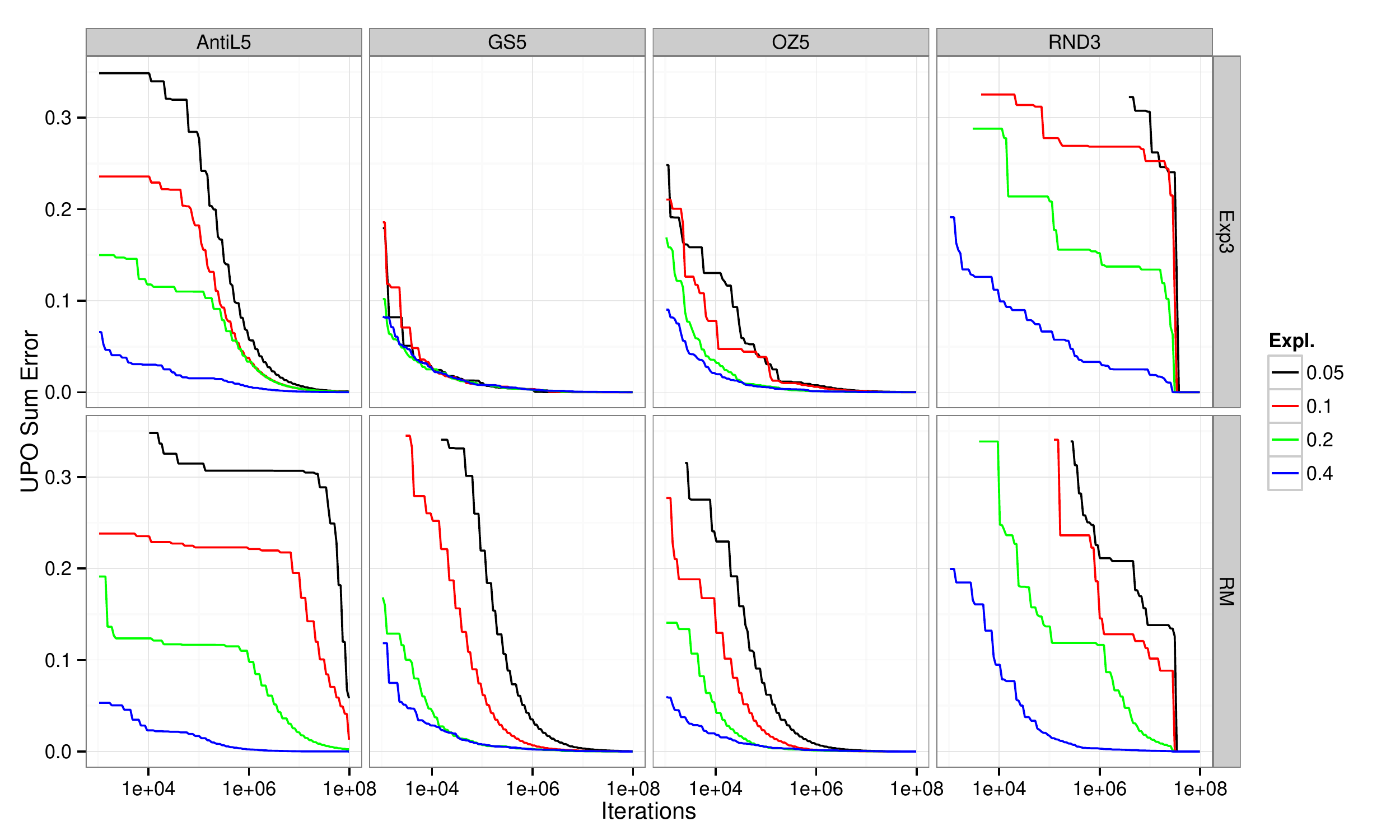}
\caption{The maximum of the bias in payoff observations in MCTS without averaging the sample values.}\label{fig:upoALL}
\end{figure}

We measured the value of this sum in the root node of the four domains described above.
Besides the random games, the depth of the game was set to 5.
For the random games, the depth and the branching factor was $B=D=3$. Figure~\ref{fig:upoALL} presents one graph for each domain and each algorithm. The x-axis is the number of iterations and the y-axis depicts the maximum value of the sum from the iteration on the x-axis to the end of the run of the algorithm. The presented value is the maximum from 50 runs of the algorithms. For all games, the difference eventually converges to zero. Generally, larger exploration ensures that the difference goes to zero more quickly and the bias in payoff observation is smaller.

\begin{figure} 
\centering
\begin{subfigure}{0.49\textwidth}
\includegraphics[width=\textwidth]{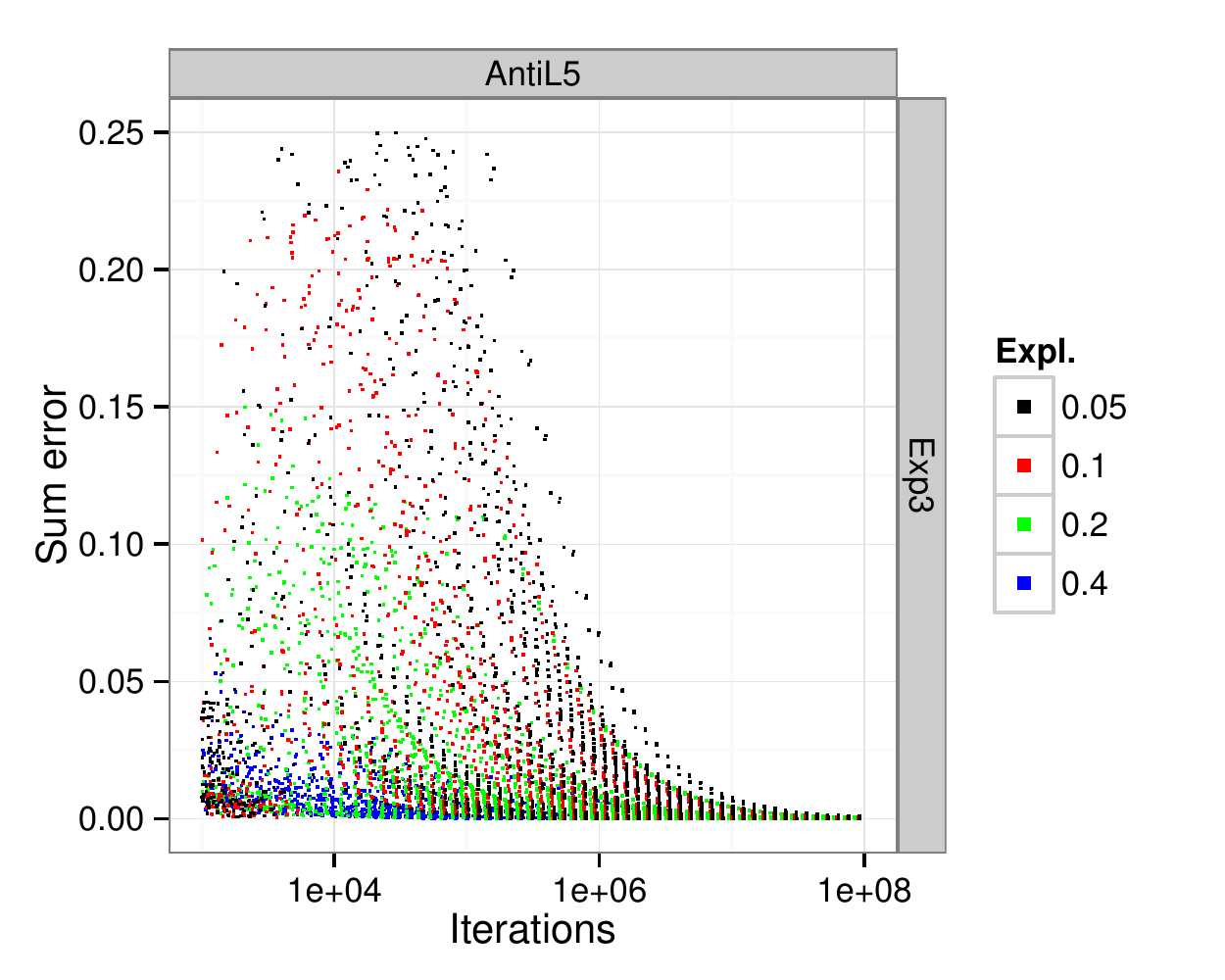}
\caption{Various exploration factors}\label{fig:upoAnti}
\end{subfigure}
\begin{subfigure}{0.49\textwidth}
\includegraphics[width=\textwidth]{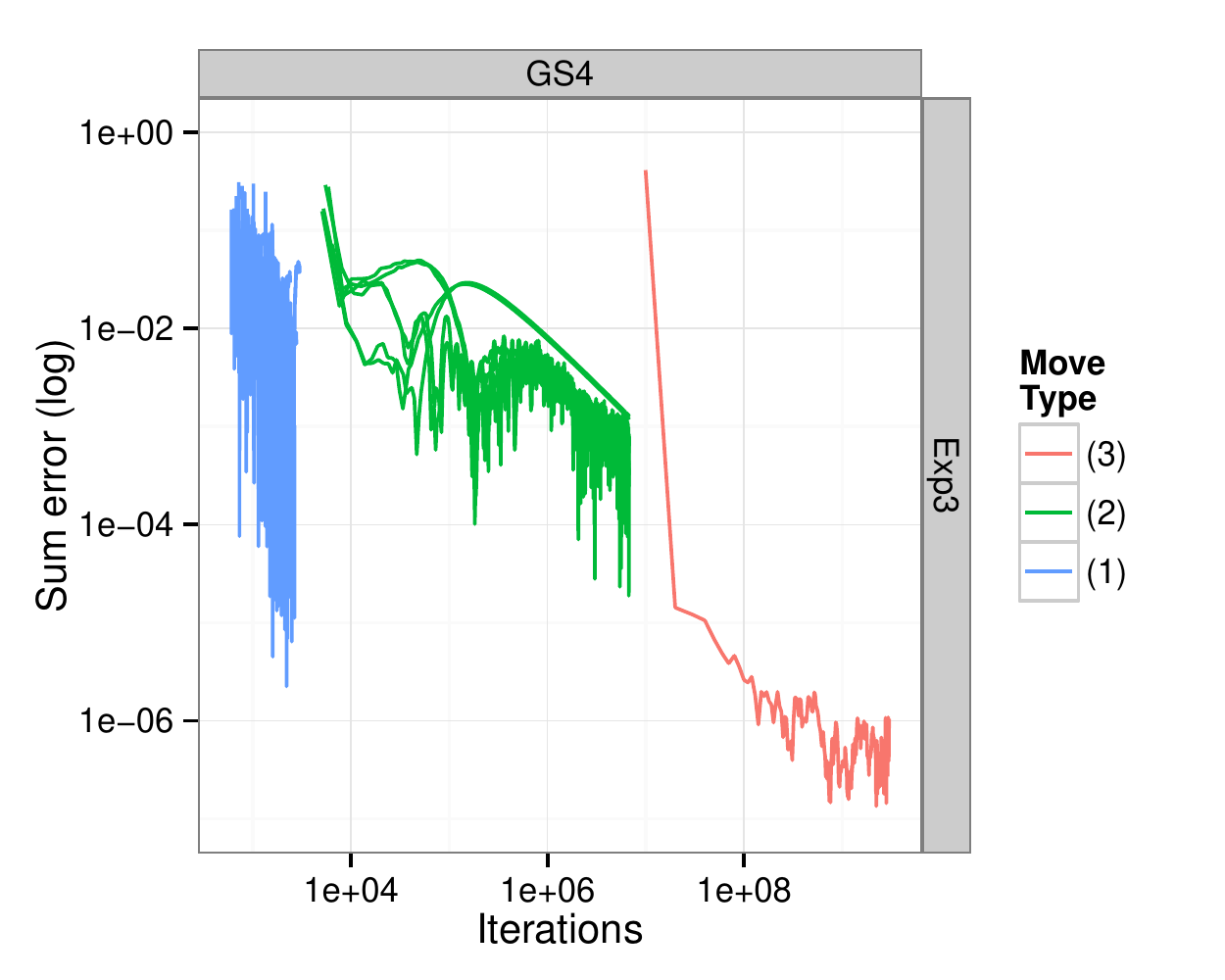}
\caption{Various joint actions}\label{fig:upoGS4}
\end{subfigure}
\caption{The dependence of the current value of $\left|\bar{s}_{ij}\left(n\right)-\tilde{s}_{ij}\left(n\right)\right|$
on the number of iterations that used the given joint action in (a) Anti game and (b) Goofspiel with 4 cards per deck.}
\end{figure}

The main reason for the bias is easy to explain in the Anti game. Figure~\ref{fig:upoAnti} presents the maximal values of the bias in small time windows during the convergence from all 50 runs. It is apparent that the bias during the convergence tends to jump very high (higher for smaller exploration) and then gradually decrease. This, however, happens only until certain point in time. The reason for this behavior is that if the algorithm learns an action is good in a specific state, it will use it very often and do the updates for the action with much smaller weight in $\tilde{s}_{ij}\left(n\right)$ than the updates for the other action. However, when the other action later proves to be substantially better, the value of that action starts increasing rather quickly. At the same time, its probability of being played starts increasing and as a result, the weights used for the received rewards start decreasing. This will cause a strong dependence between the rewards and the weights, which causes the bias. With smaller exploration, it takes more time to identify the better alternative action; hence, when it happens, the wrong action has already accumulated larger reward and the discrepancy between the right values and the probability of playing the actions is even stronger.

We also tested satisfaction of the UPO property in the root node of depth 4 Goofspiel, using  Exp3 algorithm and exploration $\epsilon=0.001$. The results in Figure~\ref{fig:upoGS4} indicate that Exp3 with exploration $0.001$ possesses the $0.001$-UPO property, however this time, much higher $n_{0}$ is required (around $5\cdot10^{6}$).

We can divide the joint actions $\left(i,j\right)$ at the root into three groups: (1) the actions which both players play (nearly) only when exploring, (2) the actions which one of the players chooses only because of the exploration, and (3) the actions which none of the players uses only because of the exploration. In Figure~\ref{fig:upoGS4}, (1) is on the left, (2) in the middle and (3) on the right. The third type of actions easily satisfied $\left|\bar{s}_{ij}\left(n\right)-\tilde{s}_{ij}\left(n\right)\right|\leq\epsilon$, while for the second type, this inequality seems to eventually hold as well. The shape of the graphs suggests that the difference between $\bar{s}_{ij}\left(n\right)$ and $\tilde{s}_{ij}\left(n\right)$ will eventually get below $\epsilon$ as well, however, the $10^9$ iterations we used were not sufficient for this to happen. Luckily, even if the inequality did not hold for these cases, it does not prevent the convergence of SM-MCTS algorithm to an approximate equilibrium. In the proof of Proposition~\ref{prop: when tilde s = bar s}, the term $\left|\bar{s}_{ij}\left(n\right)-\tilde{s}_{ij}\left(n\right)\right|$ is weighted by the empirical frequencies $T_{j}/T$ (or even $T_{ij}/T$). For an action which is only played because of exploration, this number converges to $\epsilon/\left(\mbox{number of actions}\right)$
(resp. $\left(\epsilon/\left(\mbox{number of actions}\right)\right)^{2}$ ), so even if we had $\left|\bar{s}_{ij}\left(n\right)-\tilde{s}_{ij}\left(n\right)\right|=1$, we could still bound the required term by $\epsilon$, which is needed in the proof of the respective theorem.

\subsection{Empirical convergence rate of SM-MCTS(-A) algorithms}

\begin{figure}[tp]
\centering
\includegraphics[width=0.7\textwidth]{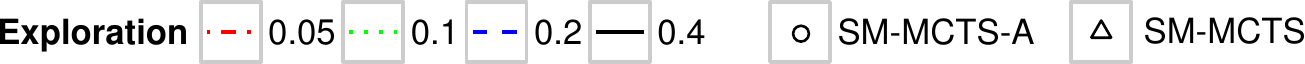}

\begin{subfigure}{0.45\textwidth}
\includegraphics[width=\textwidth]{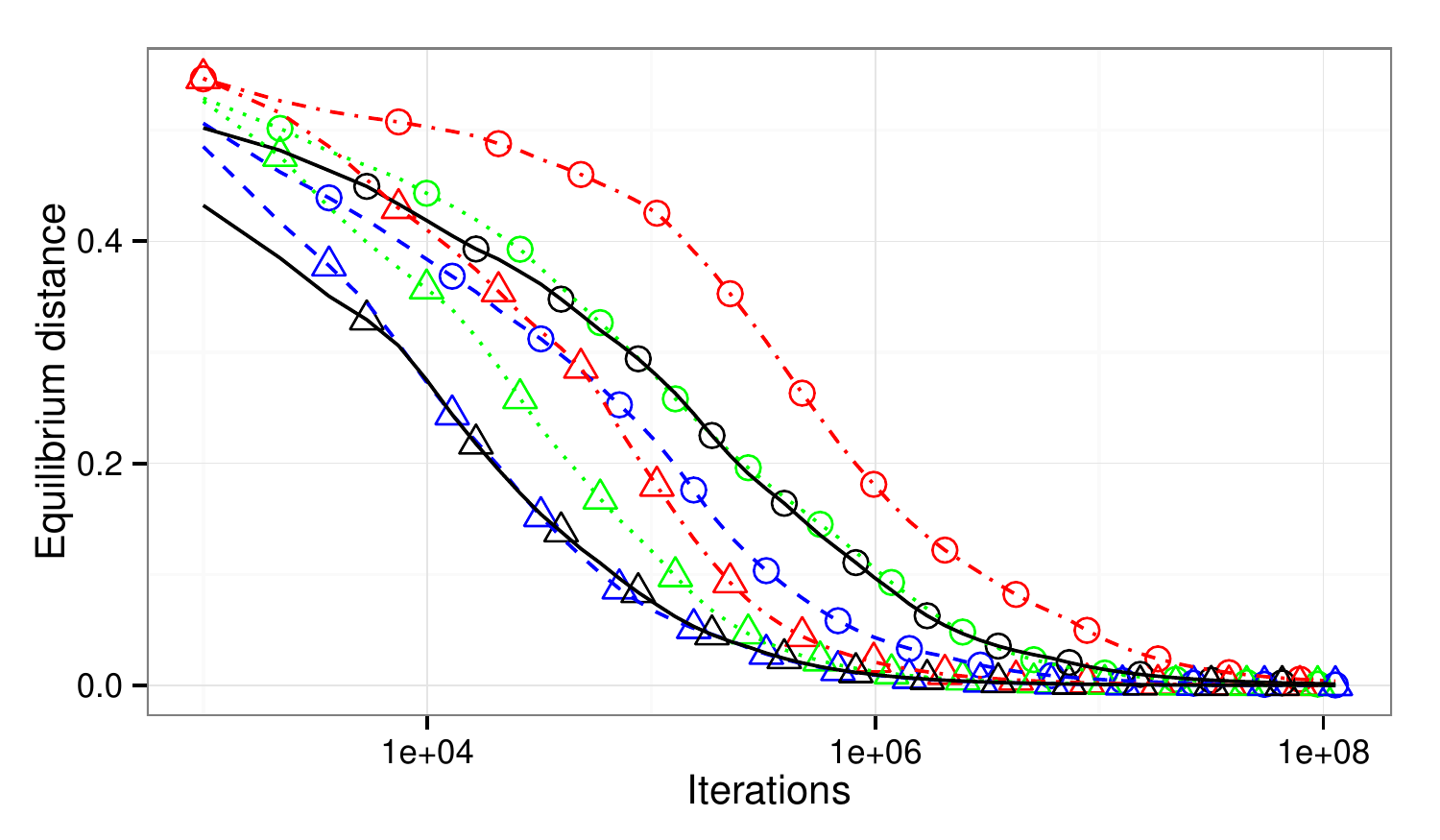}
\caption{Anti(5), Exp3}\label{fig:convAnti_Exp3}
\end{subfigure}
\begin{subfigure}{0.45\textwidth}
\includegraphics[width=\textwidth]{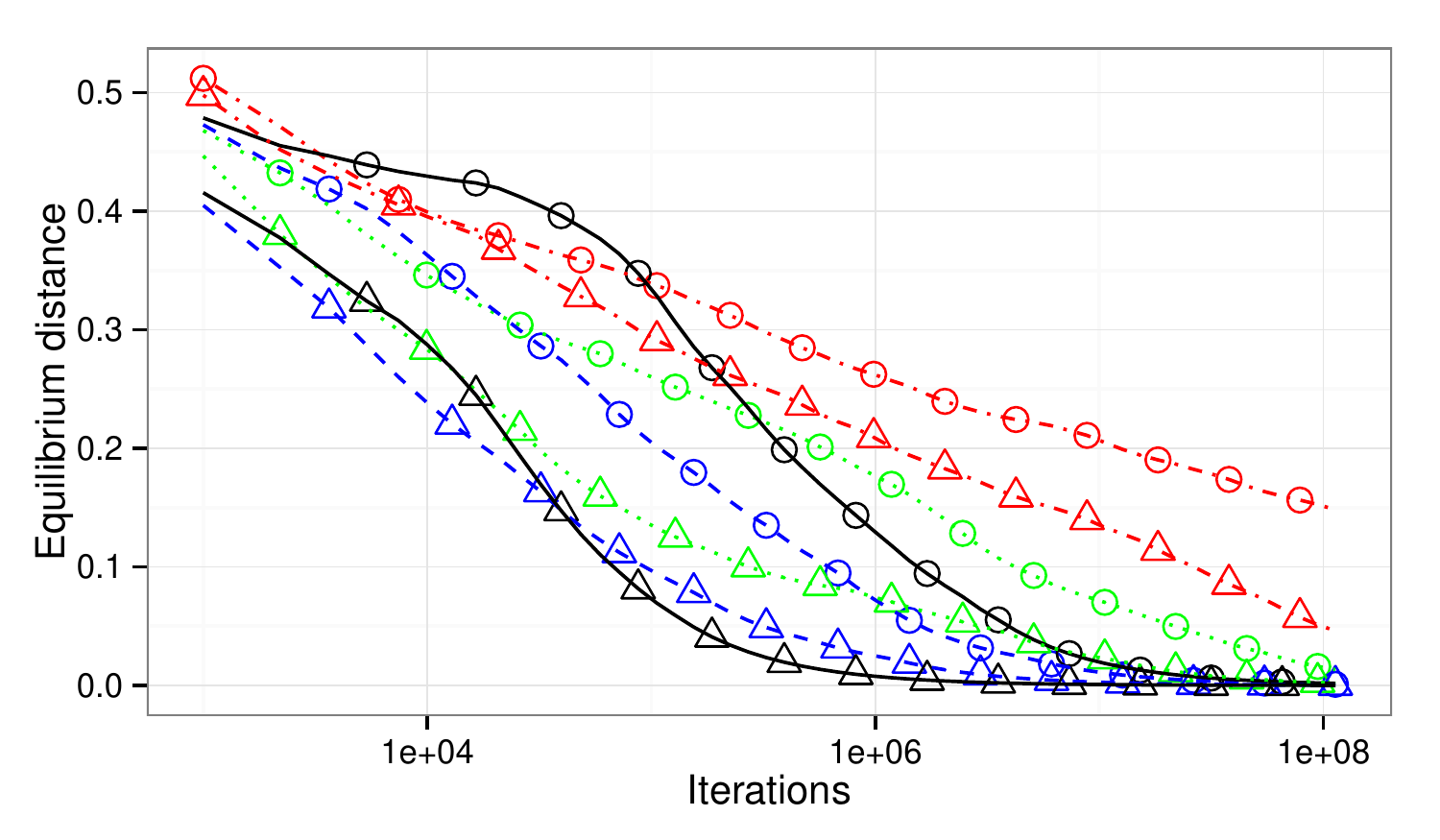}
\caption{Anti(5), RM}\label{fig:convAnti_RM}
\end{subfigure}

\begin{subfigure}{0.45\textwidth}
\includegraphics[width=\textwidth]{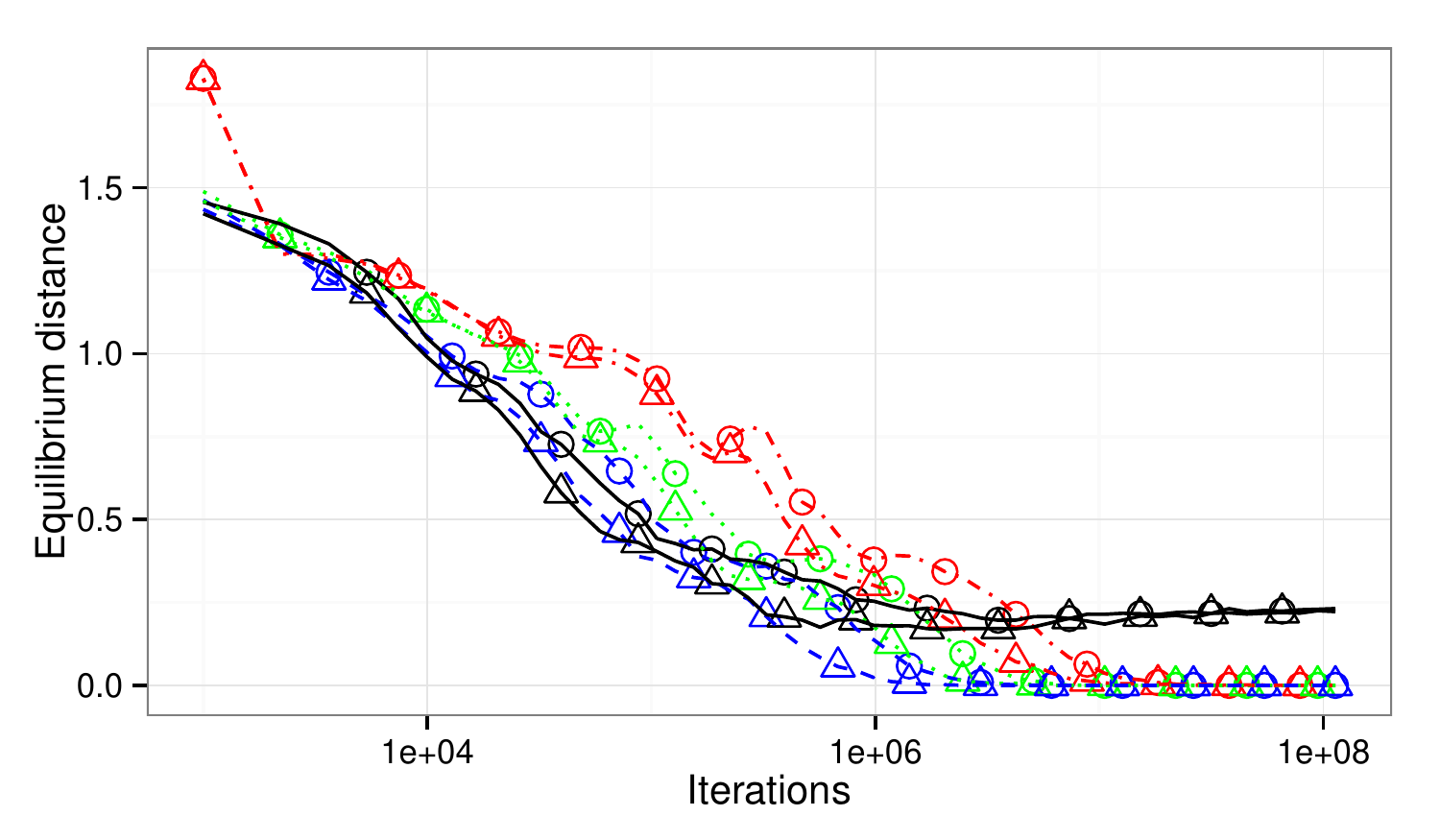}
\caption{Goofspiel(5), Exp3}\label{fig:convGS_Exp3}
\end{subfigure}
\begin{subfigure}{0.45\textwidth}
\includegraphics[width=\textwidth]{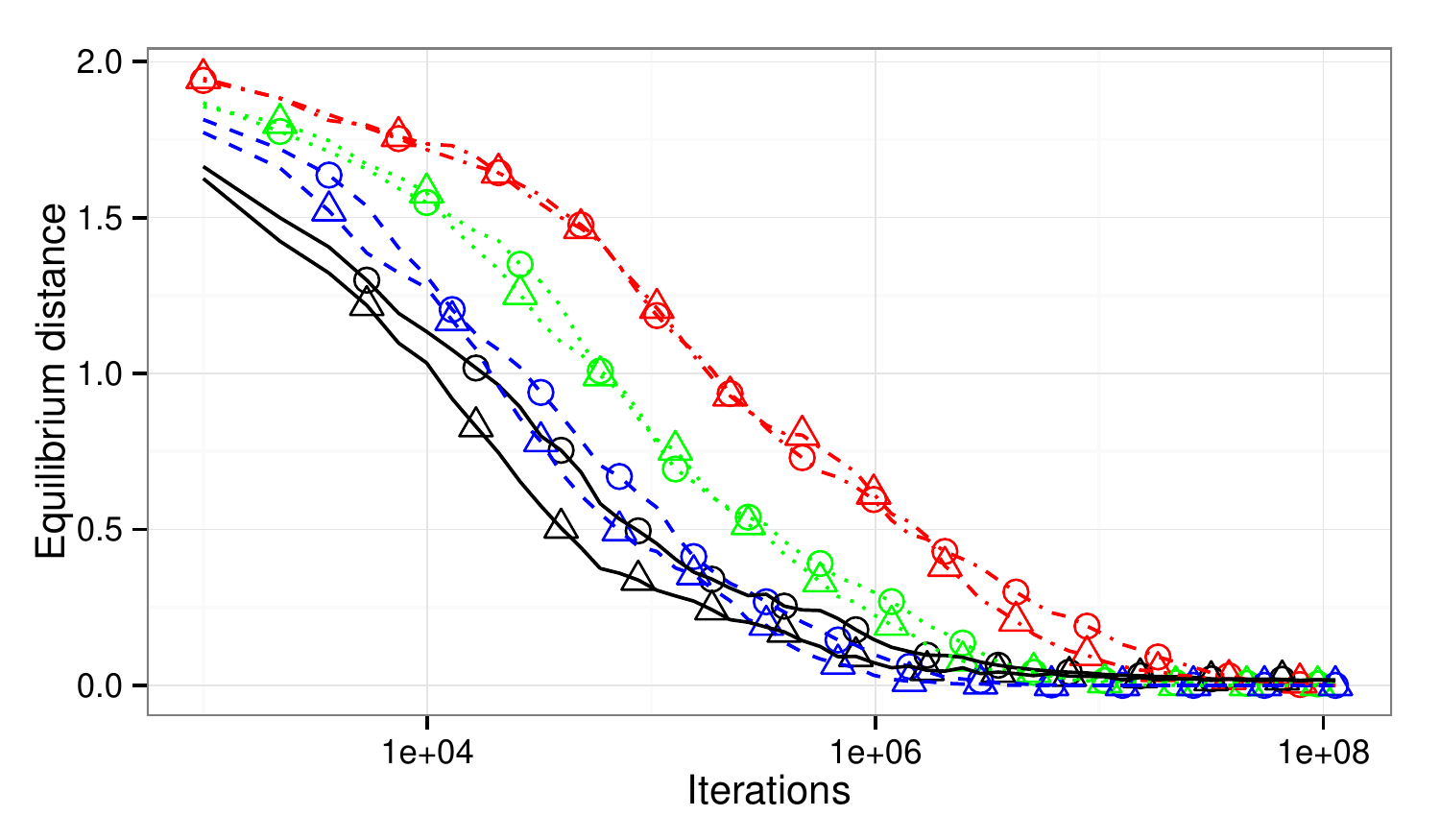}
\caption{Goofspiel(5), RM}\label{fig:convGS_RM}
\end{subfigure}

\begin{subfigure}{0.45\textwidth}
\includegraphics[width=\textwidth]{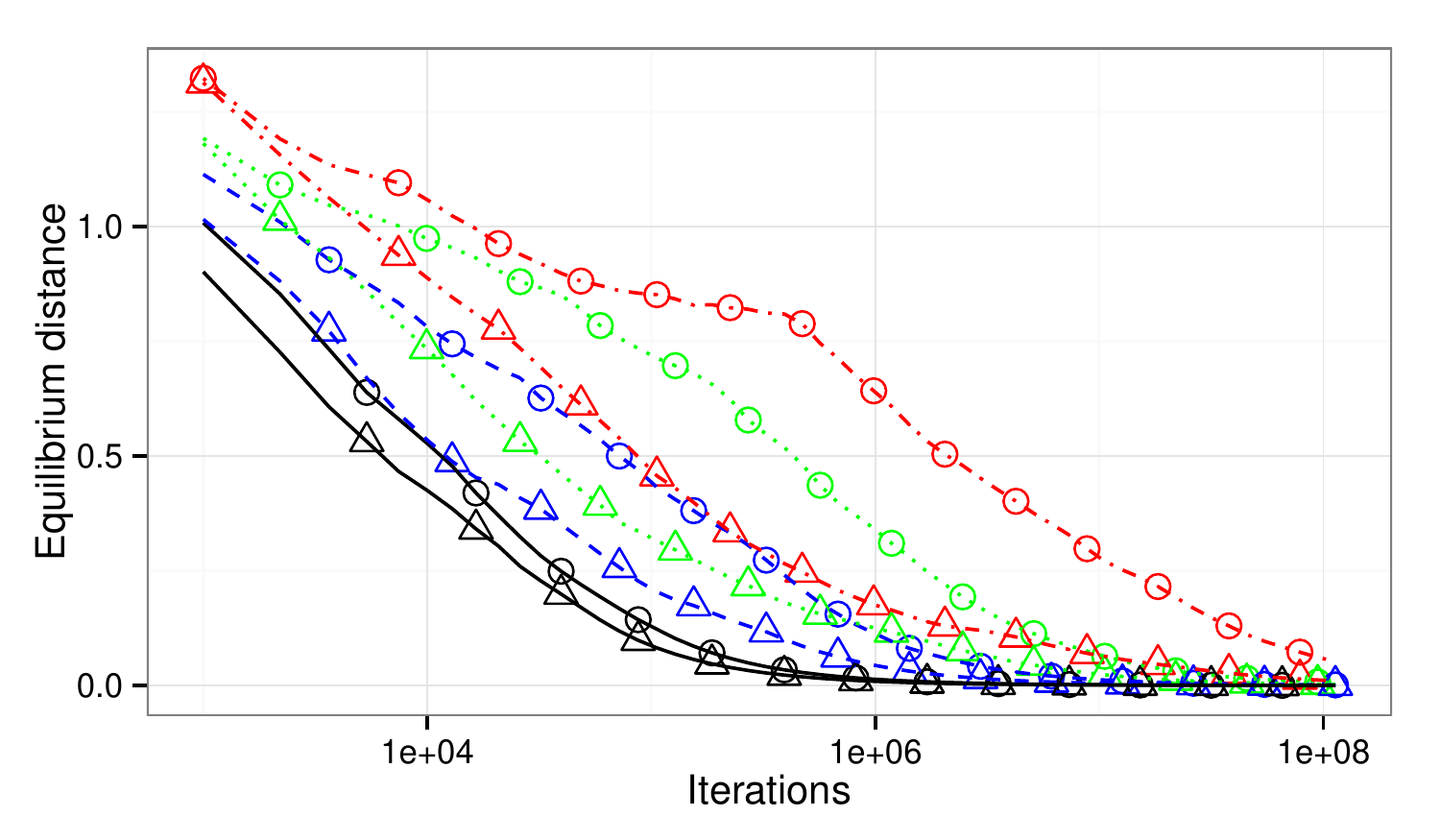}
\caption{Oshi-Zumo(5), Exp3}\label{fig:convAnti_Exp3}
\end{subfigure}
\begin{subfigure}{0.45\textwidth}
\includegraphics[width=\textwidth]{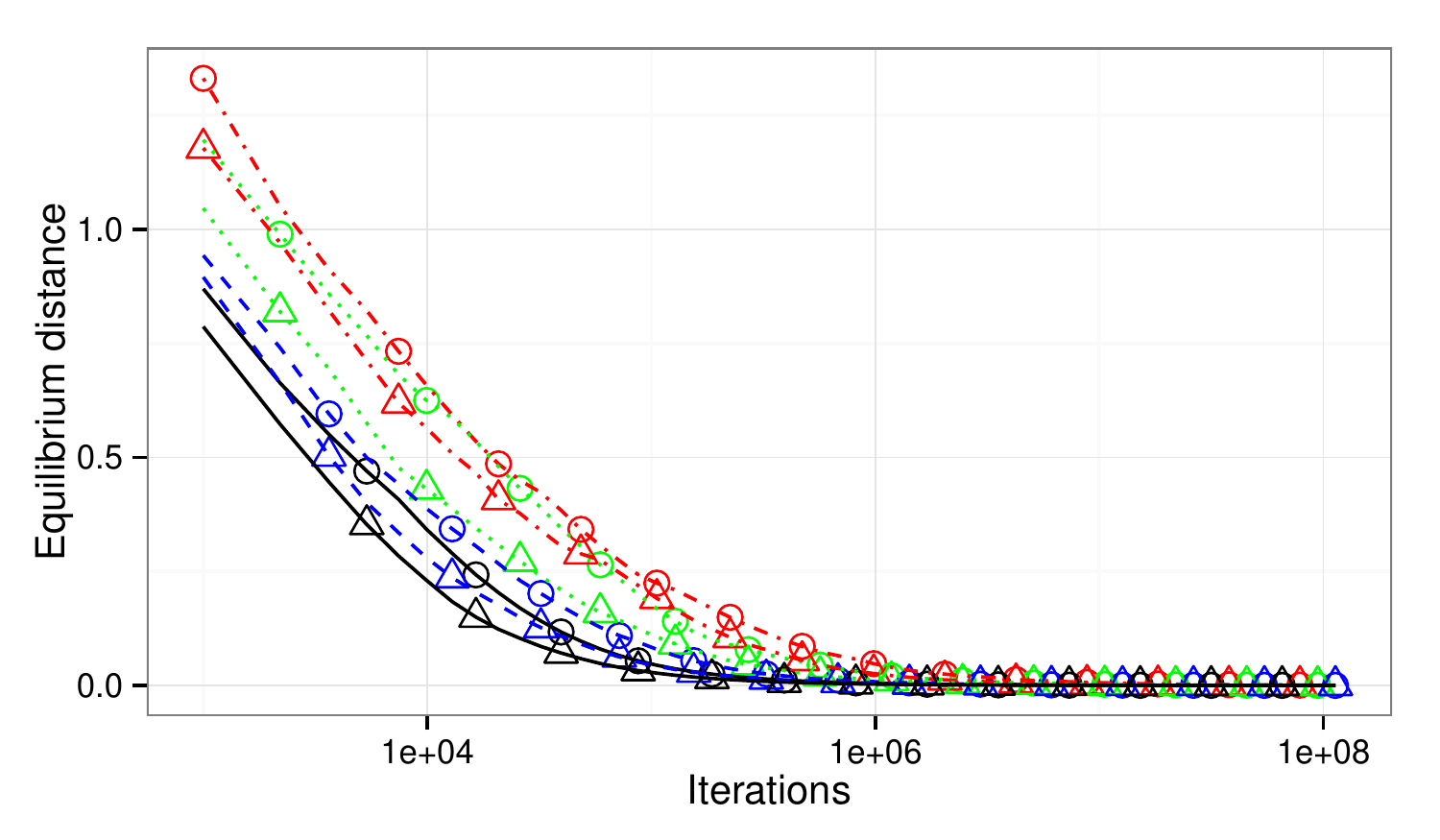}
\caption{Oshi-Zumo(5), RM}\label{fig:convAnti_RM}
\end{subfigure}

\begin{subfigure}{0.45\textwidth}
\includegraphics[width=\textwidth]{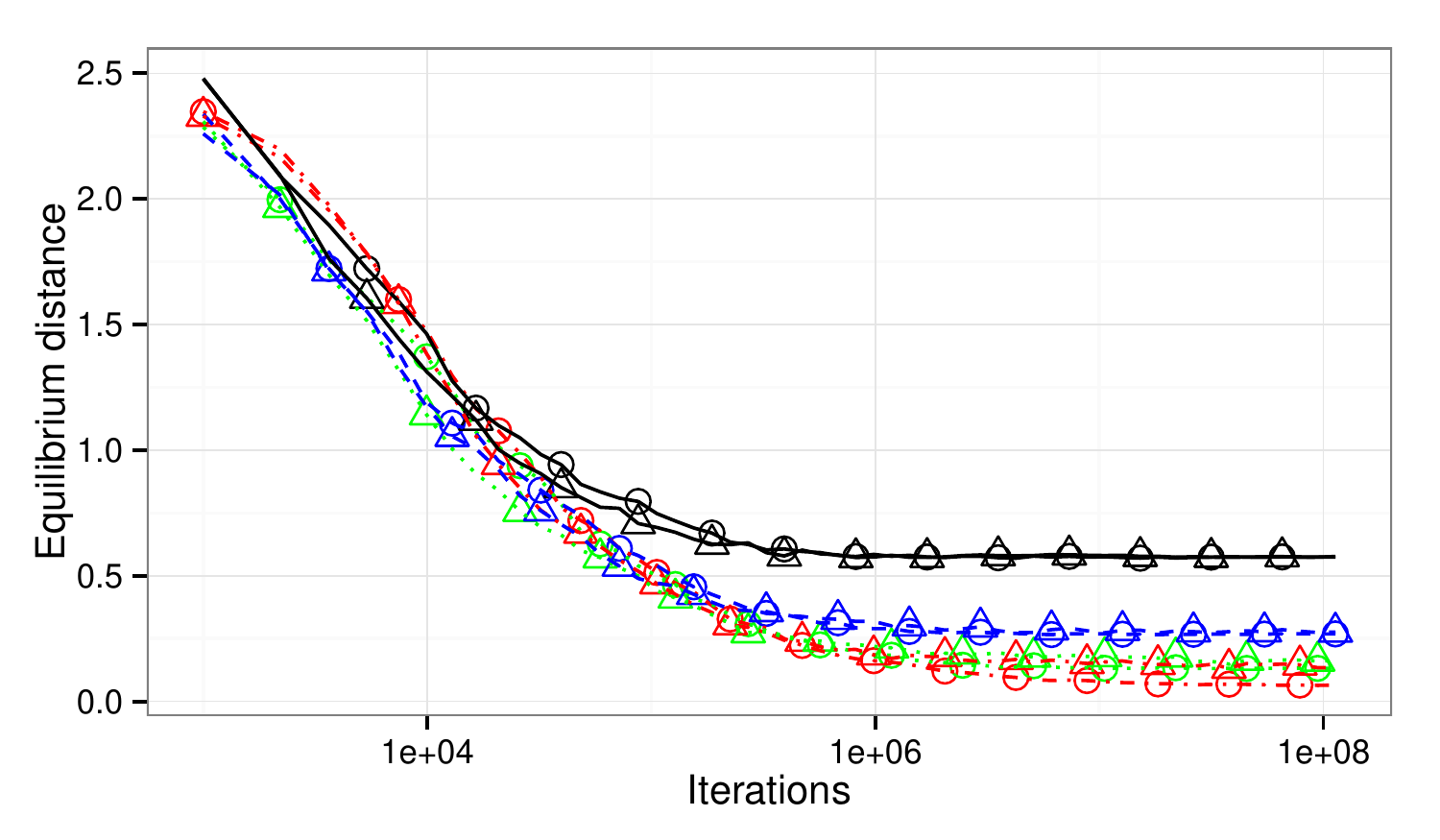}
\caption{Random(3,3), Exp3}\label{fig:convAnti_Exp3}
\end{subfigure}
\begin{subfigure}{0.45\textwidth}
\includegraphics[width=\textwidth]{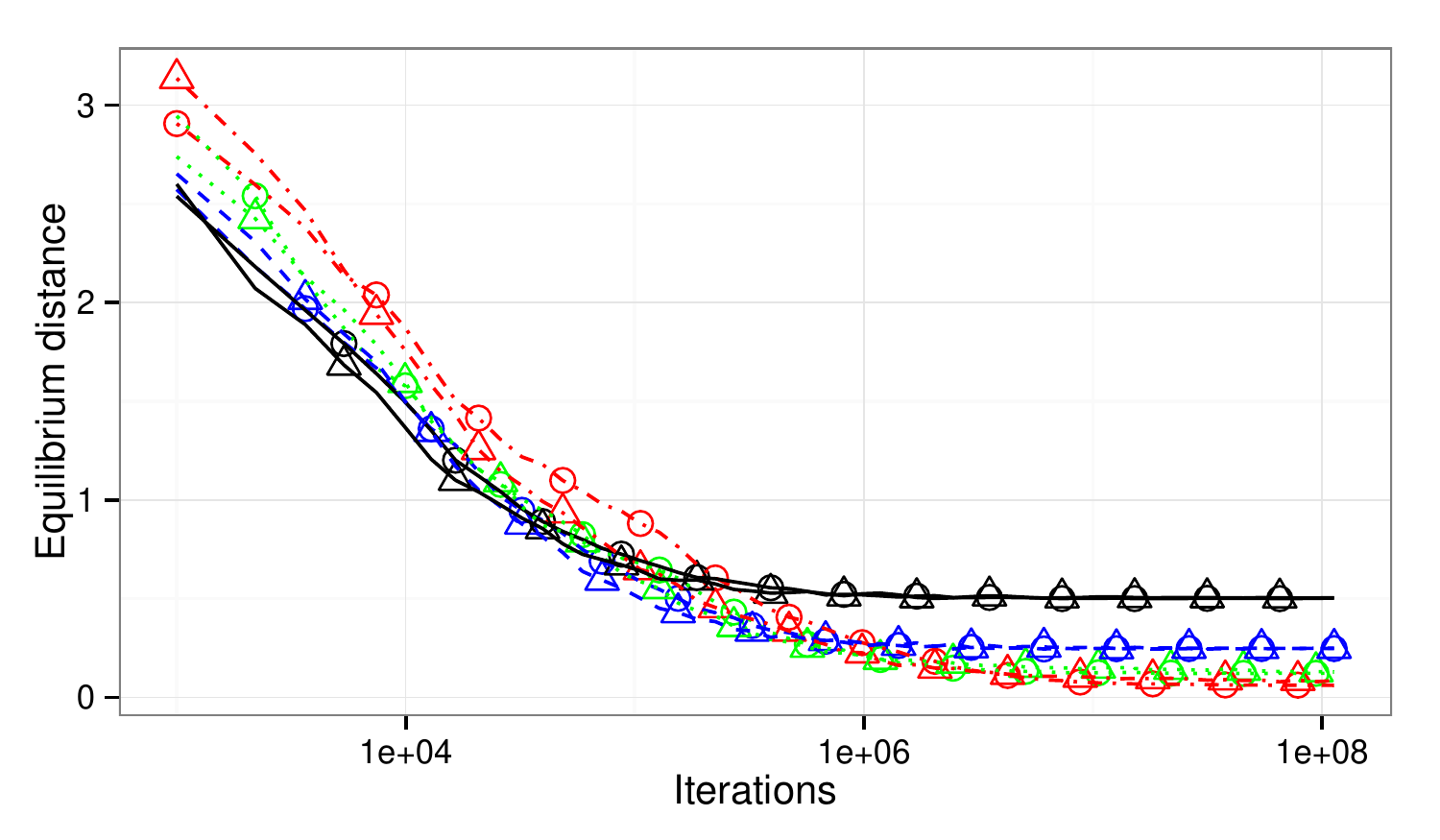}
\caption{Random(3,3), RM}\label{fig:convAnti_RM}
\end{subfigure}
\caption{Comparison of empirical convergence rates of SM-MCTS (triangles) and SM-MCTS-A (circles) with Exp3 and RM selection functions in various domains.}\label{fig:conv}
\end{figure}

In this section, we investigate the empirical convergence rates of the analyzed algorithms. We first compare the speeds of convergence of SM-MCTS-A and SM-MCTS and then investigate the dependence of the error of the eventual solution of the algorithms on relevant parameters. Finally, we focus on the effect of removing the exploration samples discussed in Section~\ref{sec: exploitability}.

\subsubsection{SM-MCTS with and without averaging}

Figure~\ref{fig:conv} presents the dependence of the exploitability of the strategies produced by the algorithms on the number of executed iterations. We removed the samples caused by exploration from the strategy, as suggested in Section~\ref{sec: exploitability}. All iterations are executed from the root of the game. The colors (and line types) in the graphs represent different settings of the exploration parameter. SM-MCTS-A (circles) seems to always converge to the same distance from the equilibrium as SM-MCTS (triangles), regardless of the used selection function.
The convergence of the variant with averaging is generally slower.
The difference is most visible in the Anti game with Exp3 selection function, where the averaging can cause the convergence to require even 10 times more iterations to reach the same distance from NE as the algorithm without averaging. The situation is similar also in Oshi-Zumo. However, the effect is much weaker with RM selection. With the exception of the Anti game, the variants with and without averaging converge at almost the same speed.

In Section~\ref{sec: convergence}, we show that the finite time convergence rate that we were able to prove is not very good. These experiments show that in our practical problems of smaller size, suitable selection of the exploration parameter allows the algorithms to converge to its eventual solution within $10^6$ iterations. This indicates that the bound can be substantially improved.

\subsubsection{Distance from the equilibrium}

Even though the depth of most games in Figure~\ref{fig:conv} was 5, even with large exploration ($0.4$), the algorithm was often able to find the exact equilibrium. This indicates that in practical problems, even the linear bound on the distance from the equilibrium from the example in Section~\ref{sec:Discussion} is too pessimistic.

If the game contains pure Nash equilibria, as in Anti and the used setting of Oshi-Zumo, exact equilibrium can often be found.
If (non-uniform) mixed equilibria are required, the distance of the eventual solution from the equilibrium increases both with the depth of the game as well as the amount of exploration.
The effect of the amount of exploration is visible in Figure~\ref{fig:convGS_Exp3}, where the largest exploration prevented the algorithm from converging to the exact equilibrium. More gradual effect is visible in Figures~\ref{fig:conv}(g,h), where the distance form the equilibrium seems to increase linearly with increasing exploration.
Note that in all cases, the exploitability (computed as the sum of exploitability of both players) was less than $2\cdot\epsilon M$, where $\epsilon$ is the amount of exploration an and M is the maximum utility value.

\begin{figure}[t]
\centering
\includegraphics[width=0.4\textwidth]{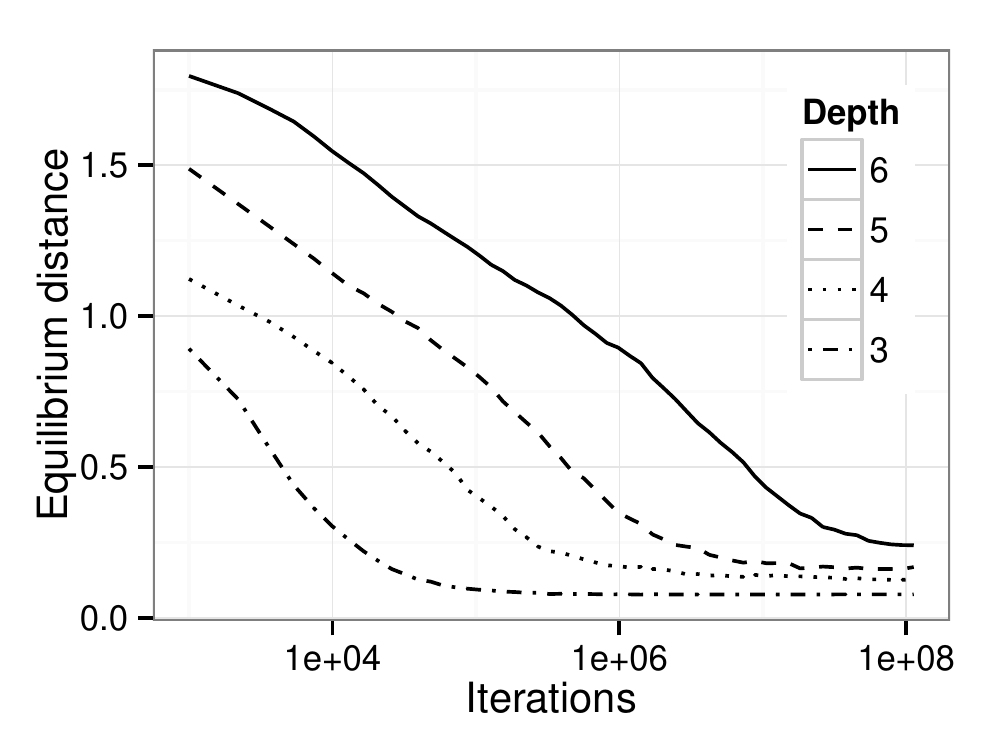}
\caption{Convergence of SM-MCTS with Exp3 selection on random games with three actions of each player in each stage and various depths.}\label{fig:convD}
\end{figure}

Figure~\ref{fig:convD} presents the average distance from the equilibrium with SM-MCTS, Exp3 selection and exploration $\epsilon=0.2$ in random games with $B=3$ and various depths. The eventual error increases with increasing depth, but even with depth of 6, the eventual error was on average around 0.25 and always less than $0.3$.

\subsubsection{Removing exploration}

\begin{figure}[t]
\centering
\includegraphics[width=0.3\textwidth]{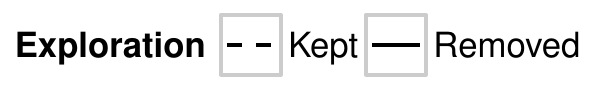}

\begin{subfigure}{0.244\textwidth}
\includegraphics[width=\textwidth]{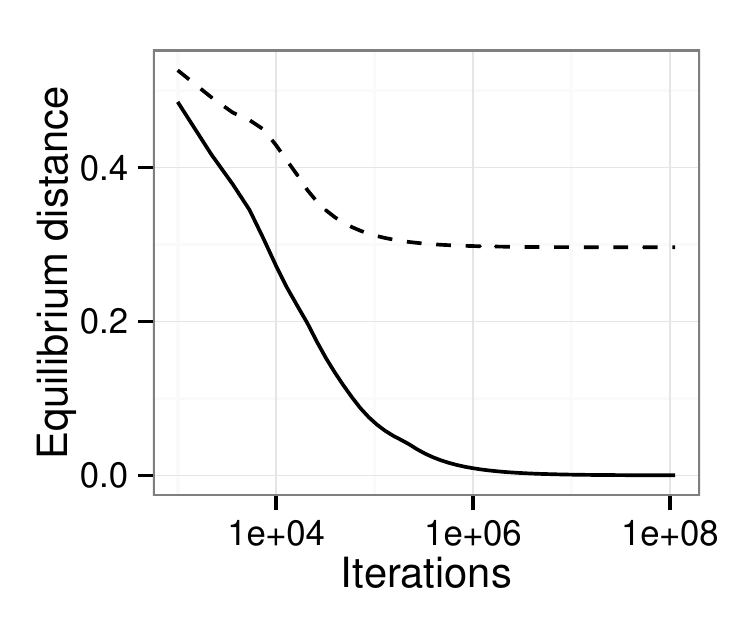}
\caption{Anti(5)}\label{fig:convRem_Anti}
\end{subfigure}
\begin{subfigure}{0.244\textwidth}
\includegraphics[width=\textwidth]{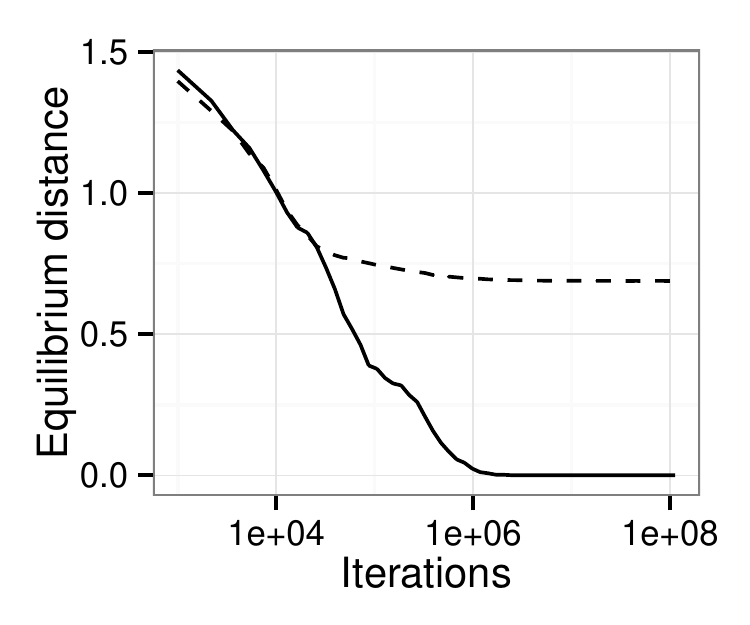}
\caption{Goofspiel(5)}\label{fig:convRem_GS}
\end{subfigure}
\begin{subfigure}{0.244\textwidth}
\includegraphics[width=\textwidth]{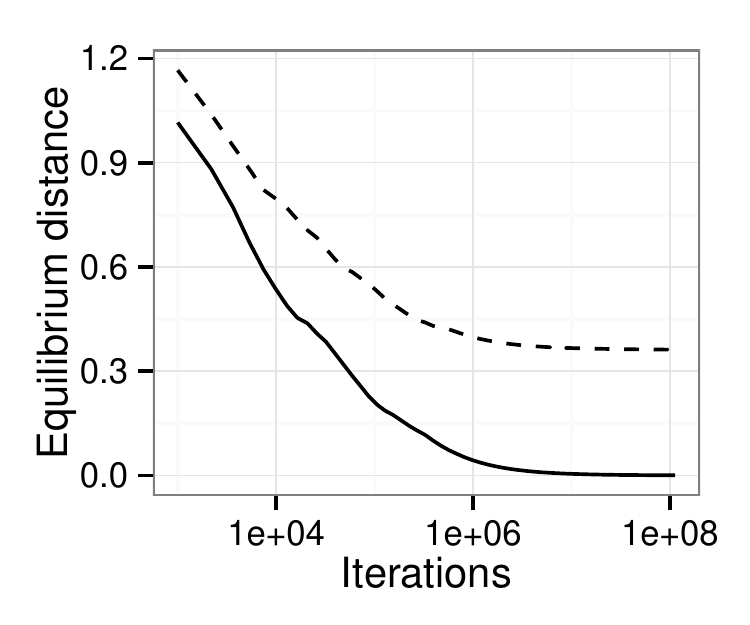}
\caption{Oshi-Zumo(5)}\label{fig:convRem_OZ}
\end{subfigure}
\begin{subfigure}{0.244\textwidth}
\includegraphics[width=\textwidth]{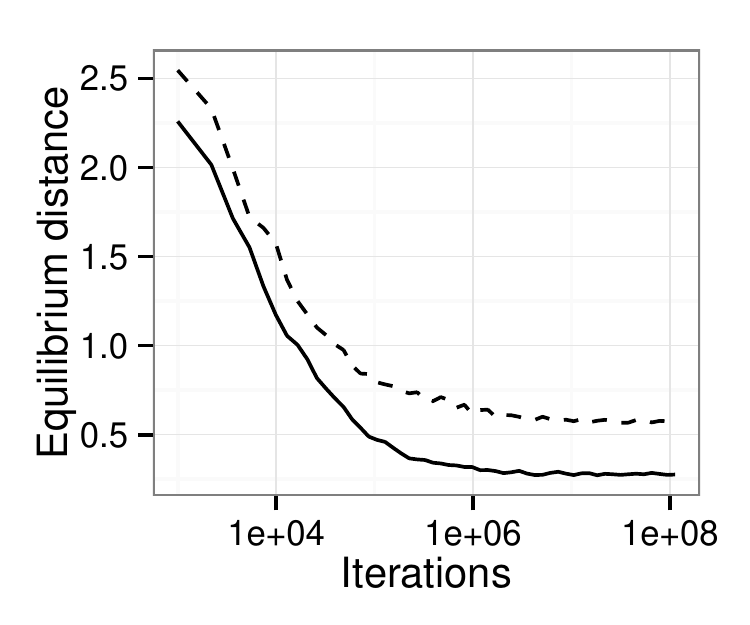}
\caption{Random(3)}\label{fig:convRem_RND}
\end{subfigure}
\caption{The effect of removing exploration samples in SM-MCTS with Exp3 selection and $\epsilon=0.2$.}\label{fig:explRem}
\end{figure}

In Section~\ref{sec: exploitability}, we show that the computed strategy cannot, in general, get worse when we disregard the samples caused by exploration. Figure~\ref{fig:explRem} shows that in practice, the strategy is usually improved from the very beginning of the convergence and the exploration should always be removed.

\section{Conclusion\label{sec:Discussion}}

Monte Carlo Tree Search has recently become a popular algorithm for creating artificial game-playing agents. Besides perfect information games, where the behavior of the algorithm is reasonably well understood, variants of the algorithm has been successful also in more complex imperfect-information games. However, there was very little pre-existing theory that would describe the behavior of the algorithms in these games and provide guarantees on their performance.

In this paper, we provide the theory and guaranteed results for the simplest, but still important, subclass of imperfect information games -- sequential zero-sum games with simultaneous moves, but otherwise perfect information. These games already include one of the major complications caused by imperfect information, which is the need to use randomized strategies to guarantee the optimal performance. We also note that while we focus on games with simultaneous moves, all presented theoretic results (apart from the SM-MCTS counterexample from Section~\ref{sub: Counterexample}) trivially apply also to perfect information games with sequential moves.

Our main results from Section \ref{sec: convergence} show that a variant of Monte Carlo Tree Search algorithm, which we call SM-MCTS-A, in combination with any Hannan consistent algorithm is guaranteed to eventually converge to Nash equilibrium of the game.
Moreover, if the used selection function, in addition to being HC, has the Unbiased Payoff Observations property, even the standard SM-MCTS algorithm is guaranteed to converge to an approximate Nash equilibrium.
On the other hand, in Section~\ref{sec:Counterexample} we present a counterexample showing that there exist HC algorithms, which converge with SM-MCTS-A, but not with SM-MCTS.

More detailed results are summarized in Table~\ref{tab:summary}. In Theorem \ref{thm: SM-MCTS-A convergence} (\ref{thm: SM-MCTS convergence}) we show that SM-MCTS-A (SM-MCTS) algorithm with $\epsilon$-HC selection function eventually converges at least to $C\epsilon$-NE of a game, where for game depth $D$, $C$ is of the order $D^2$ ($2^D$). In Section~\ref{sec:Counterexample}, we show that the worst case dependence of $C$ on $D$ cannot be sublinear, even after the exploration is removed. This gives us both lower and upper bounds on the value of $C$, but it remains to determine whether these bounds are tight. We form a hypothesis that the tight bound is a linear dependence $C=3D$ ($C=2D$ after the exploration is removed).

\begin{table}
\small
\renewcommand{\arraystretch}{1.5}
\begin{center} \begin{tabular}{ r|c|c|c| }
\multicolumn{1}{r}{}
 &  \multicolumn{1}{c}{SM-MCTS-A}
 & \multicolumn{2}{c}{SM-MCTS~~~} \\
\cline{2-4}
Assumptions & $\epsilon$-HC & $\epsilon$-HC, $\epsilon$-UPO & $\epsilon$-HC only\\
\cline{2-4}
Upper bound & $2D(D+1)\epsilon$ & $\left(12(2^D-1)-8D\right)\epsilon$ & might not converge\\
\cline{2-3}
Lower bound & $2D\epsilon$ & $2D\epsilon$ & to approx. NE at all\\
\cline{2-4}
\end{tabular}
\end{center}
\caption{Summary of the proven bounds on the worst case eventual exploitability of the strategy produced by SM-MCTS(-A) run with an $\epsilon$-Hannan consistent selection function}\label{tab:summary}
\end{table}

In Section~\ref{sec: exploitability}, we provide an analysis of previously suggested improvement of SM-MCTS algorithm, which proposes the removal of samples caused by exploration from the played strategy. We prove both formally and empirically that this modification of the algorithm is sound and generally improves the strategy.

In Theorem \ref{thm:Bound} we show that, for a fixed confidence level, SM-MCTS-A algorithm converges to the given $C\epsilon$-equilibrium at rate at least $1/T^{\tfrac{2}{D^2}}$. This estimate is most likely overly pessimistic, as suggested by the empirical results.

Finally, we provide empirical investigation of the algorithms that shows that  in practical problems, the convergence times as well as the eventual distance from the Nash equilibrium is better than the theoretic guarantees. 
We show that SM-MCTS-A converges slower than SM-MCTS with the commonly used HC selection functions, but they converge to the same distance from the equilibrium. 
Moreover, the difference in convergence speed is smaller with regret matching than with Exp3 and in many domains, it is negligible.

While this paper provides a significant step towards understanding of MCTS in simultaneous move games, it also leaves some problems open. First of all, many of the guarantees presented in the paper have not been shown to be tight so an obvious future research direction would be to improve the guarantees or show the tightness of these results. Also, better characterization of the requirements (on the selection function) which guarantee convergence with SM-MCTS algorithm could be provided. For example, it would be interesting to formally prove that the common Hannan consistent algorithms guarantee unbiased payoff observations or a similar property that is sufficient to guarantee convergence of SM-MCTS in this setting.

Furthermore, MCTS algorithms are generally used with incremental building of the search tree and a problem-specific heuristic simulation strategy outside of the portion of the search tree in memory. It would be interesting to analyze the behavior of the algorithms with respect to basic statistical properties of the simulation strategy, such as its bias and variance. Lastly, our analysis of simultaneous move games can be used as a basis for analyzing MCTS as it is used in more general classes of imperfect information games.

\bibliography{refs}

\appendix

\section{Proofs}
In this section we present the proofs for those results, which have not been already proven in the main text. We start with the proof of Lemma \ref{emp a avg}, which states that eventually, there is no difference between empirical and average strategies.

\begin{proof}[Proof of Lemma \ref{emp a avg}]
It is enough to show that $ \underset{t\rightarrow\infty}{\limsup}\, |\hat{\sigma_1}(t,i)-\bar{\sigma_1}(t,i)|=0$ holds
almost surely for any given $i$. Using the definitions of $\hat{\sigma_1}(t,i)$
and $\bar{\sigma_1}(t,i)$, we get
\[
\hat{\sigma_1}(t,i)-\bar{\sigma_1}(t,i)
  = \frac{1}{t} \left(t_i - \sum_{s=1}^t \sigma_1(s,i)\right)
  = \frac{1}{t} \sum_{s=1}^t \left(\delta_{i, i(s)} - \sigma_1(s,i) \right),
\]
 where $\delta_{i,j}$ is the Kronecker delta. Using the (martingale version of) Central Limit Theorem on the sequence of random variables $X_t=\sum_{s=1}^t \left(\delta_{i, i(s)} - \sigma_1(s,i) \right)$ gives the result (the conditions clearly hold, since $\mathbf{E}\left[\delta_{i,i(t)} - \sigma_1(t,i)| X_1,...,X_{t-1}\right]=0$ implies that $X_t$ is a martingale and $\delta_{i,i(t)} - \sigma_1(t,i)\in\left[-1,1\right]$ guarantees that all required moments are finite).
\end{proof}

Next, we prove Lemma \ref{lemma: A* je HC} - which states that $\epsilon$-Hannan consistency is not substantially affected by additional exploration.
\begin{proof}[Proof of Lemma \ref{lemma: A* je HC}]
 Denoting by {*} the variables corresponding to the algorithm $A^{*}$
we get
\begin{eqnarray*}
r^{*}(t) & = & \frac{1}{t}R^{*}(t)\leq\frac{1}{t}\left(1\cdot t_{\textrm{ex}}+R(t-t_{\textrm{ex}})\right) \\
 & = & \frac{t_{\textrm{ex}}}{t}+\frac{R(t-t_{\textrm{ex}})}{t-t_{\textrm{ex}}}\cdot\frac{t-t_{\textrm{ex}}}{t},
\end{eqnarray*}
where, for given $t\in\mathbb{N}$, $t_{\textrm{ex}}$ denotes the number of times $A^{*}$ explored up to $t$-th iteration. By Strong Law of Large Numbers we have that $\underset{t\rightarrow\infty}{\lim}\,\frac{t_\textrm{ex}}{t}=\gamma$ holds almost surely. This implies 
\begin{eqnarray*}
\limsup_{t\rightarrow\infty}r^{*}\left(t\right) & \leq & \limsup_{t\rightarrow\infty}\frac{t_{\textrm{ex}}}{t}+\limsup_{t-t_{\textrm{ex}}\rightarrow\infty}\frac{R\left(t-t_{\textrm{ex}}\right)}{t-t_{\textrm{ex}}}\cdot\limsup_{t\rightarrow\infty}\frac{t-t_{\textrm{ex}}}{t}\\
 & \leq & \gamma+\epsilon\left(1-\gamma\right)\\
 & \leq & \gamma+\epsilon,
\end{eqnarray*}
which means that $A^{*}$ is $\left(\epsilon+\gamma\right)$-Hannan
consistent. The guaranteed exploration property of $A^{*}$ is trivial. 
\end{proof}

\subsection{Proofs related to the convergence of SM-MCTS-A}
In this section we give the proofs for lemmas which were used to prove Theorem \ref{thm: SM-MCTS-A convergence}. We begin with Lemma \ref{L: HC-and-NE}, which established a connection between average payoff $g$ and game value $v$ for matrix games.

\begin{proof}[Proof of Lemma \ref{L: HC-and-NE}]
It is our goal to draw conclusion about the quality of the empirical strategy based on information about the performance of our $\epsilon$-HC algorithm. Ideally, we would like to somehow relate the utility $u(\sigma)$ to the average payoff $g(t)$. However, as this is generally impossible, we can do the next best thing:
\begin{flalign}
u\left(br,\hat{\sigma}_{2}(t)\right)=\, & \,\underset{i}{\max}\,\underset{j}{\sum}\hat{\sigma}_{2}(t,j)a_{ij}=\underset{i}{\max}\,\underset{j}{\sum}\frac{t_{j}}{t}a_{ij}=\frac{1}{t}\underset{i}{\max}\,\underset{j}{\sum}t_{j}a_{ij}\nonumber \\
=\, & \,\frac{1}{t}\underset{i}{\max}\,\underset{s=1}{\overset{t}{\sum}}a_{ij(s)}=\frac{1}{t}G_{\max}(t)=g_{\max}(t).\label{eq: u(br, )}
\end{flalign}
\textbf{Step 1:} Let $\eta>0$. Using $\epsilon$-HC property gives us the existence of such $t_{0}$ that $g_{max}(t)-g(t)<\epsilon+\frac{\eta}{2}$ holds for all $t\geq t_{0}$, which is equivalent to $g(t)>g_{max}-(\epsilon+\frac{\eta}{2})$. However, in our zero-sum matrix game setting, $g_{max}$ is always at least $v$, which implies that $g(t)>v-(\epsilon+\frac{\eta}{2})$.
Using the same argument for player $2$ gives us that $g(t)<v+\epsilon+\frac{\eta}{2}$. Therefore we have the following statement, which proves the inequalities \eqref{eq: HC a NE2}:
\begin{equation*}
\forall t\geq t_0:\ v- (\epsilon+\frac{\eta} 2 ) < g(t)<v+\epsilon+\frac{\eta}{2} \textrm{ holds almost surely.}
\end{equation*}

\vspace{5mm}
\textbf{Step 2:} We assume, for contradiction with inequalities \eqref{eq: HC a NE 1}, that with non-zero probability, there exists an increasing sequence of time steps $t_{n}\nearrow\infty$, such that $u\left(br,\hat{\sigma}_{2}(t_{n})\right)\geq v+2\epsilon+\eta$ for some $\eta>0$. Combing this with the inequalities, which we proved above, we see that 
\begin{eqnarray*}
\limsup_{t\rightarrow\infty}r\left(t\right) & \geq & \limsup_{n\rightarrow\infty}r\left(t_{n}\right)=\limsup_{n\rightarrow\infty}\left(g_{\max}\left(t_{n}\right)-g\left(t_{n}\right)\right)\\
 & = & \limsup_{n\rightarrow\infty}\left(u\left(br,\hat{\sigma_{2}}\left(t_{n}\right)\right)-g\left(t_{n}\right)\right)\\
 & \geq & v+2\epsilon+\eta-\left(v+\epsilon+\eta/2\right) = \epsilon + \eta > \epsilon
\end{eqnarray*}
holds with non-zero probability, which is in contradiction with $\epsilon$-Hannan consistency. 
\end{proof}

\begin{remark}
\label{rem: tilde notation}In the following proof, and in the proof of Proposition \ref{prop: when tilde s = bar s},  we will be working with regrets, average payoffs and other quantities related to matrix games with error, in which we have two sets of rewards - the rewards $a_{ij}$ corresponding to the matrix M and the ``observed'' rewards $a_{ij}(t)$. We will denote the variables related to the distorted rewards $a_{ij}(t)$ by normal symbols (for example $g_{\max}(t)=\max_i \frac 1 t \sum_{s=1}^t a_{i j(s)}(s)$ and use symbols with tilde for the variables related to the rewards $a_{ij}$ (for example $\tilde{g}_{\max}(t)= \max_i \frac 1 t \sum_{s=1}^{t}a_{i j(s)}$).
\end{remark}
\begin{proof}[Proof of Proposition \ref{Prop: hry s chybou}]
This proposition strengthens the result of Lemma \ref{L: HC-and-NE} and its proof will also be similar. The only additional technical ingredient is the following inequality \eqref{eq:gmax-tildegmax}:

Since $M(t)$ is a repeated game with error $c\epsilon$, there almost
surely exists $t_{0}$, such that for all $t\geq t_{0}$, $\,\left|a_{ij}(t)-a_{ij}\right|\leq c\epsilon$
holds. This leads to
\begin{equation}
\left|g_{\max}(t)-\tilde{g}_{\max}(t) \right| \leq \underset{i}{\max}\,\left|\frac{1}{t}\underset{s=1}{\overset{t}{\sum}}\left(a_{ij(s)}(s)-a_{ij(s)}\right)\right|\leq\frac{t_{0}}{t}+c\epsilon\cdot\frac{t-t_{0}}{t}\overset{t\rightarrow\infty}{\longrightarrow}c\epsilon.\label{eq:gmax-tildegmax}
\end{equation}
The remainder of the proof contains no new ideas, and it is nearly exactly the same as the proof of Lemma \ref{L: HC-and-NE}, therefore we just note what the two main steps are:

\noindent \textbf{Step 1:} Hannan consistency gives us that
\begin{equation*}
\forall \eta>0\ \exists t_0\in\mathbb N\  \forall t\geq t_0:\ g\left(t\right) \geq v-\left(\epsilon+\eta\right)-\left|g_{\max}\left(t\right)-\tilde{g}_{\max}\left(t\right)\right| \textrm{ holds a.s.},
\end{equation*} 
from which we deduce the inequalities
\begin{equation*}
v - ( \epsilon+c\epsilon) \leq \underset{t\rightarrow\infty}{\liminf}\, g(t) \leq \underset{t\rightarrow\infty}{\limsup}\, g(t) \leq v+\epsilon+c\epsilon.
\end{equation*}

\noindent \textbf{Step 2:} For contradiction we assume that there exists an increasing sequence of time steps $t_{n}\nearrow\infty$, such that 
\begin{equation}
u(br,\hat{\sigma}_{2}(t_{n}))\geq v+2(c+1)\epsilon+\eta \label{eq: contradiction}
\end{equation}
holds for some $\eta>0$. Using the identity $\tilde{g}_{\max}\left(t_{n}\right)=u\left( br,\hat{\sigma}_{2}(t_{n}) \right)$ and inequalities \eqref{eq:gmax-tildegmax} and \eqref{eq: contradiction}, we then compute that the regret $r(t_n)$ is too high, which completes the proof:
\begin{eqnarray*}
\epsilon \geq \limsup_{t\rightarrow\infty} r\left(t\right) & \geq & \limsup_{n\rightarrow\infty} r\left(t_{n}\right) = \limsup_{n\rightarrow\infty} \left(g_{\max}\left(t_{n}\right)-g\left(t_{n}\right)\right)\\
 & \overset{\eqref{eq: u(br, )}}{\geq} & \limsup_{n\rightarrow\infty} \left( u\left( br,\hat{\sigma}_{2}(t_{n}) \right) - \left| \tilde{g}_{\max}(t_{n}) - g_{\max} (t_{n}) \right| \right) - \limsup_{n\rightarrow\infty} g(t_{n}) \\
 & \overset{\eqref{eq:gmax-tildegmax},\eqref{eq: contradiction}}{\geq} & \left(v+2(c+1)\epsilon+\eta\right) - c\epsilon - \left(v+\epsilon+c\epsilon\right) \\
 & = & \epsilon+\eta > \epsilon.
\end{eqnarray*}
\end{proof}

\subsection{Proofs related to the convergence of SM-MCTS}
In Section \ref{sub: SM-MCTS-A bound} we gave the proof of convergence of  those SM-MCTS algorithms, which were based on $\epsilon$-UPO selection functions. It remains to prove Proposition \ref{prop: when tilde s = bar s}, which establishes a connection between regrets $R$ and $\tilde R$ of the selection function with respect to the observed rewards and with respect to the exact subgame values. The goal is to show that if $R(T)$ is small and algorithm $A$ is $\epsilon$-UPO, then the regret $\tilde R(T)$ is small as well.

\begin{proof}[Proof of Proposition \ref{prop: when tilde s = bar s}]
Let $A$ be a $\epsilon$-HC algorithm with $\epsilon$-UPO property. Recall here Remark \ref{rem: tilde notation} - by $\tilde R\left(T\right)$ we denote the regret of action sequence $i(t)$ chosen by $A$ against the adversary's action sequence $j(t)$ \emph{in the matrix game }$\left(v_{ij}\right)$ - that is 
\[
\tilde{R}\left(T\right)=\max_{i^*}\sum_{t=1}^{T}v_{i^*j(t)}-\sum_{t=1}^{T}v_{i(t)j(t)}=:\max_{i^*}\tilde S_{i^*}(T),
\]
and by $R\left(T\right)$ we denote the ``observed'' regret 
\[ R\left(T\right)=\max_{{i^*}}\sum_{t=1}^{T}s_{{i^*}j(t)}\left(t_{{i^*}j(t)}\right)-\sum_{t=1}^{T}s_{i(t)j(t)}\left(t_{i(t)j(t)}\right)=:\max_{i^*}S_{i^*}(T). \]
To prove the proposition it is sufficient to assume that $\limsup_{T}R\left(T\right)/T\leq\epsilon$
and show that $\limsup_{T}\tilde{R}\left(T\right)/T\leq2\left(c+1\right)\epsilon$
holds almost surely. Let $i^*$ be an action of player 1. Denote by $a)$ the fact that, by definition of $w_{ij}\left(n\right)$, $T_{j}=\sum_{m=1}^{T_{i^*j}}w_{i^*j}\left(m\right)$ holds for each $T$ and $j$, and by $b)$ the equivalence $s_{i^*j(t)}(t_{i^*j(t)})=s_{i^*j}(m)$$\iff (j(t)=j\ \& \ t_{i^*j(t)}=m)$. We can rewrite $S_{i^*}\left(T\right)$ as follows:
\begin{eqnarray*}
S_{i^*}\left(T\right) & = & \sum_{t=1}^{T}s_{i^*j(t)}\left(t_{i^*j(t)}\right)-\sum_{t=1}^{T}s_{i(t)j(t)}\left(t_{i(t)j(t)}\right)\\
 & \overset{b)}{=} & \sum_{j}\sum_{m=1}^{T_{i^*j}}s_{i^*j}\left(m\right)\left|\left\{ t\leq T|\, t_{i^*j(t)}=m\,\&\, j(t)=j\right\}\right| -\sum_{i,j}\sum_{m=1}^{T_{ij}}s_{ij}\left(m\right)\\
 & \overset{a)}{=} & \sum_{j}\frac{T_{j}}{\sum_{m=1}^{T_{i^*j}}w_{i^*j}\left(m\right)}\sum_{m=1}^{T_{i^*j}}s_{i^*j}\left(m\right)w_{ij}\left(m\right)-\sum_{i,j}\frac{T_{ij}}{T_{ij}}\sum_{m=1}^{T_{ij}}s_{ij}\left(m\right)\\
 & = & \sum_{j}T_{j}\tilde{s}_{i^*j}\left(T_{i^*j}\right)-\sum_{i,j}T_{ij}\bar{s}_{ij}\left(T_{ij}\right)\\
 & = & \sum_{j}T_{j}\left(v_{i^*j}+\tilde{s}_{i^*j}\left(T_{i^*j}\right)-v_{i^*j}\right)-\sum_{i,j}T_{ij}\left(v_{ij}+\bar{s}_{ij}\left(T_{ij}\right)-v_{ij}\right)\\
 & =: & \sum_{j}\sum_{m=1}^{T_{j}}v_{i^*j}-\sum_{i,j}\sum_{m=1}^{T_{ij}}v_{ij}+X_{i^*}\left(T\right)\\
 & = & \sum_{t=1}^{T}v_{i^*j(t)}-\sum_{t=1}^{T}v_{i(t)j(t)}+X_{i^*}\left(T\right),\\
 & = & \tilde{S}_{i^*}\left(T\right)+X_{i^*}\left(T\right)
\end{eqnarray*}
where 
\[ X_{i^*}\left(T\right) := \sum_{j}T_{j}\left(\tilde{s}_{i^*j}\left(T_{i^*j}\right)-v_{i^*j}\right)-\sum_{i,j}T_{ij}\left(\bar{s}_{ij}\left(T_{ij}\right)-v_{ij}\right). \]
In particular, we can bind the regret $R(T)$ as
\[
\tilde R(T)=\max_{i^*} \tilde S_{i^*}(T) =\max_{i^*} \left( S_{i^*}(T) - X_{i^*}(T) \right) \leq R(T) + \max_{i^*} \left|X_{i^*}(T)\right|. \]
Clearly $X_{i^*}\left(T\right)$ satisfies 
\begin{eqnarray*}
\left|\frac{X_{i^*}\left(T\right)}{T}\right| & \leq & \sum_{j}\frac{T_{j}}{T}\left|\tilde{s}_{i^*j}\left(T_{i^*j}\right)-v_{i^*j}\right|+\sum_{i,j}\frac{T_{ij}}{T}\left|\bar{s}_{ij}\left(T_{ij}\right)-v_{ij}\right|.
\end{eqnarray*}
Using the $\epsilon$-UPO property and the assumption that $\limsup_{n}\,\left|\bar{s}_{ij}\left(n\right)-v_{ij}\right|\leq c\epsilon$
holds a.s. for each $i,j$, we get 
\begin{eqnarray*}
\limsup_{T\rightarrow\infty}\left|\frac{X_{i^*}\left(T\right)}{T}\right| & \leq & \limsup_{T\rightarrow\infty}\sum_{j}\frac{T_{j}}{T}\left|\tilde{s}_{i^*j}\left(T_{i^*j}\right)-v_{i^*j}\right|+\\
 & & +\sum_{i,j}\frac{T_{ij}}{T}\left|\bar{s}_{ij}\left(T_{ij}\right)-v_{ij}\right|\\
 & \leq & \left(c+1\right)\epsilon\limsup_{T\rightarrow\infty}\sum_{j}\frac{T_{j}}{T}+\\
  & & +c\epsilon\limsup_{T\rightarrow\infty}\sum_{i,j}\frac{T_{ij}}{T}=\left(2c+1\right)\epsilon.
\end{eqnarray*}
 Consequently, this implies that 
\begin{eqnarray*}
\limsup_{T\rightarrow\infty} \tilde R\left(T\right)/T & \leq & \limsup_{T\rightarrow\infty} R\left(T\right)/T+\limsup_{T\rightarrow\infty}\max_{i^*}X_{i^*}\left(T\right)/T \\
 & \leq & \epsilon+\left(2c+1\right)\epsilon=2\left(c+1\right)\epsilon
\end{eqnarray*}
 holds almost surely, which is what we wanted to prove.\end{proof}

\subsection{Details related to the counterexample for Theorem \ref{thm: SM-MCTS convergence}}

In Section \ref{sub: Counterexample} (Lemma \ref{lemma: modification of A}) we postulated the existence of algorithms, which behave similarly to those from Example \ref{ex: ideal case}, but unlike those from Example \ref{ex: ideal case}, the new algorithms are $\epsilon$-HC. First, we define these algorithms and then we prove their properties in Lemma \ref{lem: A_I properties}. Lemma \ref{lemma: modification of A}, which was needed in Section \ref{sub: Counterexample}, follows directly from Lemma \ref{lem: A_I properties}.

\begin{remark}
In the following description of algorithm  $A_{1}^{J}$, $ch$ denotes how many times the other player cheated,
while $\bar{ch}$ is the average ratio of cheating in following the
cooperation pattern. The variables $\tilde{ch}$ and $\tilde{\bar{ch}}$
then serve as the estimates of $ch$ and $\bar{ch}$. We present the precise definitions below. The nodes I, J, actions X, Y, L, R, U and D and the respective payoffs refer to the game $G$ from Figure \ref{fig: hra}.
\end{remark}
\textbf{Definition of the algorithm $A_{1}^{J}$:}

Fix an increasing sequence of integers $b_{n}$ and repeat for $n\in\mathbb{N}$:

\begin{enumerate}
\item \textbf{Buffer building $B_{n}$: }Play according to some $\epsilon$-HC
algorithm for $b_{n}$ iterations (continuing with where we left of
in the $\left(n-1\right)$-th buffer building phase).
\item \textbf{Cooperation $C_{n}$:} Repeat $U,U,D,D$ for $t=1,2,...$
and expect the other player to repeat $L,R,R,L$. At each iteration
$t$, with probability $\epsilon$, check whether the other player
is cooperating - that is play $D$ instead of $U$ resp. $U$ instead
of $D$, and if the payoff does not correspond to the expected pattern
the second player should be following, set $\tilde{ch}\left(t\right)=\frac{1}{\epsilon}$.
If the other player passes this check, or if we did not perform it,
set $\tilde{ch}\left(t\right)=0$.
\item \textbf{End of cooperation }(might not happen)\textbf{:} While executing
step 2, denote $\tilde{\bar{ch}}\left(t\right):=\frac{1}{t}\sum_{s=1}^{t}\tilde{ch}\left(s\right)$.
Once $t$ satisfies 
\[
\frac{\epsilon\cdot b_{n}}{b_{n}+t}+\frac{1\cdot t}{b_{n}+t}\geq2\epsilon,
\]
we check at each iteration whether the estimate $\tilde{\bar{ch}}\left(t\right)$
threatens to exceed $2\epsilon$ during the next iteration or not.
If it does, we end the cooperation phase, set $n:=n+1$ and continue
by the next buffer building phase.
\item \textbf{Simulation of the other player:} While repeating steps 1, 2 and 3, we simulate the other player's algorithm $A_{2}^{J}$ (this is possible, since from the knowledge of our action and the received payoff, we can recover the adversary's action).
If it ends the cooperation phase and starts the next buffer building
phase, we do the same.
\item Unless the cooperation phase is terminated, we stay in phase $C_n$ indefinitely.
\end{enumerate}

\noindent \textbf{Definition of $A_2^J$:} The algorithm $A_{2}^{J}$ is identical to $A_{1}^{J}$, except for
the fact that it repeats the pattern U, D, D, U instead of L, R, R, L and expects
the other player to repeat L, R, R, L.

\noindent \textbf{Definition of $A^I$:} The algorithm $A^{I}$ is a
straightforward modification of $A_1^{J}$, with $2\epsilon$ in place
of $\epsilon$ - it repeats the sequence Y, X, X, Y and expects to
receive payoffs 0, 0, 0, ... whenever playing $X$ and payoffs 1, 0, 1, 0, ...
when playing $Y$. However, whenever it deviates from the Y, X, X, Y
pattern in order to check whether these expectations are met, it plays
the same action once again (in order to avoid disturbing the payoff
pattern for $Y$).

\begin{remark}\label{rem: A_IJ is correct}
The steps 2 and 3 from the algorithm description are correctly defined, because the opponent's action choice can be recovered from the knowledge of our action choice and the resulting payoff. Regarding step 3, we note that the condition here is trivially satisfied for $t=1,...,\epsilon b_{n}$, so the length $t_n$ of the cooperation phases $C_{n}$ tends to infinity as $n\rightarrow\infty$, regardless of the opponent's actions.
\end{remark}

\begin{lemma}\label{lem: A_I properties}
$\left(1\right)$ When facing each other, the average strategies of algorithms $A^{I},\, A_{1}^{J}$ and $A_{2}^{J}$ will converge to the suboptimal
strategy $\sigma^{I}=\sigma_{1}^{J}=\sigma_{2}^{I}=\left(\frac{1}{2},\frac{1}{2}\right)$. However the algorithms will suffer regret no higher than $C\epsilon$ for some $C>0$ (where $C$ is independent of $\epsilon$).

$\left(2\right)$ There exists a sequence $b_{n}$ (controlling the length of phases $B_{n}$), such that when facing a different adversary, the algorithms
suffer regret at most $C\epsilon$. Consequently $A^{I},\, A_{1}^{J}\textrm{ and }A_{2}^{J}$
are $C\epsilon$-Hannan consistent.
\end{lemma}
\begin{proof}
\textbf{Part $\left(1\right)$:} Note that, disregarding the checks made by the algorithms, the average strategies in cooperation phases converge to $\left(\frac{1}{2},\frac{1}{2}\right)$. Furthermore, the probability of making the checks is the same at any iteration, therefore if the algorithms eventually settle in cooperative phase, the average strategies converge to $\left(\frac{1}{2},\frac{1}{2}\right)$.

\vspace{5mm}
\noindent \textbf{Step $\left(i\right)$: The conclusion of (2) holds for $A_{i}^{J}$, $i=1,2$.}

We claim that both algorithms $A_{i}^{J}$, $i=1,2$ will
eventually settle in the cooperative mode, thus generating the payoff
sequence $1,0,1,0$ during at least $\left(1-\epsilon\right)^{2}$-fraction
of the iterations and generating something else at the remaining $1-\left(1-\epsilon\right)^{2}<2\epsilon$
iterations. It is then immediate that the algorithms $A_{i}^{J}$
suffer regret at most $2\epsilon$ (since $\sigma^{J}$ is $2\epsilon$-equilibrium
strategy). 

\noindent  \textbf{Proof of $\left(i\right)$:} If the other player uses the same algorithm,
we have $\mathbf{E}\left[\tilde{ch}\left(t\right)\right]=\epsilon$
and $\mathbf{Var}\left[\tilde{ch}\left(t\right)\right]\leq\frac{1}{\epsilon}<\infty$,
thus by the Strong Law of Large Numbers $\tilde{\bar{ch}}\left(t\right)\rightarrow\epsilon$ almost surely.
In particular, there exists $t_{0}\in\mathbb{N}$ such that
\[ \mathbf{Pr}\left[\forall t\geq t_{0}:\,\tilde{\bar{ch}}\left(t\right)\leq2\epsilon\right]>0. \]
In Remark \ref{rem: A_IJ is correct} we observed that the cooperative
phase always lasts at least $\epsilon b_{n}$ steps. Therefore once
we have $b_{n}\geq\frac{1}{\epsilon}t_{0}$, there is non-zero probability
both players will stay in $n$-th cooperative phase forever. If they
do not, they have the same positive probability of staying in the next cooperative
phase and so on - by Borel-Cantelli lemma, they will almost surely
stay in $C_{n_{0}}$ for some $n_{0}\in\mathbb{N}$.

\vspace{5mm}
\noindent \textbf{Step $\left(ii\right)$: $A^{I}$ will eventually settle in the cooperative mode.}

\noindent \textbf{Proof of $\left(ii\right)$:} This statement can be proven by a similar argument as $\left(i\right)$. The only difference is that instead of checking whether the other player is cheating, we check whether the payoff sequence is the one expected by $A^I$. The fact that $A^{I}$ settles in cooperative mode then immediately implies that $A^{I}$ will repeat the Y, X, X, Y pattern during approximately $\left(1-4\epsilon\right)$-fraction
of the iterations (we check with probability $2\epsilon$ and we always check twice). It is then also immediate that $A^{I}$ will receive
the expected payoffs during at least $\left(1-\epsilon\right)$-fraction of the iterations (the payoff is always $0$, as expected, when $X$ is played and it is correct in at least $1-2\epsilon$ cases when playing $Y$. $X$ and $Y$ are both played with same frequency, which gives the result.). 

\vspace{5mm}
\noindent \textbf{Step $\left(iii\right):$ If $A_{i}^{J}$, $i=1,2$ and $A^{I}$ settle in cooperative mode, then $A^{I}$ suffers regret at most $C\epsilon$.}

\noindent \textbf{Proof of $\left(iii\right)$:} We denote by $a\left(t\right)$ the $t$-th action chosen by $A^{I}$ and by $(a^*(s))$ the 4-periodic of sequence of actions starting with Y, X, X, Y. Moreover we denote by $s_{Y}\left(n\right)$ the $n$-th payoff generated by $A_{1}^{J}$ and $A_{2}^{J}$ and by $x_{Y}\left(t\right)$ the number $s_{Y}\left(t_{Y}\right)$. Set also $x_{X}\left(t\right)=0$ for each $t\in\mathbb{N}$. Finally we denote by  $(x^*_B(s))$ the 4-periodic of sequence of payoffs starting with 1, 0, 0, 0. Our goal is to show that the average regret 
\[
r\left(t\right)=\frac{1}{t}\left(\sum_{s=1}^{t}x_{Y}\left(s\right)-\sum_{s=1}^{t}x_{a\left(s\right)}\left(s\right)\right)=
\]
is small. As we already observed in Example \ref{ex: ideal case}, if neither the payoff pattern $1,0,1,0,...$ from node $J$, nor the action pattern Y, X, X, Y, ... at node $I$ are disturbed, then we have $\left(a(s)\right)=\left(a^*(s)\right)$ and $\left(x_Y(s)\right)=\left(x^*_Y(s)\right)$, and therefore $A^{I}$ suffers no regret. Consequently, we have 
\begin{flalign}
r\left(t\right) & = \frac 1 t \left| \{1,...,t\} \setminus \{s\leq t|\ a(s)=a^*(s)\ \& \ x_Y(s)=x^*_Y(s)\} \right| \nonumber \\
 & \leq \frac 1 t \left| \{s\leq t|\ a(s)\neq a^*(s) \} \cup \{s\leq t|\ x_Y(s)\neq x^*_Y(s) \} \right|. \label{eq: r}
\end{flalign}

Firstly recall that when  $A^{I}$ deviates from its action pattern, it does so twice in
a row. Since the sequence $s_{Y}$ is 2-periodic, any disturbance
of $a\left(t\right)$ and $a\left(t+1\right)$ might change the payoffs $x_{Y}\left(t\right),x_{Y}\left(t+1\right)$
but it will affect none other. By $\left(ii\right)$ this change of action
$a\left(t\right)$ concerns at most a $4\epsilon$-fraction of the iterations and so we have
\begin{eqnarray}
\lim_{t\rightarrow\infty} \frac 1 t \left| \{s\leq t|\ a(s)\neq a^*(s) \} \right| \leq 4\epsilon. \label{eq: a*}
\end{eqnarray}

Secondly when $s_{Y}\left(n\right)$ deviates from the expected pattern,
this affects those iterations $t$ for which $t_{Y}=n$. This will
typically be $2$ iterations, unless $A^{I}$ was doing its checks
- those are done with probability $2\epsilon$, therefore with probability
$2\epsilon$, $4$ iterations are affected, with probability $\left(2\epsilon\right)^{2}$,
6 iterations are affected and so on... Since we are interested in
the limit behavior, we can assume that at average no more than $\sum_{k=0}^{\infty}2\cdot\left(2\epsilon\right)^{k}<4$
iterations are affected by each disturbance to $s_{Y}$ (assuming,
of course, that $\epsilon<\frac{1}{4}$). By $\left(i\right)$ we know
that no more than $2\epsilon$-fraction of numbers $s_{Y}(s)$ will be changed.
Consequently no more than $4\cdot2\epsilon$-fraction of payoffs $x_{Y}(s)$
will be changed and we have
\begin{eqnarray}
\lim_{t\rightarrow\infty} \frac 1 t \left| \{s\leq t|\ x_Y(s)\neq x^*_Y(s) \} \right| \leq 8\epsilon. \label{eq: p*}
\end{eqnarray}

Putting these two information together, we see that
\begin{flalign*}
\limsup_{t\rightarrow\infty}r\left(t\right)  \overset {\eqref{eq: r}}{\leq} & \limsup_{t\rightarrow\infty} \frac 1 t \left| \{s\leq t|\ a(s)\neq a^*(s) \} \right| + \\
& + \limsup_{t\rightarrow\infty} \frac 1 t \left| \{s\leq t|\ x_Y(s)\neq x^*_Y(s) \} \right| \\
\overset {\eqref{eq: a*}, \ \eqref{eq: p*}} {\leq} & 4\epsilon+8\epsilon=12\epsilon,
\end{flalign*}
 which is what we wanted to prove.

\vspace{5mm}
\noindent \textbf{Part $\left(2\right)$: Algorithms $A^J_i$, $i=1,2$ are $C\epsilon$-HC.}

\noindent \textbf{a)} Firstly, we assume that both players stick to their assigned patterns during at least $\left(1-2\epsilon\right)$-fraction of iterations (in other words, assume that $\limsup\,\bar{ch}\left(t\right)\leq2\epsilon$). By the same argument as in $\left(1\right)$, we can show that the algorithms will then suffer regret at most $C\epsilon$.

\noindent \textbf{b)} On the other hand, if $\limsup\,\bar{ch}\left(t\right)>2\epsilon$, the algorithm will almost surely keep switching between $B_{n}$ and
$C_{n}$ (consequence of Strong Law of Large Numbers). Denote by $r_{b,n}$ and $r_{c,n}$ the regret from phases $B_{n}$ and $C_{n}$, recall that $b_n, \ t_n$ are the lengths of these phases and set 
\[
r_{n}:=\frac{b_{n}}{b_{n}+t_{n}}r_{b,n}+\frac{t_{n}}{b_{n}+t_{n}}r_{c,n}=\textrm{overall regret in }B_{n}\textrm{ and }C_{n}\textrm{ together}.
\]
Finally, let $r=\limsup \,r\left(t\right)$ denote the bound on the limit of regret of $A_{i}^{J}$. We need to prove that $r\leq C\epsilon$. To do this, it is sufficient to show that $\limsup_{n}\, r_{n}$ is small - thus our goal will be to prove that if the sequence $b_{n}$ increases quickly enough, then $\limsup\, r_{n}\leq C\epsilon$ holds almost surely. Denote by $\left(F_{n}\right)$ the formula
\[
\forall t\geq\frac{1}{\epsilon}b_{n}:\,\tilde{\bar{ch}} \left(t\right) \leq 2\epsilon\implies\bar{ch} \left(t\right)\leq3\epsilon.
\]
We know that $\left|\tilde{\bar{ch}}-\bar{ch}\right|\rightarrow0$ a.s., therefore we can choose $b_{n}$ such that 
\[
\mathbf{Pr}\left[\left(F_{n}\right)\textrm{ holds}\right]\geq1-2^{-n}
\]
holds. Since $\sum2^{-n}<\infty$, Borel-Cantelli lemma gives that $\left(F_{n}\right)$ will hold for all but finitely many $n\in\mathbb{N}$. Note that if $\left(F_{n}\right)$ holds for both players, then their empirical strategy is at most $3\epsilon$ away from the NE strategy, and thus $r_{c,n}\leq C\epsilon$. Since $r_{n}$ is a convex combination of $r_{b,n}$ and $r_{c,n}$ and in $B_{n}$ we play $\epsilon$-consistently, we can compute 
\begin{equation*}
r\leq\limsup\, r_{n}\leq\max\left\{ \limsup\, r_{b,n},\limsup\, r_{c,n}\right\} \leq\max\left\{ \epsilon,C\epsilon\right\} =C\epsilon.
\end{equation*}
The proof of $C\epsilon$-Hannan consistency of $A^{I}$ is analogous.
\end{proof}

\subsection{Proofs related to the finite time bound for SM-MCTS-A}
In order to get a bound on finite time behavior of SM-MCTS-A, we need the following finite time analogy of Proposition \ref{Prop: hry s chybou}:
\begin{lemma}
\label{lemma: Variation on L4.9}Let $\epsilon>0$, $c>0$, $\delta>0$ be real numbers and let $\left(a_{ij}(t)\right)$ be a repeated game with error
$c\epsilon$ played by $\epsilon$-Hannan consistent players (using the same algorithm $A$). Denote by $t_{1}$ the time needed for algorithm
$A$ to have average regret $r(t)$ bounded by $\epsilon$ for every $t\geq t_{1}$ with probability at least $1-\delta$, by $t_{2}$ the time such that $\left|a_{ij}(t)-a_{ij}\right|<c\epsilon$ for every $t\geq t_{2}$.

Set $t_{0}=\max\left\{ t_{1},\epsilon^{-1}t_{2}\right\} $. Then with probability at least $1-2\delta$ for each $t\geq t_{0}$ we have
\begin{equation*}
v-\left(c+2\right)\epsilon\leq g\left(t\right)\leq v+\left(c+2\right)\epsilon
\end{equation*}
and no player can gain more than $2\left( c+2\right) \epsilon$ utility by deviating from strategy $\hat{\sigma}\left(t\right)$ in the matrix game $M$.
\end{lemma}

\begin{remark}
The proof is a simple modification of Proposition \ref{Prop: hry s chybou}.
The difference between this lemma and Proposition \ref{Prop: hry s chybou} is
the introduction of $\delta$ and $t_{0}$. The important part
of the claim is the value of the constant $t_{0}$.
\end{remark}
\begin{proof}
As in equation \eqref{eq:gmax-tildegmax} of the mentioned proof, we get for $t\geq t_2$
\begin{equation*}
\left|g_{\max}(t)-\tilde{g}_{\max}(t) \right| \leq \underset{i}{\max}\,\left|\frac{1}{t}\underset{s=1}{\overset{t}{\sum}}\left(a_{ij(s)}-a_{ij(s)}(s)\right)\right|\leq 1 \cdot \frac{t_{2}}{t}+c\epsilon\cdot\frac{t-t_{2}}{t}.
\end{equation*}
If we now take $t\geq t_0$, we have $t\geq \epsilon^{-1}t_2$, and thus
\begin{equation*}
\left|g_{\max}(t)-\tilde{g}_{\max}(t) \right| \leq\frac{t_{2}}{\epsilon^{-1}t_2}+c\epsilon\cdot\frac{t}{t}=(c+1)\epsilon.
\end{equation*}
Next step is  the same as in the previous proof, except that we replace $c\epsilon$ by $(c+1)\epsilon$. This leads to the following analogy of the inequality \eqref{eq: HC a NE2}, which holds with probability at least $1-2\delta$:
\begin{equation*}
\forall t \geq t_1:\ v-\left(c+2\right)\epsilon\leq g\left(t\right)\leq v+\left(c+2\right)\epsilon.
\end{equation*}
Remainder of the proof also proceeds as before, thus we only present the main steps: Assume that
\begin{equation}
u\left(br,\hat{\sigma}_{2}\left(t\right)\right)\geq v+2\left(c+2\right)\epsilon+\eta \label{eq: A}
\end{equation}
 holds for some $t\geq t_{0} \geq t_1$ and $\eta>0$. We get the inequality
 \[
 g_{\max}\left(t\right)\geq v+\left(c+2\right)\epsilon+\epsilon+\eta,
 \]
which we combine with inequality
\[
g(t)\leq v+(c+2)\epsilon
\]
in order to get 
\begin{equation}
r\left(t\right) = g_{\max}\left(t\right) - g\left(t\right) \geq \left[ v+\left(c+2\right)\epsilon+\epsilon+\eta \right] - \left[ v+(c+2)\epsilon \right] = \epsilon + \eta > \epsilon. \label{eq: B}
\end{equation}
Since both players are $\epsilon$-HC and $t\geq t_1$, we know that the inequality \eqref{eq: B} (or its analogy for the second player) cannot hold with probability higher than $2\delta$, and thus neither can the inequality \eqref{eq: A} (or its analogy for the second player). This concludes the proof.
\end{proof}

In the proof of Theorem  \ref{thm:Bound}, we will need to guarantee that each of nodes in depth $d$ of game tree gets visited at least $T$-times with high probability, for some $T\in\mathbb N$. Because of this, we include the following technical notation, and related Lemma \ref{lemma:t(T,d)}. Firstly, we present the notation and the corresponding lemma, then we proceed to give the necessary details.
\begin{notation}
Let $\delta,\gamma>0$ and $b\in\mathbb N$. For integers $T$ and $d$, we denote by $t(T,d)$ the smallest number, such that if algorithm with exploration rate $\gamma$ is used in a game with branching factor $b$ for $t(T,d)$ iterations, then with probability at least $1-\delta$, each of the nodes in depth $d$ will be visited at least $T$-times.
\end{notation}
\begin{lemma}\label{lemma:t(T,d)}
Let $\delta,\ \gamma>0$ and $b\in\mathbb N$ and assume that $T\in\mathbb N$ satisfies $T\geq4\log\delta^{-1}+4\log2$. Then we have
\[
t(T,d)\leq16\left(\frac{1}{\gamma}\right)^{d-1}\log2b^{d-1}\cdot b^{d-1}T.
\]
\end{lemma}
\begin{remark}
Consider a game with branching factor $b$ and depth $D$ and an algorithm $A$ which explores with probability $\gamma>0$. For the purposes of this section, we will say that the root is in depth $d=1$, there are at most $b$ nodes in depth $d=2$, $b^{d-1}$ nodes in depth $d$, $d\leq D$ and the whole game tree $\mathcal{H}$ has got at most $\left|\mathcal{H}\right|=1+b+...b^{D-1}$ nodes. For simplicity
we assume that the game tree is already built. At each iteration, there is probability $\gamma^{d-1}$ that the algorithm $A$ will explore on every level of
the game tree between the root and $d$-th level. This means that with probability $\gamma^{d-1}$ one of the nodes on $d$-th level will be chosen with respect to the uniform distribution over all of these $b^{d-1}$ nodes.

Let $n,\ t \in \mathbb N$. We denote by $U(n)$ the uniform distribution over the set $\{1,...,n\}$, let $X_{s}\overset{\textrm{iid}}{\sim}U\left(n\right)$ for $s=1,...,t$ be independent random variables with the same distribution as $U\left(n\right)$ and set $mU(n,t)=\underset{1\leq i\leq n}{\min}\sum_{s=1}^{t}\mathbb I \left( X_{s}=i\right)$. Moreover, let $S\in\mathbb N$ and consider $S$ independent random choices whether to explore with probability $\gamma^{d-1}$ (or else play accordingly to some other strategy with probability $1-\gamma^{d-1}$). Out of these $S$ iterations, we will choose to explore $\tilde{S}$ many times. Using this notation, the number $t(T,d)$ satisfies
\[ t(T,d)=\min \left\{ S\in\mathbb N|\ \mathbf{Pr}[mU(b^{d-1},\tilde{S})\geq T]\geq1-\delta. \right\}. \]
 \end{remark}

\begin{proof}[Proof of Lemma \ref{lemma:t(T,d)}]
Denote $\tilde{T}=8\log\left(2n\right)nT$. Firstly we show that
$\mathbf{Pr}[mU(n,\tilde{T})\geq T]\geq1-\delta/2$, then we follow
with the inequality $\mathbf{Pr}\left[\tilde{S}\geq\frac{1}{2}\gamma^{d-1}S\right]\geq1-\delta/2$.
Putting these two inequalities together with $n=b^{d-1}$, $S=2\left(\frac{1}{\gamma}\right)^{d-1}\tilde{T}$
gives the result.

\noindent \textbf{Step 1:} For the first part, we find $u\in\mathbb{N}$, such that $p=\mathbf{Pr}[mU(n,u)\geq1]\geq\frac{1}{2}$.
Using elementary combinatorics we get that $1-p=\mathbf{Pr}[mU(n,u)=0]\leq1-n\cdot\left(\frac{n-1}{n}\right)^{u}$,
and thus 
\begin{eqnarray*}
p\geq\frac{1}{2} & \iff & n\cdot\left(\frac{n-1}{n}\right)^{u}\geq\frac{1}{2}\\
 & \iff & \exp\left(u\log\frac{n-1}{n}\right)\geq\frac{1}{2n}\\
 & \iff & u\log\frac{n-1}{n}\geq\log\frac{1}{2n}\\
 & \iff & u\geq\log2n\cdot\log\left(1+\frac{1}{n-1}\right)\\
 & \Longleftarrow & u\geq\log2n\cdot\frac{1}{\frac{1}{2}\cdot\frac{1}{n-1}}\\
 & \Longleftarrow & u\geq2n\log2n.
\end{eqnarray*}

Dividing $\tilde{T}$ into $4T$ blocks of length $2n\log2n$ gives
us $4T$ independent ``trials'' with success probability at least
$\frac{1}{2}$, where success increases of $mU$ by at least one in
given block. This means that $\mathbf{Pr}[mU(n,\tilde{T})\geq T]\geq\mathbf{Pr}[B(4T,\frac{1}{2})\geq T]$,
where $B(\cdot,\cdot)$ denotes binomial distribution. Chernoff bound for binomial distribution gives us 
\[
\mathbf{Pr}[B(4T,\frac{1}{2})\geq T]\geq1-\exp\left(-\frac{1}{2\cdot\frac{1}{2}}\frac{\left(2T-T\right)^{2}}{4T}\right)=1-\exp\left(-T/4\right).
\]
Therefore, the conclusion holds for integers $T\in\mathbb N$ satisfying $\exp\left(-T/4\right)\leq\delta/2$, which is equivalent to $T\geq4\log\delta^{-1}+4\log2$.

\noindent \textbf{Step 2:} For the second part, we know that $\tilde{S}\sim B(S,\gamma^{d-1})$ is a binomially distributed variable. Using Chernoff bound, we have (writing $p=\gamma^{d-1}$) 
\begin{eqnarray*}
\mathbf{Pr}[\tilde{S}\geq\frac{p}{2}S] & = & 1-\mathbf{Pr}\left[B(S,p) \leq \frac{p}{2}S\right] \\
& \geq & 1-\exp\left(-\frac{1}{2p}\cdot\frac{\left(pS-\frac{p}{2}S\right)^{2}}{S}\right) \\
& = & 1-\exp\left(-\frac{\gamma^{d-1}}{8}S\right).
\end{eqnarray*}
 Choosing $S\geq2\left(\frac{1}{\gamma}\right)^{d-1}\tilde{T}$, we
get both $\frac{p}{2}S\geq\tilde{T}$ and 
\[
1-\exp\left(-\frac{\gamma^{d-1}}{8}S\right)\geq1-\exp\left(-\tilde{T}/4\right)\geq1-\exp\left(-T/4\right)\geq1-\delta/2.
\]

\end{proof}

We are now ready to give the following proof:
\begin{proof}[Proof of Theorem \ref{thm:Bound}]
Fix $\delta>0$ and let $T_{A}$ denote the time needed for $A$
to have average regret smaller than $2\epsilon$' with probability
at least $1-\delta$.

Set $T_{1}=T_{A}$. For a fixed node $h$ in depth $D$, we have by Lemma \ref{lemma: Variation on L4.9} that the inequalities $v_{h}-2\epsilon\leq g_{h}(t)\leq v_{h}+2\epsilon$ (since $d_h=D$, we can represent the situation at $h$ as a repeated game with error $c\epsilon=0$ and $t_{2}=0$) hold for $t\geq T_{1}=\max\left\{ T_{1},\epsilon^{-1}\cdot0\right\} $
with probability at least $1-2\delta$. However, we need these inequalities to hold for all $b^{D-1}$ nodes in depth $D$ at once. To guarantee that we visit all of them at least $T_{1}$ many times with probability at least $1-2\delta$, we need $t\geq T_{2}$, where $T_{2}=t(\epsilon^{-1}T_{1},D)$. Taking product of all these probabilities, we get that all of the mentioned conditions hold at the same time with probability at least $\left(1-2\delta\right)^{b^{D-1}}(1-\delta)$. We are done with $D$-th level of the game tree.

Assume that everything above holds. Note that surely $T_{2}\geq T_{A}$.
By Lemma \ref{lemma: Variation on L4.9} ($t_{0}=\max\left\{ T_{A},\epsilon^{-1}T_{2}\right\} =\epsilon^{-1}T_{2}$,
$c=1$) we know that if we visit a node $h$ in depth $D-1$ at
least $\epsilon^{-1}T_{2}$ many times, the inequalities $v_{h}-4\epsilon\leq g_{h}(t)\leq v_{h}+4\epsilon$
will hold with probability at least  $1-2\delta$.
Again we find such a number $T_{3}$ that will guarantee that we
visit each of these $b^{D-2}$ nodes at least $\epsilon^{-1}T_{2}$
many times with probability at least $1-\delta$ (as in paragraph
above $T_{3}=t(\epsilon^{-1}T_{2},D-2)$). The probability of all
conditions mentioned in this paragraph being satisfied at once will
be $\left(1-2\delta\right)^{b^{D-2}}(1-\delta)$.

We continue by induction and receive numbers $T_{4},\, T_{5},...,T_{D}$, which satisfy $T_{d+1}=t(\epsilon^{-1}T_{d},d)$. We now calculate the
exact numbers for probability, equilibrium distance and time bounds:

Taking product of probabilities $\left(1-2\delta\right)^{b^{d-1}}(1-\delta)$
for $d=1,...,D$ gives us 
\[
\left(1-2\delta\right)^{2\left(b^{D-1}+...+b^{2}+b+1\right)}(1-\delta)^D=\left(1-2\delta\right)^{2\left|\mathcal{H}\right|}(1-\delta)^D\geq 1-\left(2\left|\mathcal{H}\right|+D\right)\delta.
\]
On $\left(D+1-d\right)$-th level of the tree, we have $v_{h}-\left(c_{d}+2\right)\epsilon\leq g_{h}(t)\leq v_{h}+\left(c_{d}+2\right)\epsilon$, where $c_{d}=2d-2$ (easily checked by induction, as $c_{d+1}=c_{d}+2$ and $c_{1}=0$). Lemma \ref{lemma: Variation on L4.9} implies that no player can gain more than $2\left(c_{d}+2\right)\epsilon=4d\epsilon$ by changing his action on given level only. Note that these possible deviations sum to 
\begin{equation}
\sum_{d=1}^{D}4d\epsilon=2D\left(D+1 \right)\epsilon. \label{eq: sum}
\end{equation}

Finally we calculate the value of $T_{D}$ by substituting from Lemma \ref{lemma:t(T,d)}:
\begin{align*}
T_{D}= & T_{A}\cdot\prod_{d=2}^{D}\left(\epsilon^{-1}\cdot16\left(\frac{1}{\gamma}\right)^{d-1}\log2b^{d-1}\cdot b^{d-1}\right)\\
= & T_{A}\cdot\left(16\epsilon^{-1}\right)^{D-1}\left(\frac{b}{\gamma}\right)^{\frac{D-1}{2}\left(D-1+1\right)}\log\left(2b+...+2b^{D-1}\right)\\
= & T_{A}16^{D-1}\epsilon^{-\left(D-1\right)}\left(\frac{b}{\gamma}\right)^{\frac{D}{2}\left(D-1\right)}\log\left(2\left|\mathcal{H}\right|-2\right).
\end{align*}

With probability at least $1-\left(2\left|\mathcal{H}\right|+D\right)\delta$ we get (by \eqref{eq: sum}) that for every $t\geq T_{D}$ no player can gain more than $2D\left(D+1 \right)\epsilon$ utility by choosing a different action in any (or all at once) node of the tree. Multiplying this result by two, we get that the empirical frequencies will form $4D\left(D+1 \right)\epsilon$-equilibrium for every $t\geq T_{D}$ with probability at least $1-\left(2\left|\mathcal{H}\right|+D\right)\delta$ - the proof is finished.

Note that in the proof above, we assumed that SM-MCTS-A algorithm starts when the game tree is already built. If we wanted to start with an empty game tree, we would have to increase the time bound by additional constant, such that with high probability, we first visit all $b$ nodes one level below the root at least once, then we visit all $b^2$ nodes in the next level, and so on. By Lemma~\ref{lemma:t(T,d)} this constant would be equal to $t(1,1)+...+t(1,D)$, a number which is negligible when compared to other parts of the finite time bound.
\end{proof}

\end{document}